\documentclass[a4paper,fleqn,usenatbib]{mnras}

\usepackage{newtxtext,newtxmath}

\usepackage[T1]{fontenc}
\usepackage{ae,aecompl}

\usepackage{graphicx}	
\usepackage{amsmath}	
\usepackage{amssymb}	
\usepackage{booktabs}
\usepackage{subfig}
\usepackage{enumitem}
\usepackage{lscape}
\usepackage{adjustbox}
\usepackage{rotating}
\usepackage{multicol}
\usepackage[dvipsnames]{xcolor}

\title[Ionised gas in SBS 1415+437]{Small-scale chemical abundance analysis in a blue compact dwarf galaxy SBS 1415+437}

\author[Kumari et al.]{
Nimisha Kumari$^{1}$\thanks{E-mail: nkumari@ast.cam.ac.uk(NK)},
Bethan L. James$^{2}$,
Mike J. Irwin$^{1}$,
Alessandra Aloisi$^{2}$\\
$^{1}$Institute of Astronomy, University of Cambridge, CB3 0HA UK \\
$^{2}$Space Telescope Science Institute, 3700 San Martin Dr, Baltimore, MD 21218}

\date{Accepted XXX. Received YYY; in original form ZZZ}

\pubyear{2015}

\newcommand{\pGuseva}{\textcolor{blue}{(G03)~}}
\newcommand{\tGuseva}{\textcolor{blue}{G03}}	
\newcommand{\tPerezMontero}{\textcolor{blue}{P17}}

\begin{document}

\label{firstpage}
\pagerange{\pageref{firstpage}--\pageref{lastpage}}
\maketitle

\begin{abstract}
We use integral field spectroscopic (IFS) observations from Gemini Multi-Object Spectrograph-North (GMOS-N) to analyse the ionised gas in the principal star-forming region in the blue compact dwarf galaxy SBS 1415+437. The IFS data enable us to map the weak auroral line [O \textsc{iii}] $\lambda$4363 at a spatial scale of $\sim$6.5 pc across a region of $\sim$143 $\times$ 143 pc$^2$. This in turn allows us to use the robust direct T$_e$-method to map the ionic and elemental abundances of nitrogen (N) along with the alpha-elements, oxygen (O), neon (Ne), sulphur (S) and argon (Ar). We utilise these abundances to map the relative abundances of N, Ne, S and Ar with respect to O. We segment this predominantly photoionised region of study into elliptical annuli on the basis of the H$\alpha$ flux distribution to study the variation of chemical abundances and their ratios, and find no significant chemical variation. We also perform chemical abundance analysis on the integrated spectra of the region under study and elliptical annuli within it. We find that the inferred abundances are in agreement with the median of the abundances obtained from the chemical abundance maps of the principal star-forming region and the mapped values within annuli.The finding has important implications for direct comparison with high-redshift observations, where spatial resolution is not available, and for a consistent approach to track chemical evolution across cosmic time.    
\end{abstract}

\begin{keywords}
galaxies: individual: SBS 1415+437 -- galaxies: dwarfs -- stars: formation -- galaxies: abundances -- ISM: HII regions
\end{keywords}



\section{Introduction}

\indent Our understanding of the chemical evolution of the Universe relies on our knowledge of the origin and distribution of elements in the nearby and distant Universe. The study of abundance of oxygen (O) and its distribution is important because it is the third most abundant element in the Universe after hydrogen and helium, and hence acts as a good proxy to the total metal-content of a star-forming system. The abundances of other elements and their relative abundance with respect to oxygen tell us about the nucleosynthetic origin of these elements. For example, $\alpha$-elements such as neon (Ne), sulphur (S) and argon (Ar) are thought to be produced in massive stars along with oxygen \citep[see e.g.][]{WoosleyWeaver1995}, while nitrogen (N) is thought to form in massive, as well as intermediate mass stars, though there have been many debates on its origin \citep[see e.g.][and references therein]{Kumari2018}. Various observational studies have explored the distribution of these elements in stars and gas within star-forming and quiescent systems \citep[see e.g.][]{Vilchez1998, Henry1999, Lopez2007, Garcia-Benito2010, Monreal-Ibero2011, Lind2011, Berg2013, Lopez-Hernandez2013, Berg2015}. These studies show that the distribution of chemical abundances may vary over both small sub-galactic scales or large galactic scales within galaxies \citep[see e.g.][]{Henry1999, Kewley2010, Pilkington2012, James2015}. The abundance pattern tells us about various constituents of chemical evolutionary models such as star-formation history, chemical enrichment, merger events, gas dynamics and stellar-populations.

\begin{table}
	\centering
	\caption{General Properties of SBS 1415+437}
	\label{tab properties}
	\begin{tabular}{@{}lc@{}}
		\toprule
		Parameter                    & SBS 1415+437          \\  \midrule
		Morphological Type           & BCD             \\
		R.A. (J2000.0)               & 14h17m01.4s      \\
		DEC (J2000.0)                & +43d30m05s       \\
		Redshift (z)$^a$                 & $\sim$0.00203 \\
		Distance (Mpc)$^b$              & 13.6      \\
		Helio. Radial Velocity(km s$^{-1}$)$^a$ & 609 $\pm$    2           \\
		E(B-V)$^c$                       & 0.0077 $\pm$  0.0003    \\
		M$_{\rm{stellar}}$ (10$^7$ M$_{\odot}$)$^d$ & 17 $\pm$ 3 \\
		M$_{\rm{molecular}}$ (10$^7$ M$_{\odot}$)$^d$ & 7.6 $\pm$ 2.3 \\
		M$_{\rm{H \textsc{i}}}$ (10$^7$ M$_{\odot}$)$^d$ & 6.8 $\pm$ 0.7 \\
		\bottomrule
		\multicolumn{2}{l}{$^a$ Taken from NED}\\
		\multicolumn{2}{l}{$^b$ \citet{Aloisi2005}}\\
		\multicolumn{2}{l}{$^c$ Foreground galactic extinction \citep{Schlafly2011}}\\
		\multicolumn{2}{l}{$^d$ \citet{Lelli2014}}\\
	\end{tabular}
\end{table}

\begin{table*}
	\centering
	\caption{GMOS-N IFU observing log for SBS 1415+437}
	\label{log}
	\begin{tabular}{@{}cccccc@{}}
		\toprule
		Grating      & \begin{tabular}[c]{@{}c@{}}Central wavelength\\ (\AA)\end{tabular} & \begin{tabular}[c]{@{}c@{}}Wavelength Range\\ (\AA)\end{tabular} & \begin{tabular}[c]{@{}c@{}}Exposure Time\\ (s)\end{tabular} & Average Airmass & Standard Star \\ \midrule
		B600+\_G5307 & 4700                                                               &    3250 -- 6118                                          & 2$\times$1850                                                        &  1.098, 1.092   & Wolf1346      \\
		R600+\_G5304 & 6900                                                               &    5443 -- 8361                                           & 2$\times$1800                                                        & 1.161, 1.121           & Wolf1346      \\
		\bottomrule
	\end{tabular}
\end{table*}

\indent  Studies of large and small-scale chemical variation have been revolutionised with the advent of integral field spectroscopy (IFS) which allows us to derive spatially-resolved spectroscopic properties of the star-forming systems \citep{Perez-Montero2011, Sanchez2012, Westmoquette2013,  Kehrig2013, Ho2015, Belfiore2017}. For example the spectrum of a star-forming galaxy, or an H \textsc{ii} region, contains a plethora of emission lines. IFS enables us to map emission lines across the star-forming galaxy, or region within a galaxy, depending on the field-of-view (FOV) of instruments, and therefore map the physical properties encoded in those emission lines. This simultaneous observation in wavelength and space is key to an efficient study of the distribution of chemical abundances within a star-forming system \citep{James2009, James2010, James2013a, James2013b, Lagos2012, Lagos2014, Lagos2016, Kumari2018}.

\indent In this paper, we explore small-scale chemical variation within a blue compact dwarf (BCD) galaxy SBS1415+437 by using IFS observations obtained from the integral field unit (IFU) installed on the Gemini Multi-Object Spectrograph (GMOS; \citet{Hook2004}). This is the first ever IFS study of SBS1415+437, whose general properties are tabulated in Table \ref{tab properties}. SBS1415+437 is a suitable target to study the small-scale chemical abundance variation for several reasons. The oxygen abundance derived previously from long-slit observations of this galaxy shows that its metallicity is $\sim$ Z$_{\odot}$/21 \citep[]{Guseva2003, Thuan1999}. Such a low-metallicity ensures the detection of weak auroral line such as [O \textsc{iii}] $\lambda$4363 with enough signal-to-noise (S/N) to allow us to map the chemical abundance of various elements and their ratios via the robust direct T$_e$-method. Moreover, this galaxy is relatively nearby ($\sim$ 13.6 Mpc), which enables us to map the variation on scales as small as $\sim$ 6.5 pc with the high spatial sampling (0.1\arcsec) of GMOS-IFU. Though this galaxy was initially thought to be quite young with the first burst of star-formation occurring in the last 100 Myr \citep{Thuan1999}, a photometric analysis has shown the presence of asymptotic giant branch and red giant branch stars which are at least $\sim$ 1 Gyr old \citep{Aloisi2005}. Being a BCD at extremely low-metallicity, this galaxy is a local analogue of high-redshift galaxies despite hosting older stellar population, and hence a study of chemical distribution within this galaxy provides insight into the characteristic physical processes of the high-redshift Universe.

\indent This paper is the third in a series of spatially-resolved analyses of star-forming regions in BCDs \citep[see][]{Kumari2017, Kumari2018} where we aim to better understand the physical and chemical properties of such systems at very small scales. The performed analysis is in the framework of the following three questions: (a) Is the ionised gas in a star-forming region chemically homogeneous? (b) What is the main source of excitation of the ionised gas in and around a star-forming region? (c) What is the distribution in age of the population of stars responsible for ionising the gas?

\indent The paper is structured as follows. In Section \ref{section:data}, we present the observations along with a brief discussion of the major steps of data-reduction. In Section \ref{section:results}, we present the results and a discussion of the maps of emission line fluxes, dust attenuation, gas kinematics, electron temperature and density, ionic and elemental abundances of various elements and the abundance ratios and stellar properties. In this section, we also perform a radial profile analysis on the elemental abundance maps and abundance ratio maps to investigate any systematic signatures of chemical inhomogeneities. Section \ref{section:summary} summarises our results.

\begin{figure}
	\centering
	\includegraphics[width=0.45\textwidth]{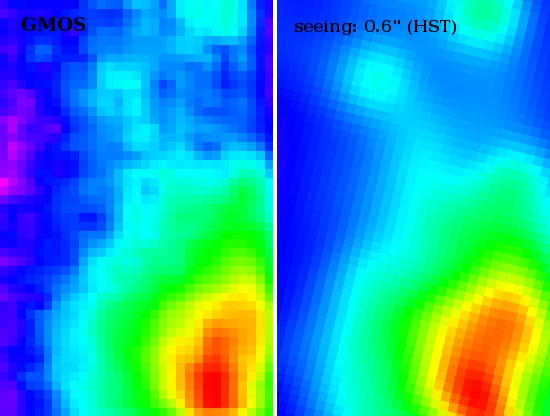}
	\caption{Left: R-band continuum image from GMOS-IFU. Right: HST/ACS (F606W filter) image convolved by a bidimensionsal gaussian corresponding to a seeing FWHM of 0.6 arcsec and binned to the GMOS pixel size of 0.1 arcsec.}
	\label{figure:seeing}
\end{figure}

\begin{figure*}
	\centering
	
	\includegraphics[width = \textwidth]{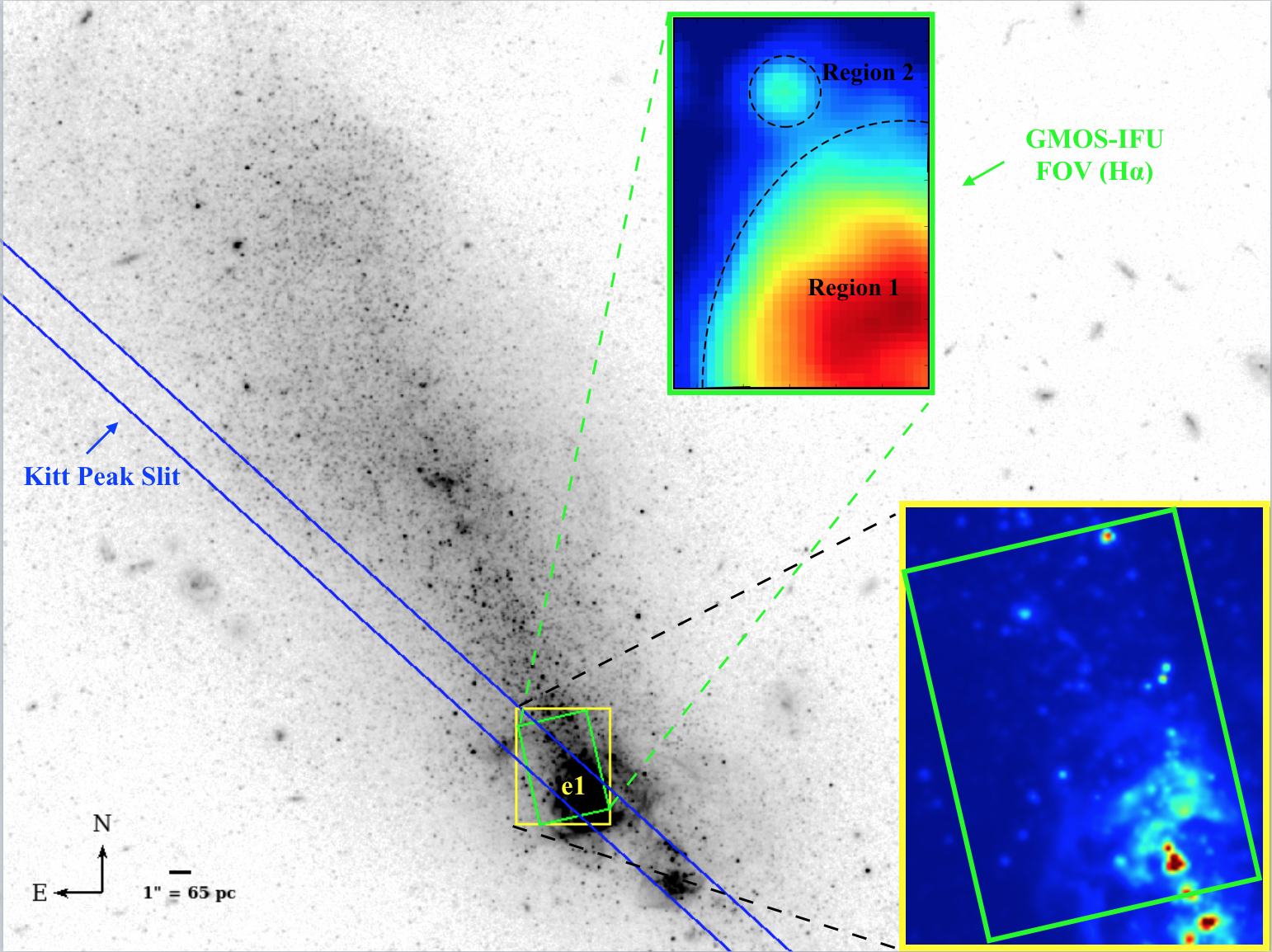}
	\caption{HST/ACS image (0.05 arcsec per pixel) of SBS1415+437 taken in the F606W (V) filter from \citet{Aloisi2005}. The small green rectangular box inside the yellow rectangular box shows the GMOS aperture (3.5 $\times$ 5 arcsec$^2$) centered at (RA, Dec) : (214.2559958, 43.50158528). The lower-right inset denotes the zoomed-in version of the small yellow and green boxes. The upper inset is the H$\alpha$ image created from GMOS-IFU data presented here, where we mark principal emission regions as Region 1 and Region 2. The colour scale and stretch of background and inset HST images are set to be different for better visibility. The blue parallel lines represent the long-slit position of spectroscopic observation with Kitt Peak 4m  Mayall Telescope, used by \citet{Guseva2003} to study various regions, including "e1" which coincides with our GMOS data. The compass on the bottom-left of the figure shows North and East on the HST image. At the distance of this galaxy, 1 arcsec corresponds to 65 pc.}
	\label{fig:hst}
\end{figure*}

\begin{figure*}
	\centering
	\includegraphics[width = \textwidth,trim={3.5cm 0 4.5cm 0}, clip]{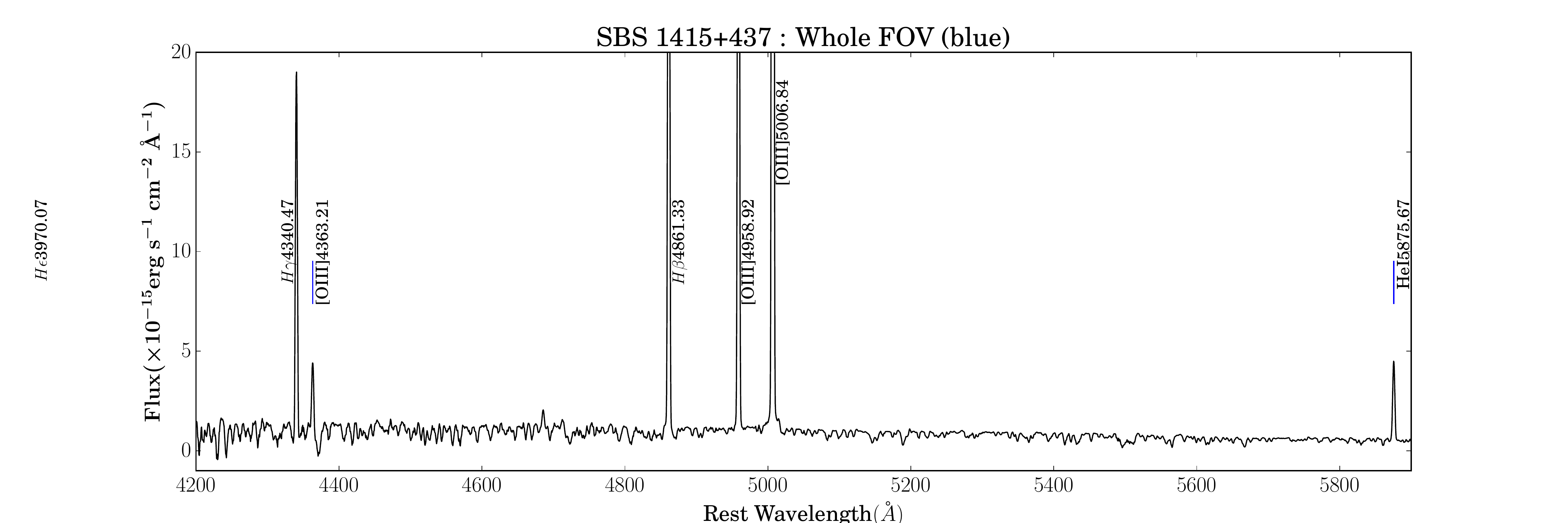}
	\includegraphics[width = \textwidth,trim={3.5cm 0 4.5cm 0}, clip]{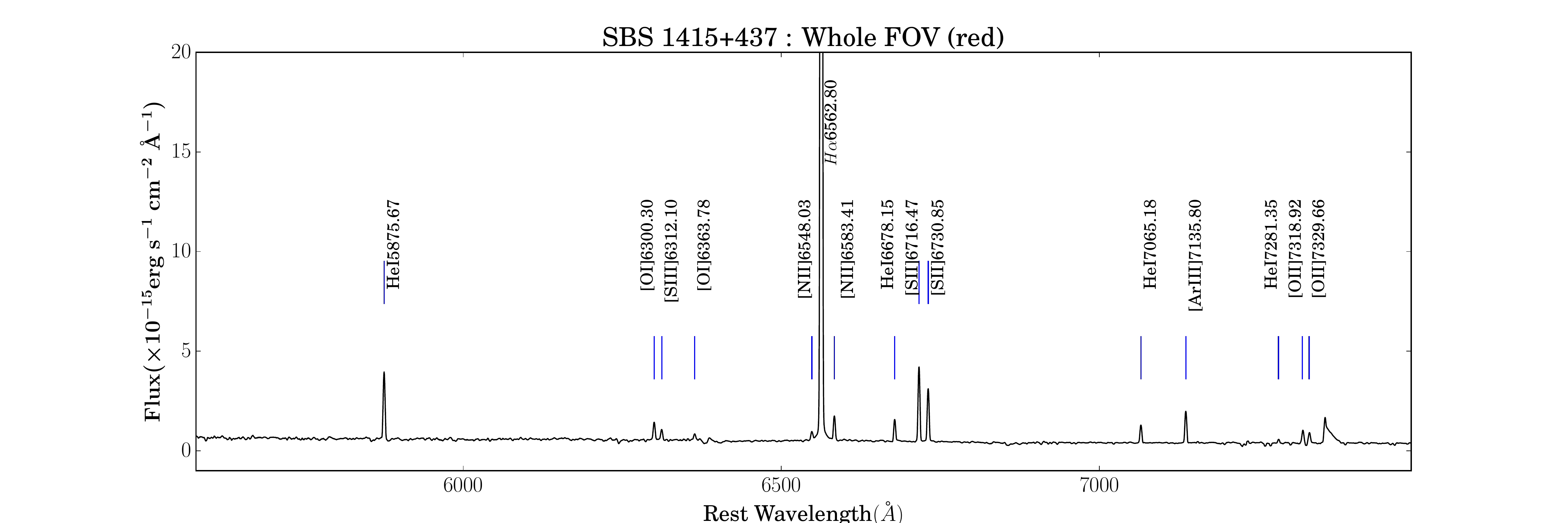}
	\caption{GMOS-IFU integrated spectra of SBS1415+437 integrated over the entire FOV. The principal emission lines are marked as blue lines at their rest wavelengths. The spectra are smoothed using a one-dimensional box kernel with an effective smoothing of 7 pixels.}
	\label{spectra}
\end{figure*}

\begin{figure*}
	\centering
	\includegraphics[width = 0.48\textwidth]{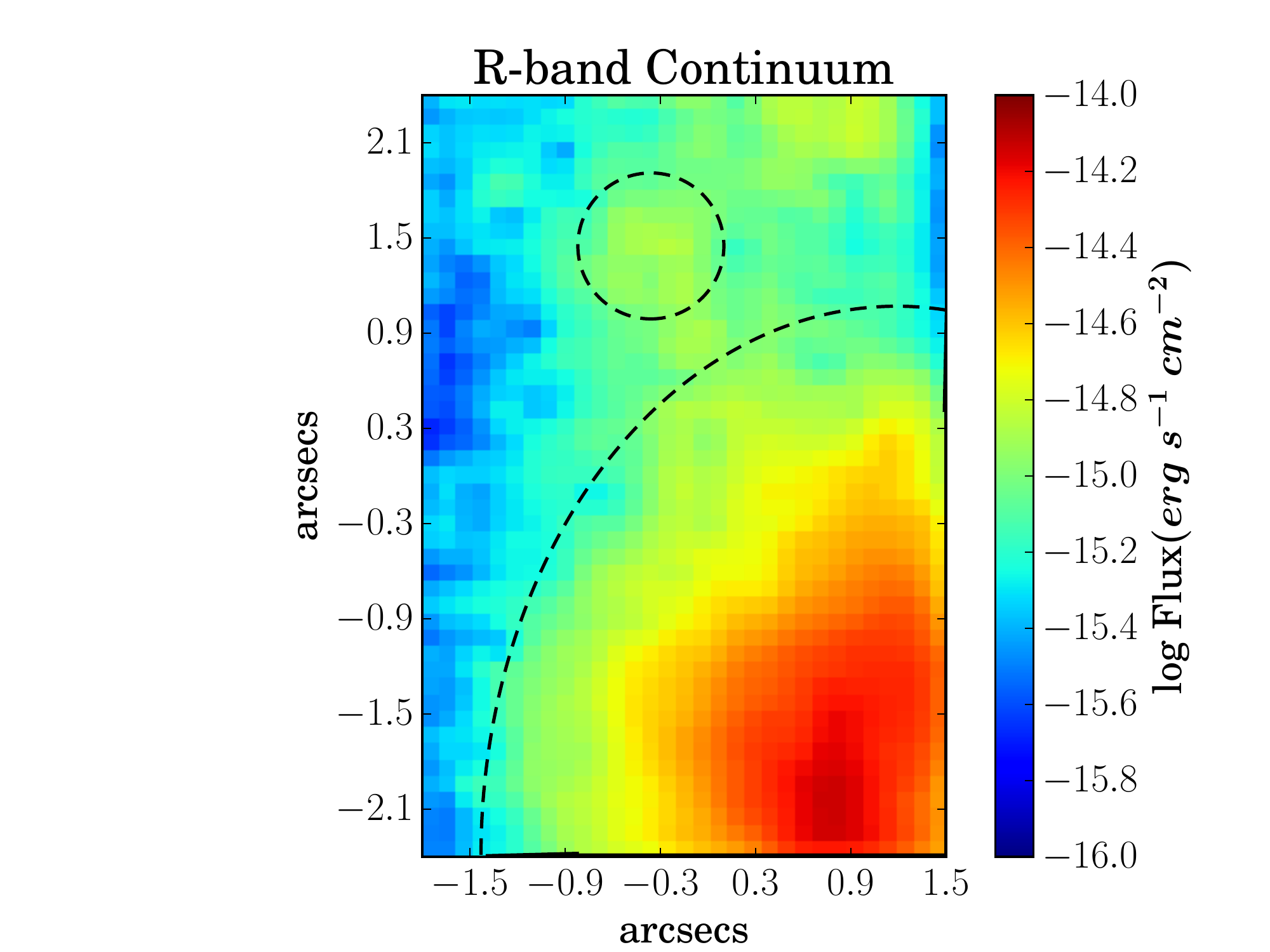}
	\includegraphics[width = 0.48\textwidth]{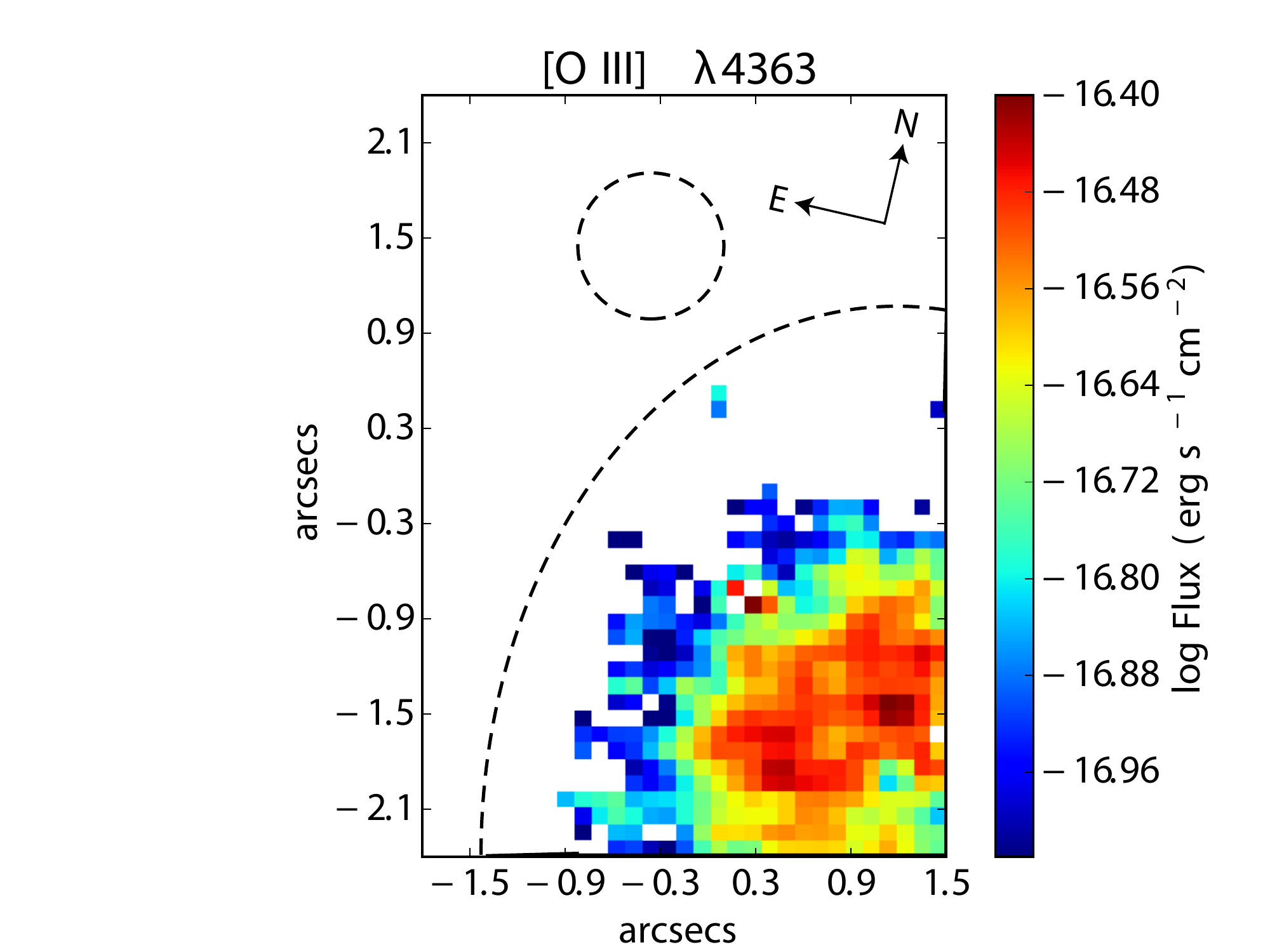}
	\includegraphics[width = 0.48\textwidth]{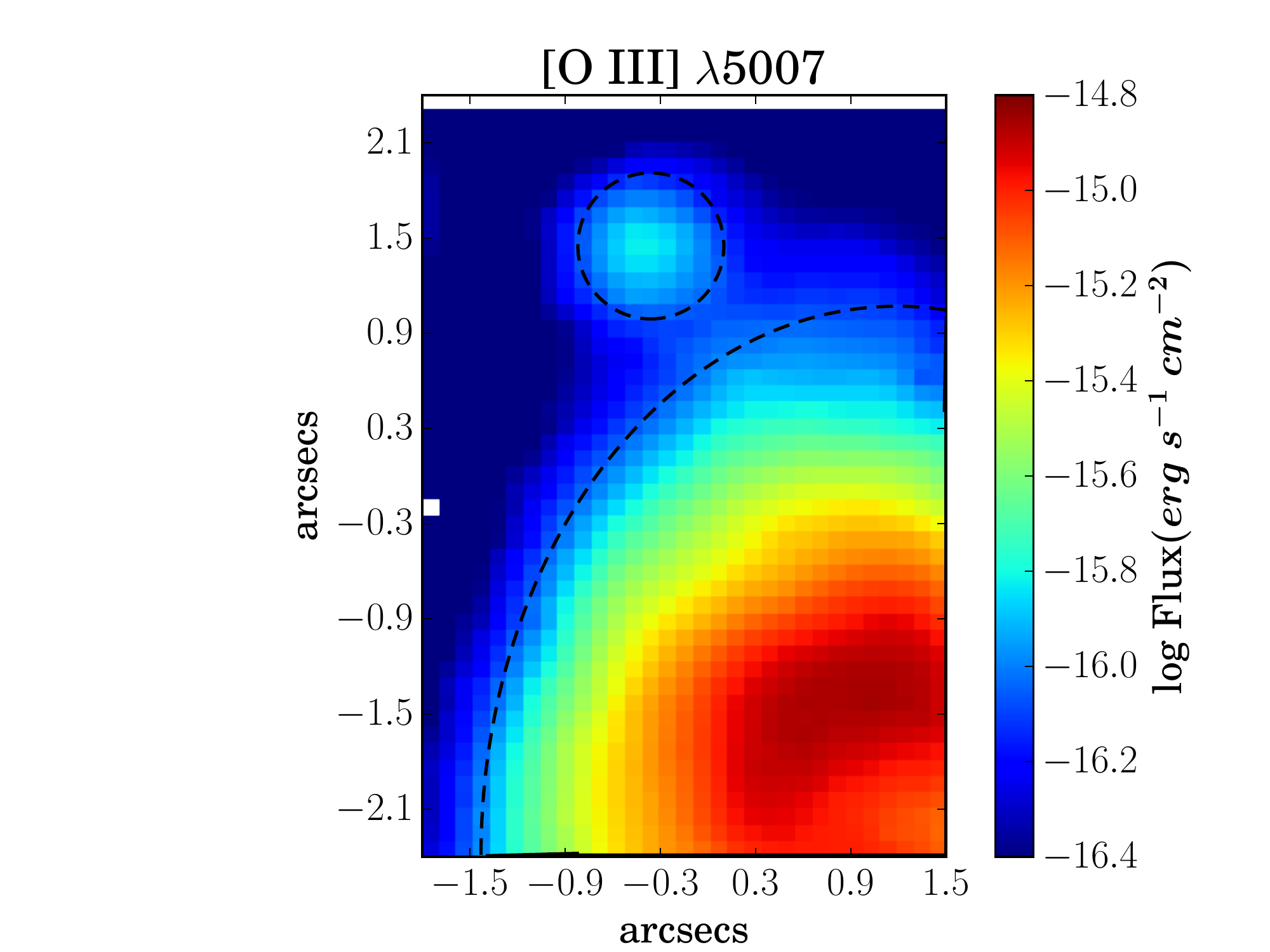}
	\includegraphics[width = 0.48\textwidth]{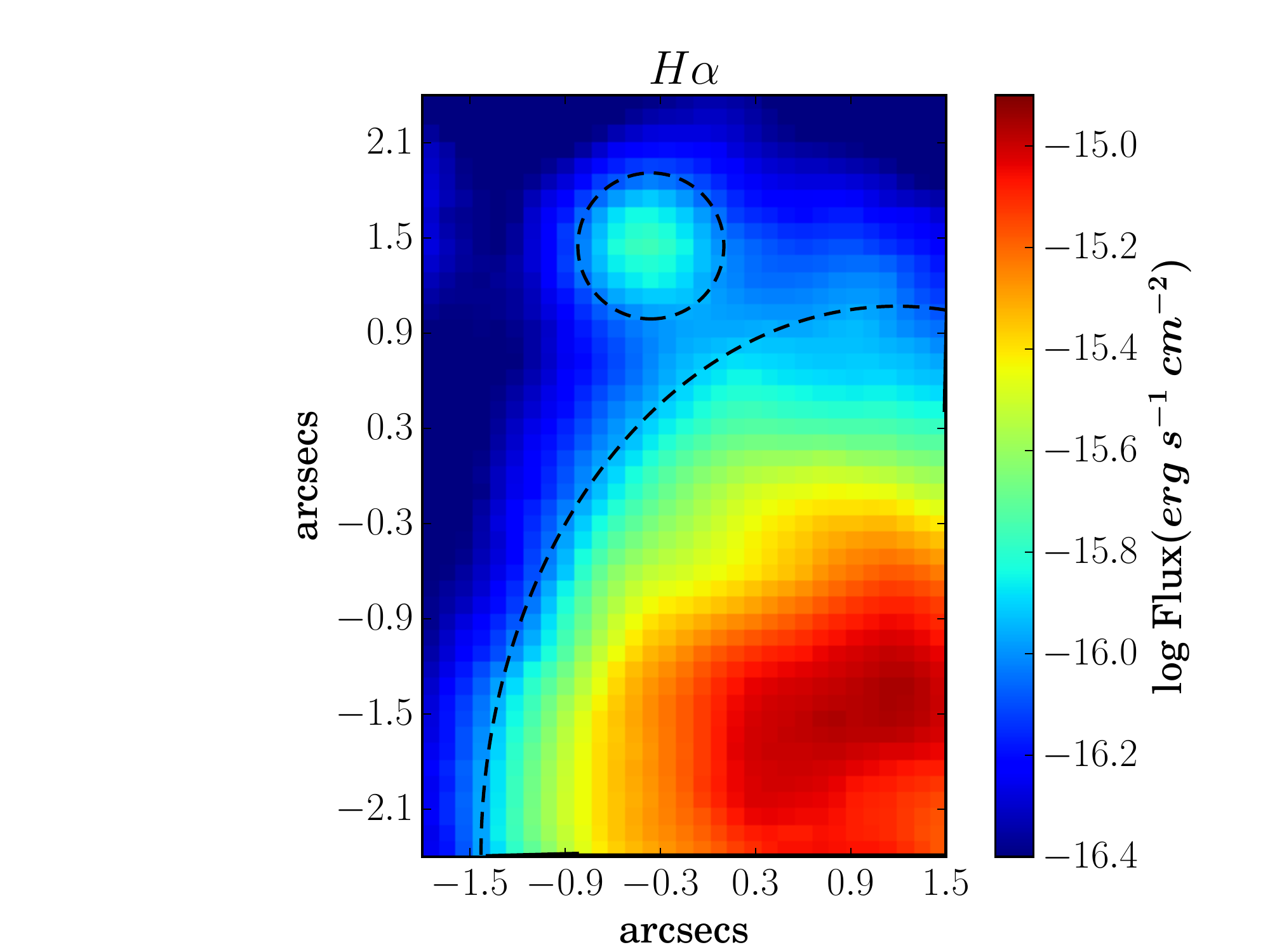}
	\includegraphics[width = 0.48\textwidth]{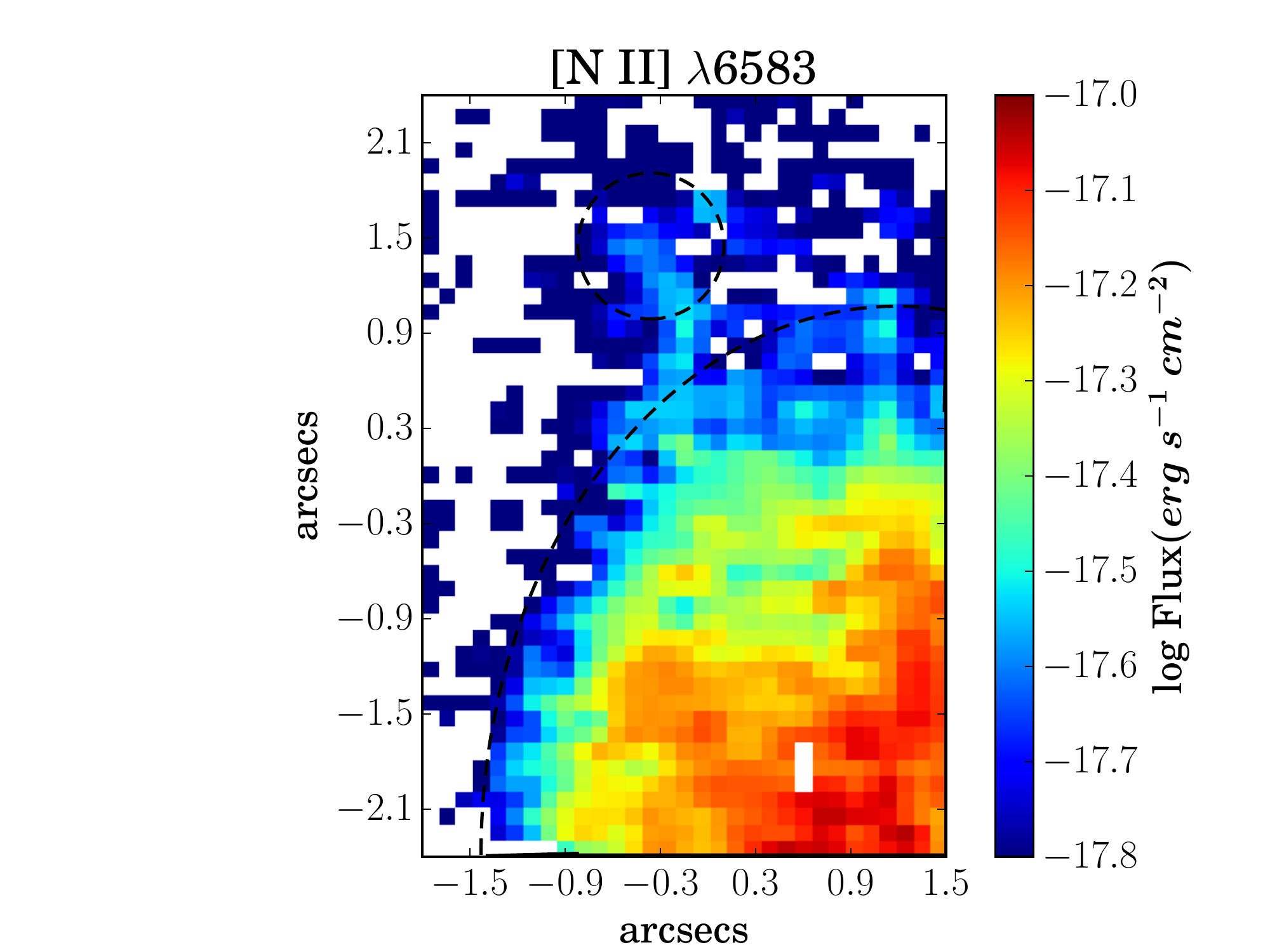}
	\includegraphics[width = 0.48\textwidth]{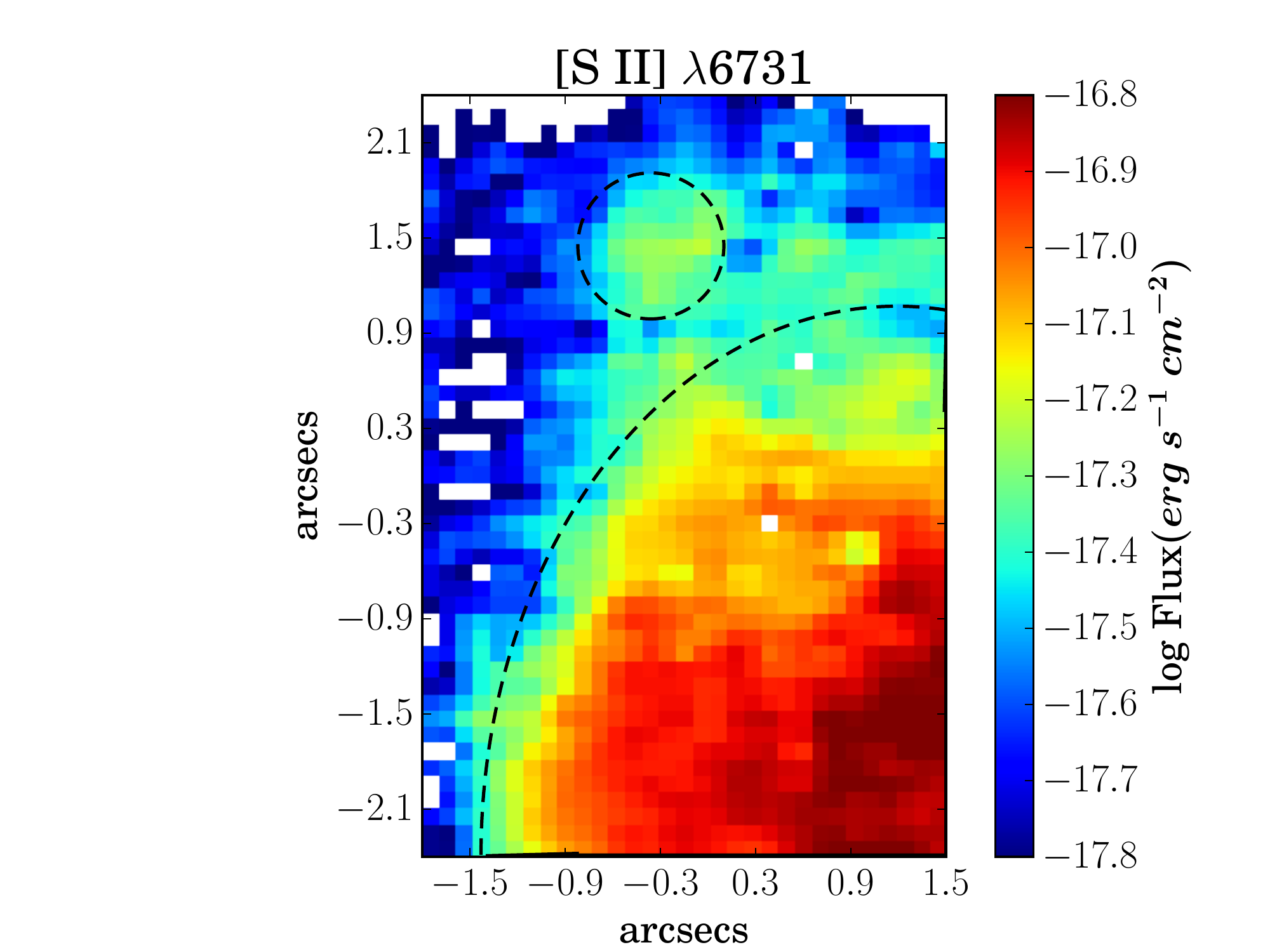}
	\caption{Observed R-band continuum map and the emission line flux maps ([O $\textsc{iii}$]$\lambda$4363,  [O $\textsc{iii}$]$\lambda$5007, H$\alpha$, [N $\textsc{ii}$]$\lambda$6583 and [S$\textsc{ii}$]$\lambda$6731) of SBS 1415+437. The dashed quarter ellipse and circle indicate the H $\textsc{ii}$ regions identified by a relatively higher flux values, and are referred as Region 1 and Region 2, respectively, in the rest of the paper. The spaxels in which emission line fluxes had S/N $<$ 3, are shown in white. }
	\label{fig:observed flux}
\end{figure*}

\section{Observation \& Data Reduction}
\label{section:data}
The target region of SBS1415+437 was observed with GMOS and the IFU unit \citep[GMOS-N IFU;][]{Allington-Smith2002} at Gemini-North telescope in Hawaii. The observation was taken in one-slit queue-mode in 2012 as a part of GMOS-IFU spectroscopy programme (PI: B James) for seven star-forming galaxies. The field-of-view covered in this mode of observation is 3.5\arcsec $\times$ 5\arcsec, and is sampled by 750 hexagonal lenslets of projected diameter of 0.2\arcsec, of which 250 lenslets are used for background determination (see Table \ref{log} for information on data observation). The blue and red regions of the optical spectrum were observed by using gratings B600+\_G5307 and R600+\_G5304. A set of standard observations of GCAL flats, CuAr lamp for wavelength calibration and standard star for flux calibration were taken with each grating. 

\indent We used the standard GEMINI reduction pipeline written in Image Reduction and Analysis Facility (IRAF)\footnote{IRAF is distributed by the National Optical Astronomy Observatory, which is operated by the Association of Universities for Research in Astronomy (AURA) under a cooperative agreement with the National Science Foundation.} to perform the basic steps of data reduction which included bias subtraction, flat-field-correction, wavelength calibration, sky subtraction, differential atmospheric correction and for conversion of observed spectra into three-dimensional data cubes. We found that the standard pipeline did not give satisfactory results for some procedures, for which we developed and implemented our own code \citep[see][]{Kumari2017, Kumari2018}. For both data cubes, we chose a spatial-sampling of 0.1\arcsec which is suitable to preserve the hexagonal sampling of the GMOS-IFU lenslets. We corrected the data cubes corresponding to the two gratings for a spatial-offset of 0.2\arcsec on both spatial-axes, and finally converted them to the row-stacked spectra for further analysis. We estimated the instrumental broadening (Full Width Half Maximum, FWHM $\sim$ 1.7~\AA) for both gratings, by fitting a Gaussian profile to several emission lines of the extracted row stacked spectra of the arc lamp. We estimated the seeing FWHM of $\sim$0.6 arcsec by comparing the R-band continuum image from GMOS-IFU with the convolved and binned image of this galaxy obtained from the Hubble Space Telescope (HST) (Figure \ref{figure:seeing}). The original HST Advanced Camera Surveys (ACS) image was taken in the equivalent of the F606W(V) filter and had a resolution of 0.05 arcsec per pixel (program \# 9361, PI: Aloisi).

\begin{table}
	\centering
	\caption{Emission line measurements (relative to H$\beta$ = 100) for the integrated spectrum of Region 1 (Quarter Ellipse) (see Section \ref{section:fluxes}). Line fluxes ($F_{\lambda}$) are extinction corrected using E(B-V) to calculate $I_{\lambda}$.} 
	\resizebox{0.5\textwidth}{!}{%
		\begin{tabular}{cccc}
			\toprule
			Line & $\lambda_{air}$ & $F_{\lambda}$ (Region 1) & $I_{\lambda}$ (Region 1) \\
			\midrule
			$H\gamma$ & 4340.47 & $47.15 \pm 0.56 $ & $48.36 \pm 1.37 $  \\
			$[OIII]$ & 4363.21 & $8.87 \pm 0.60 $ & $9.09 \pm 0.66 $ \\
			$H\beta$ & 4861.33 & $100.00 \pm 0.47 $ & $100.00 \pm 1.76 $  \\
			$[OIII]$ & 4958.92 & $119.89 \pm 0.80 $ & $119.40 \pm 2.94 $ \\
			$[OIII]$ & 5006.84 & $356.30 \pm 2.29 $ & $354.03 \pm 8.66 $  \\
			$HeI$ & 5875.67 & $10.72 \pm 0.20 $ & $10.30 \pm 0.29 $   \\
			$[OI]$ & 6300.3 & $2.09 \pm 0.06 $ & $1.98 \pm 0.07 $ \\
			$[SIII]$ & 6312.1 & $1.47 \pm 0.05 $ & $1.39 \pm 0.06 $  \\
			$H\alpha$ & 6562.8 & $303.55 \pm 1.62 $ & $286.00 \pm 6.08 $  \\
			$[NII]$ & 6583.41 & $2.92 \pm 0.40 $ & $2.75 \pm 0.38 $  \\
			$HeI$ & 6678.15 & $2.84 \pm 0.05 $ & $2.67 \pm 0.07 $  \\
			$[SII]$ & 6716.47 & $8.59 \pm 0.08 $ & $8.06 \pm 0.18 $  \\
			$[SII]$ & 6730.85 & $6.14 \pm 0.07 $ & $5.76 \pm 0.14 $  \\
			$[ArIII]$ & 7135.8 & $4.14 \pm 0.05 $ & $3.84 \pm 0.09 $  \\
			$[OII]$ & 7318.92 & $1.64 \pm 0.10 $ & $1.51 \pm 0.10 $  \\
			$[OII]$ & 7329.66 & $1.31 \pm 0.09 $ & $1.22 \pm 0.08 $  \\
			\\
			E(B-V)  && 0.057$\pm$0.005 &\\
			F(H$\beta$) && 117.42 $\pm$ 0.55& 141.97 $\pm$ 2.50\\
			\bottomrule
		\end{tabular}%
	}
	Notes: F(H$\beta$) in units of $\times$ 10$^{-15}$ erg cm$^{-2}$ s$^{-1}$\\
	\label{flux table}
\end{table}

\section{Results \& Discussion}
\label{section:results}
\subsection{Observed and Intrinsic Fluxes}
\label{section:fluxes}

\begin{figure}
	\centering
	\includegraphics[width = 0.48\textwidth]{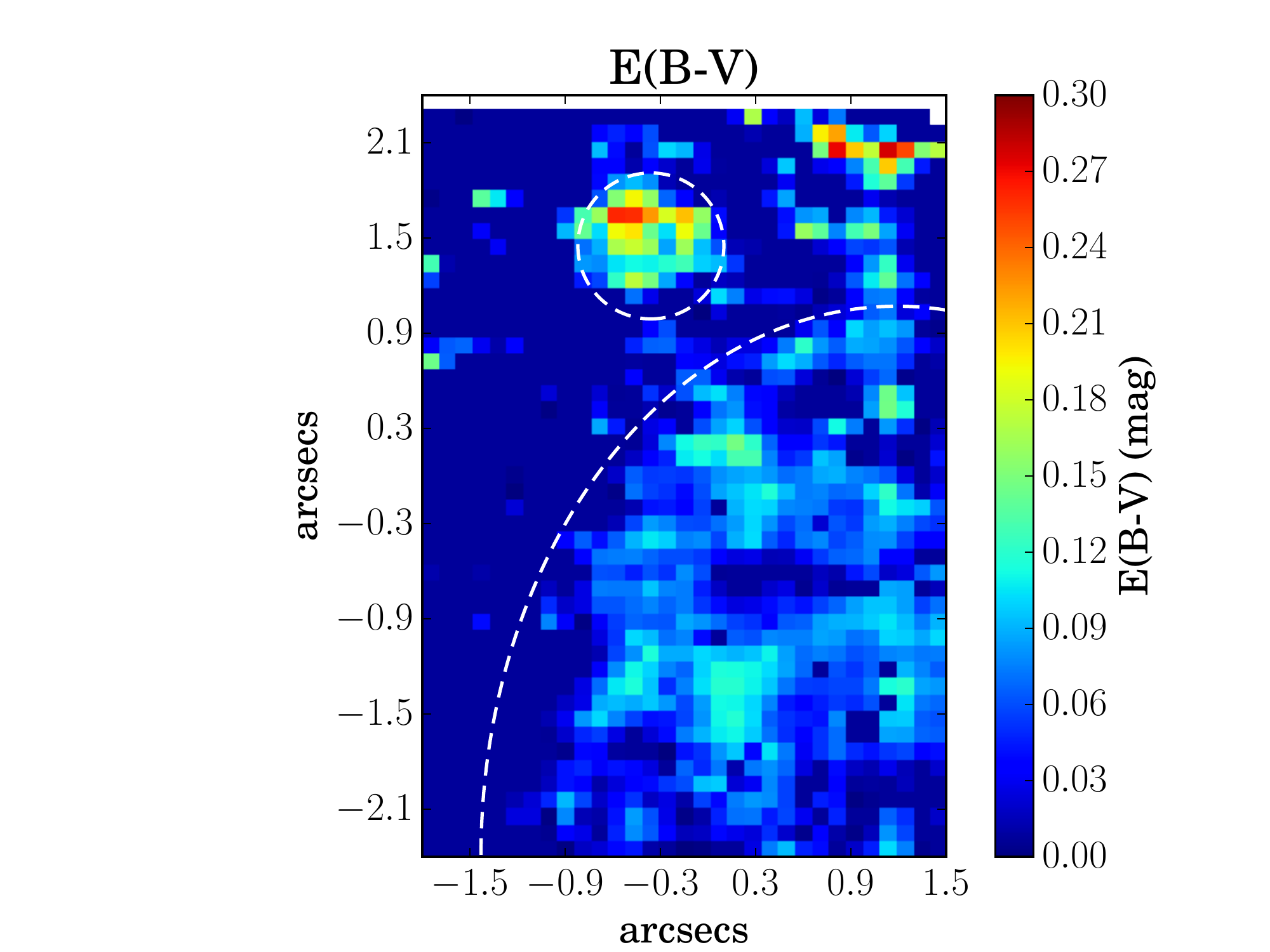}
	\caption{E(B-V) map created assuming the LMC extinction curve. Spaxels with E(B-V) $<$ 0 are set to E(B-V) of the Milky Way foreground along the line-of-sight. The spaxels in which emission line fluxes had S/N $<$ 3, are shown in white. The dashed quarter ellipse and circle indicate Region 1 and Region 2, respectively.}
	\label{fig:ebv}
\end{figure}

\begin{figure}
	\centering
	\includegraphics[width = 0.48\textwidth]{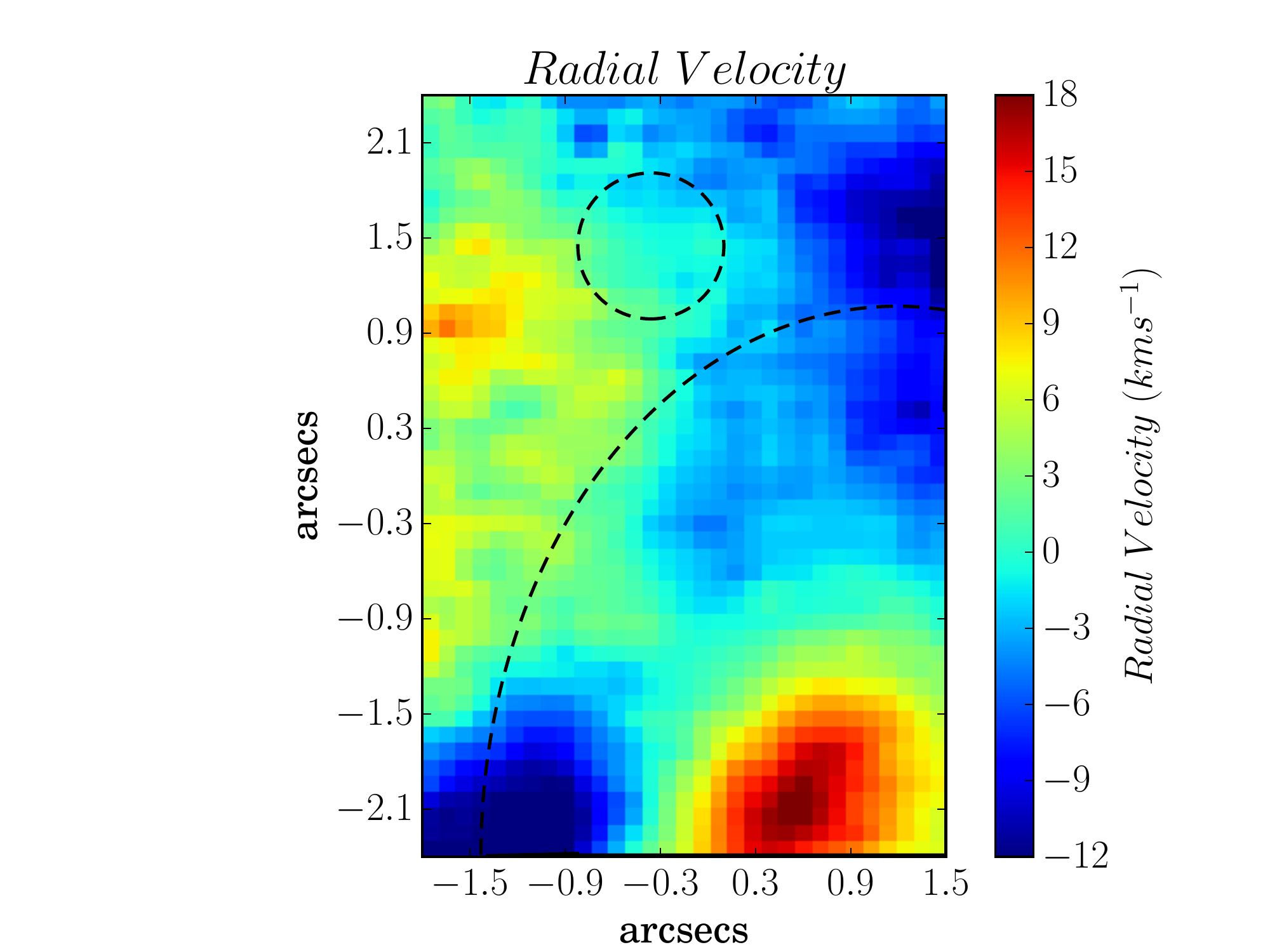}
	\includegraphics[width = 0.48\textwidth]{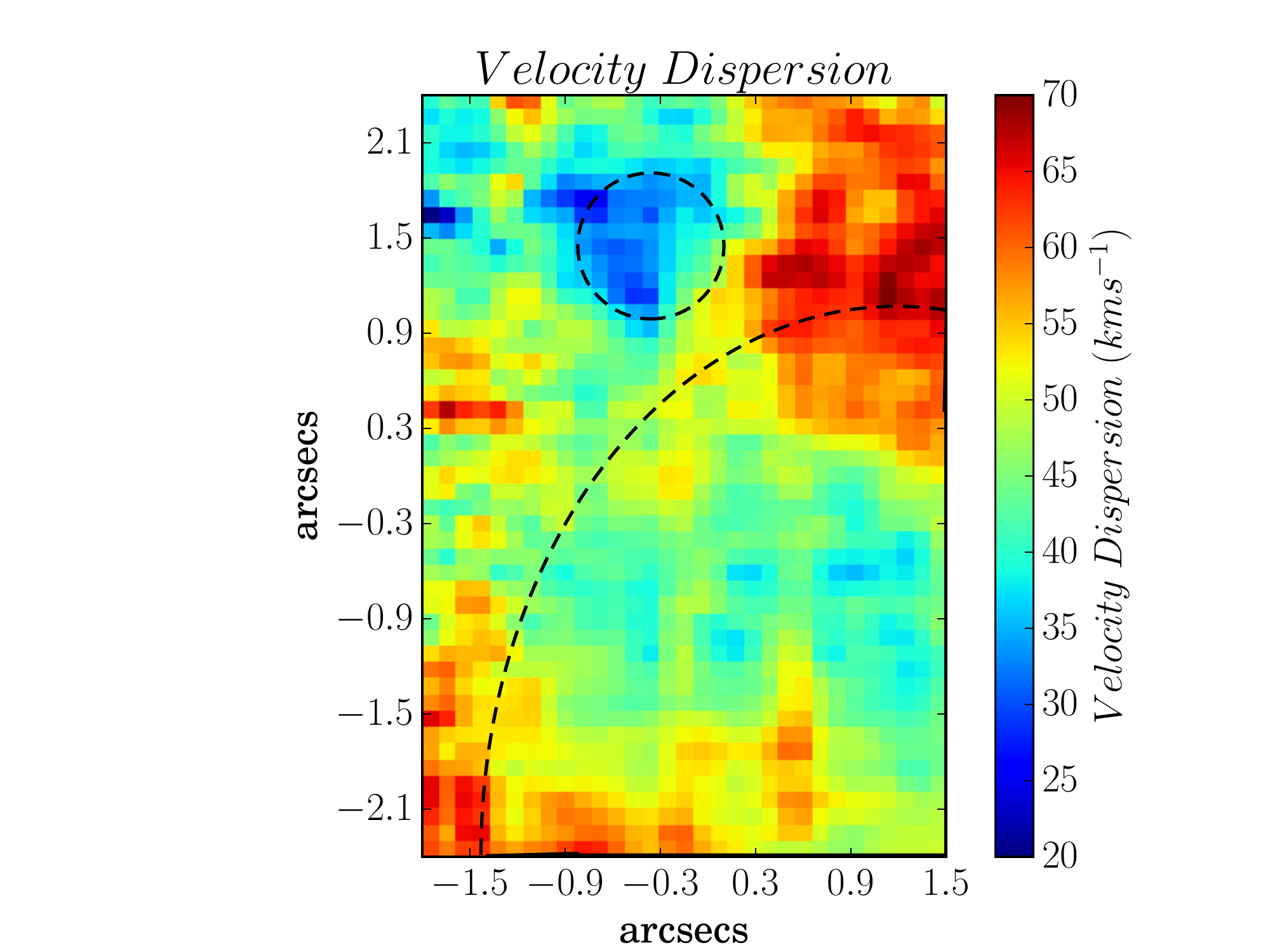}
	\caption{Radial velocity (upper panel) and velocity dispersion (bottom panel) maps of the ionised gas obtained from the H$\alpha$ emission line. Radial velocity is corrected for systemic ($\sim$ 609 km s$^{-1}$) and barycentric ($\sim$ $-$6.90 km s$^{-1}$) velocities. Velocity dispersion is corrected for instrumental broadening. The dashed quarter ellipse and circle indicate Region 1 and Region 2, respectively.}
	\label{fig:kinematics maps}
\end{figure}

\indent Figure \ref{fig:hst} shows the HST/ACS image of SBS1415+437 taken in the F606W (V) filter from \citet{Aloisi2005}. The green rectangular box represents the GMOS aperture (3.5$\arcsec$ $\times$ 5$\arcsec$). The upper-inset shows the H$\alpha$ image obtained from GMOS IFU data where we mark Region 1 and Region 2 selected on the basis of an isophotal H$\alpha$ emission. The blue parallel lines indicate the long-slit position of the Kitt Peak 4m Mayall Telescope observations used in the analysis of \citet[][hereafter \tGuseva]{Guseva2003}. Figure \ref{spectra} shows the GMOS-IFU integrated spectra of the entire FOV in the blue and red parts of the optical spectrum. We have overplotted the principal emission lines at their rest wavelengths in air. 

\indent We estimate the emission line fluxes for the recombination and collisionally excited lines within the spectra by fitting Gaussian profiles after subtracting the continuum and absorption features in the spectral region of interest. We give equal weight to flux in each spectral pixel while fitting Gaussians. The fitting uncertainties on the three  Gaussian parameters (amplitude, centroid and FWHM) are propagated to estimate the uncertainty in fluxes. These uncertainty estimates are consistent with those calculated from Monte Carlo simulations. We have propagated these estimated uncertainties in the fluxes to other quantities  using Monte Carlo simulations in the subsequent analysis. 
   
\indent Figure \ref{fig:observed flux} shows the observed fluxes of R-band continuum, [O~$\textsc{iii}$] $\lambda$4363,  [O $\textsc{iii}$] $\lambda$5007, H$\alpha$, [N $\textsc{ii}$]  $\lambda$6583 and [S$\textsc{ii}$] $\lambda$6731 over the FOV.  We obtained the R-band continuum by integrating the red cube in the wavelength range of 5890--7270 \AA~ (in the rest wavelength). In all maps, white spaxels correspond to spaxels in which emission lines have S/N < 3. The [O $\textsc{iii}$] $\lambda$4363 has S/N $>$ 3 for a region extending over 143 $\times$ 143 pc$^2$, though all pixels do not show detection (Figure \ref{fig:observed flux}, upper-right panel).  Note here that we could not detect the [O \textsc{ii}] $\lambda\lambda$3727,3729 doublet in our data because of the low sensitivity of the blue wavelength end of the GMOS-IFU. In all flux maps, there are two distinct regions with elevated fluxes, represented by a quarter-ellipse and a circle, which we refer as Region 1 and Region 2 in subsequent analysis (also see Figure \ref{fig:hst}). We detect [O $\textsc{iii}$] $\lambda$4363 in the integrated spectrum of Region 1, but not in Region 2, so analysis involving this emission line is not performed for Region 2.  The observed fluxes of the main emission lines in the integrated spectrum of Region 1 are presented in Table \ref{flux table}.   

\indent To estimate dust attenuation, we used the attenuation curve of Large Magellanic Cloud \citep[LMC][]{Fitzpatrick1999}\footnote{Note here that attenuation curves corresponding to LMC, Small Magellanic Cloud and the Milky Way have similar values in the wavelength region of interest.} along with the observed H$\alpha$/H$\beta$ ratio to first estimate the nebular emission line colour excess E(B--V), at an electron temperature and density of 10000 K and 100 cm$^{-3}$, respectively (Case B recombination). While mapping E(B-V), we found that some spaxels away from the bright star-forming regions have negative values of E(B--V) which is most likely due to stochastic error and shot noise \citep{Hong2013}. For such spaxels, we forced E(B--V) to that of the Galactic Foreground \citep[0.0077,][]{Schlafly2011}.  Figure \ref{fig:ebv} shows the E(B-V) map of the FOV, which varies between 0.0077--0.30 mag. The increased value of E(B--V) in Region 1 and Region 2 shows that the star-forming regions are dustier than the rest of the FOV. The E(B--V) map obtained was used to deredden the observed flux maps\footnote{See \citet{Kumari2017} for the formulae used.}. From the integrated spectrum of Region 1, we obtain a value of E(B--V) = 0.057$\pm$0.005, which lies in the range of values found in the literature for this region (0.08 by \citet{Thuan1999} and 0.00 by \tGuseva). The E(B--V) estimated here was used to deredden the observed emission line fluxes of Region 1. The dereddened/intrinsic fluxes for the main emission lines are shown in Table \ref{flux table}. 


\begin{figure*}
	\centering
	\includegraphics[width = 0.45\textwidth, trim={0 0 0 1.0cm}, clip]{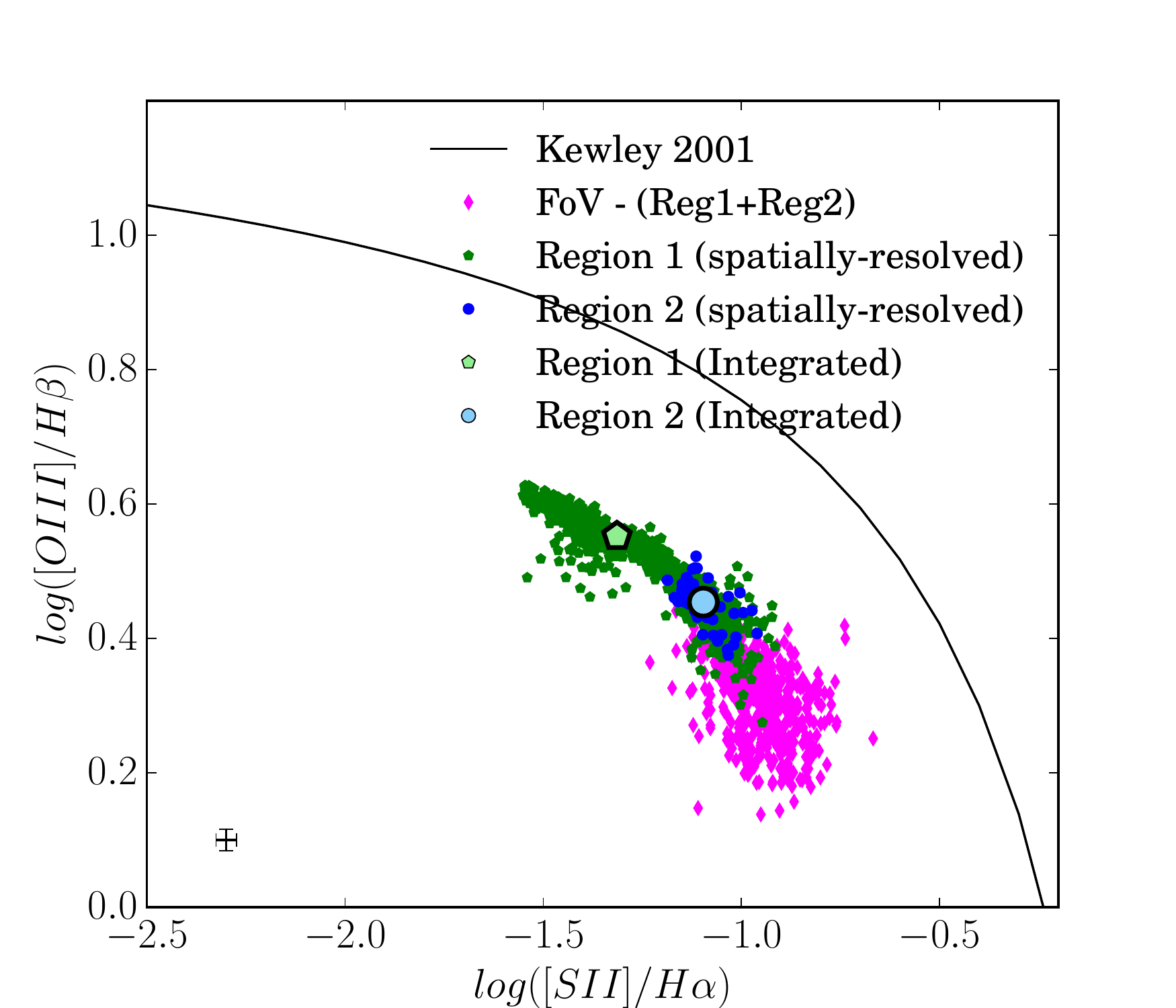}
	\includegraphics[width = 0.45\textwidth, trim={0 0 0 1.0cm}, clip]{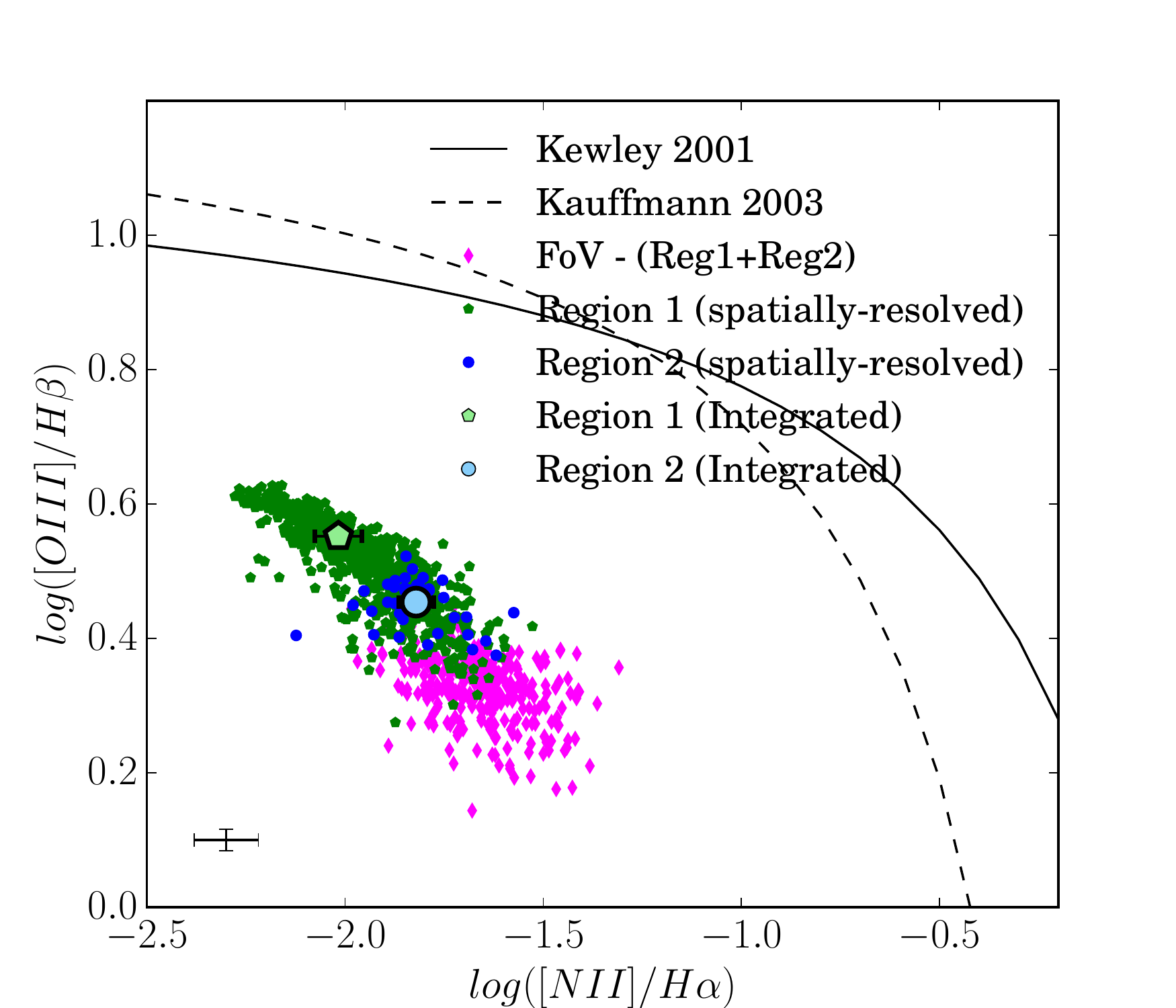}
	\caption{Emission line ratio diagnostic diagrams: [O \textsc{iii}]/H$\beta$ versus [S \textsc{ii}]/H$\alpha$ (left panel),  and [O \textsc{iii}]/H$\beta$ versus [N \textsc{ii}]/H$\alpha$ (right panel).  Black solid curve and dashed curve represent the theoretical maximum starburst line from \citet{Kewley2001} and \citet{Kauffmann2003}, respectively, showing a classification based on excitation/ionisation mechanisms. The line ratios of the two H $\textsc{ii}$ regions are colour-coded as follows: Region 1: green pentagon, Region 2: blue circle. Smaller dark-coloured markers denote the spatially-resolved (spaxel-by-spaxel) line-ratios and the bigger light-coloured markers in black contours denote the line-ratios obtained from the integrated spectrum of the corresponding regions. Magenta coloured markers denote the spatially-resolved line-ratios of the regions of FOV excluding the two H $\textsc{ii}$ regions. The size of error bars varies for line ratios and the median error bars are shown in left corner of each panel.}
	\label{BPT}
\end{figure*}

\begin{figure*}
	\centering
	\includegraphics[width = 0.48\textwidth]{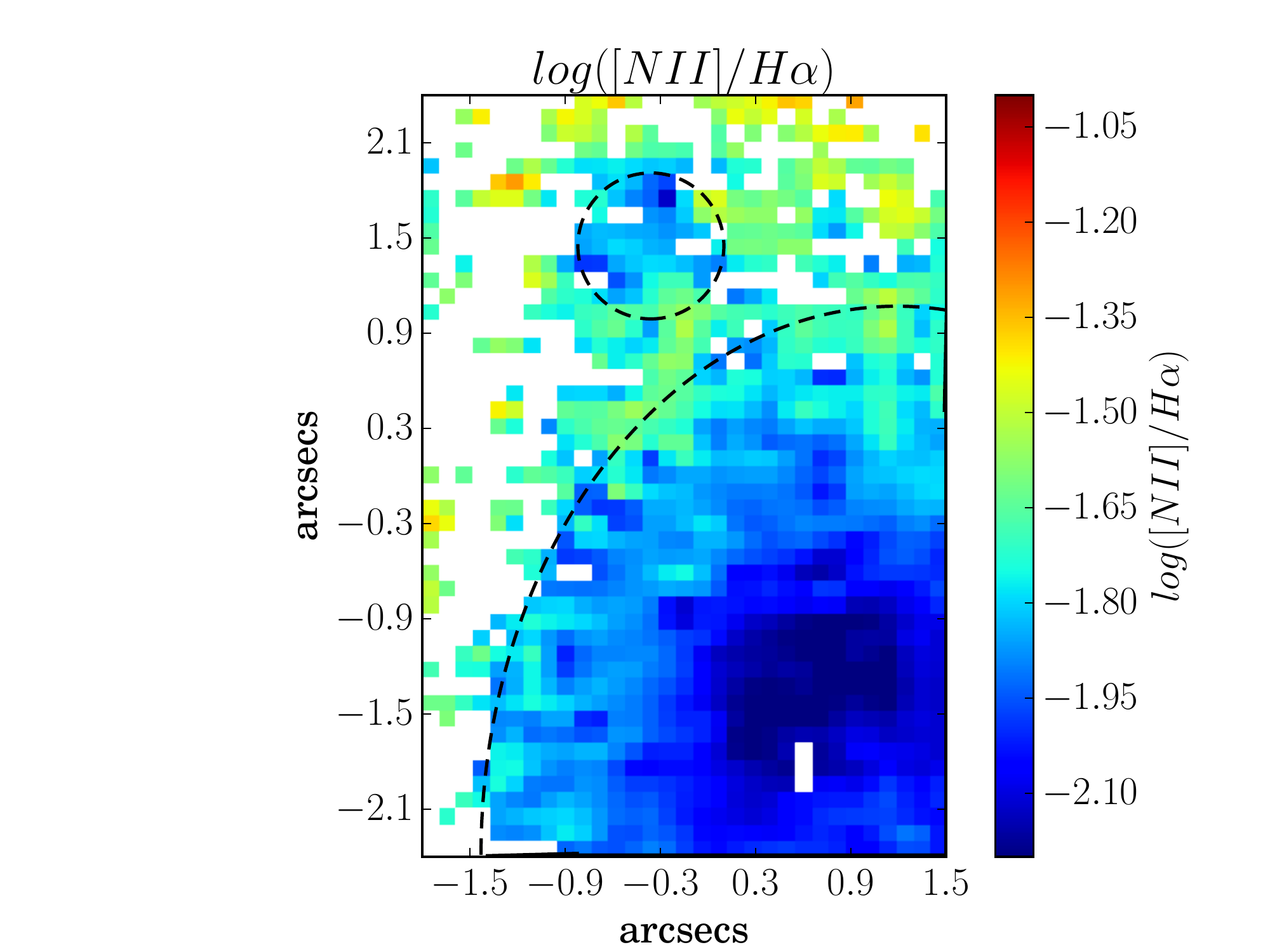}
	\includegraphics[width = 0.48\textwidth]{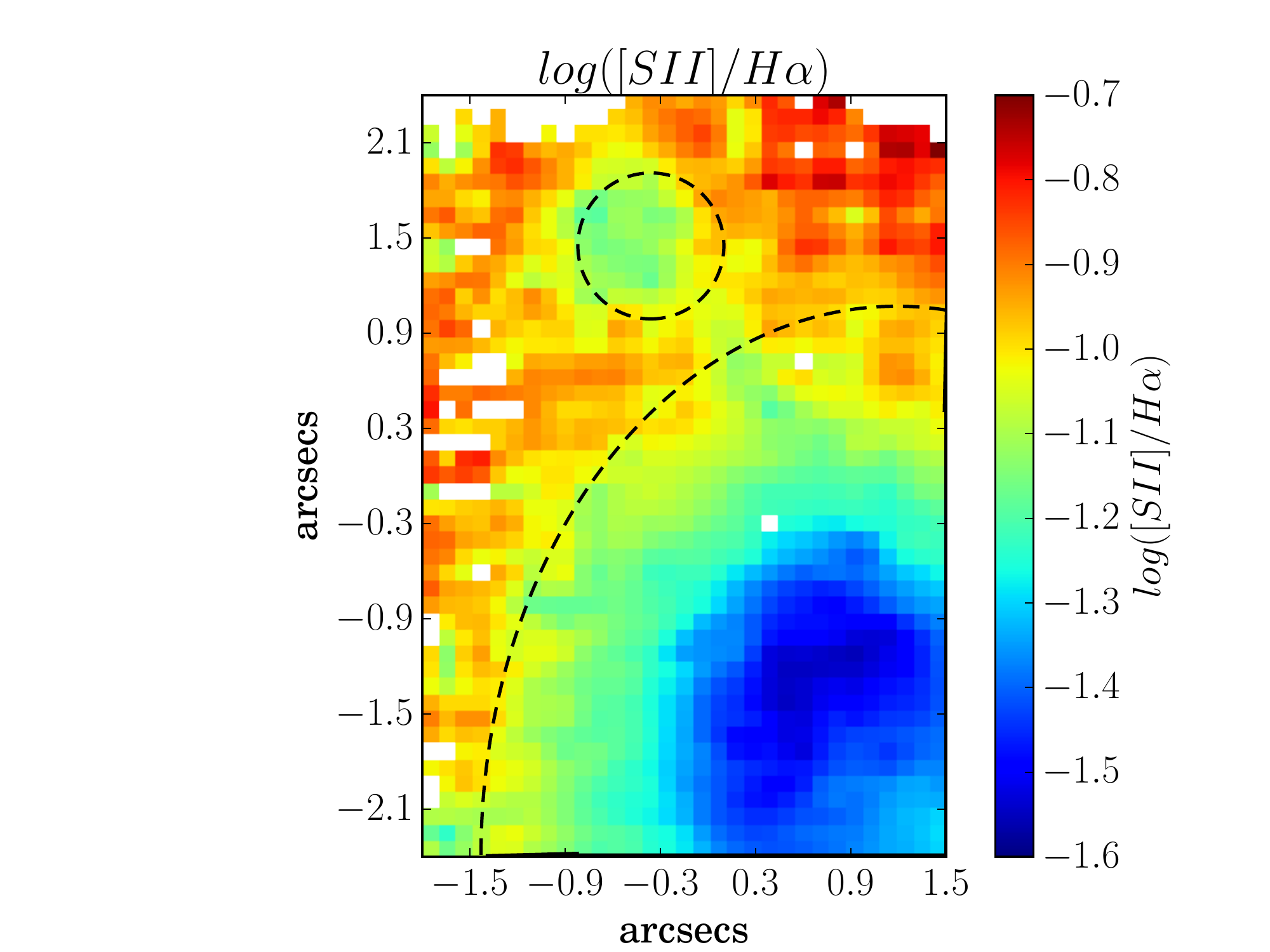}
	\includegraphics[width = 0.48\textwidth]{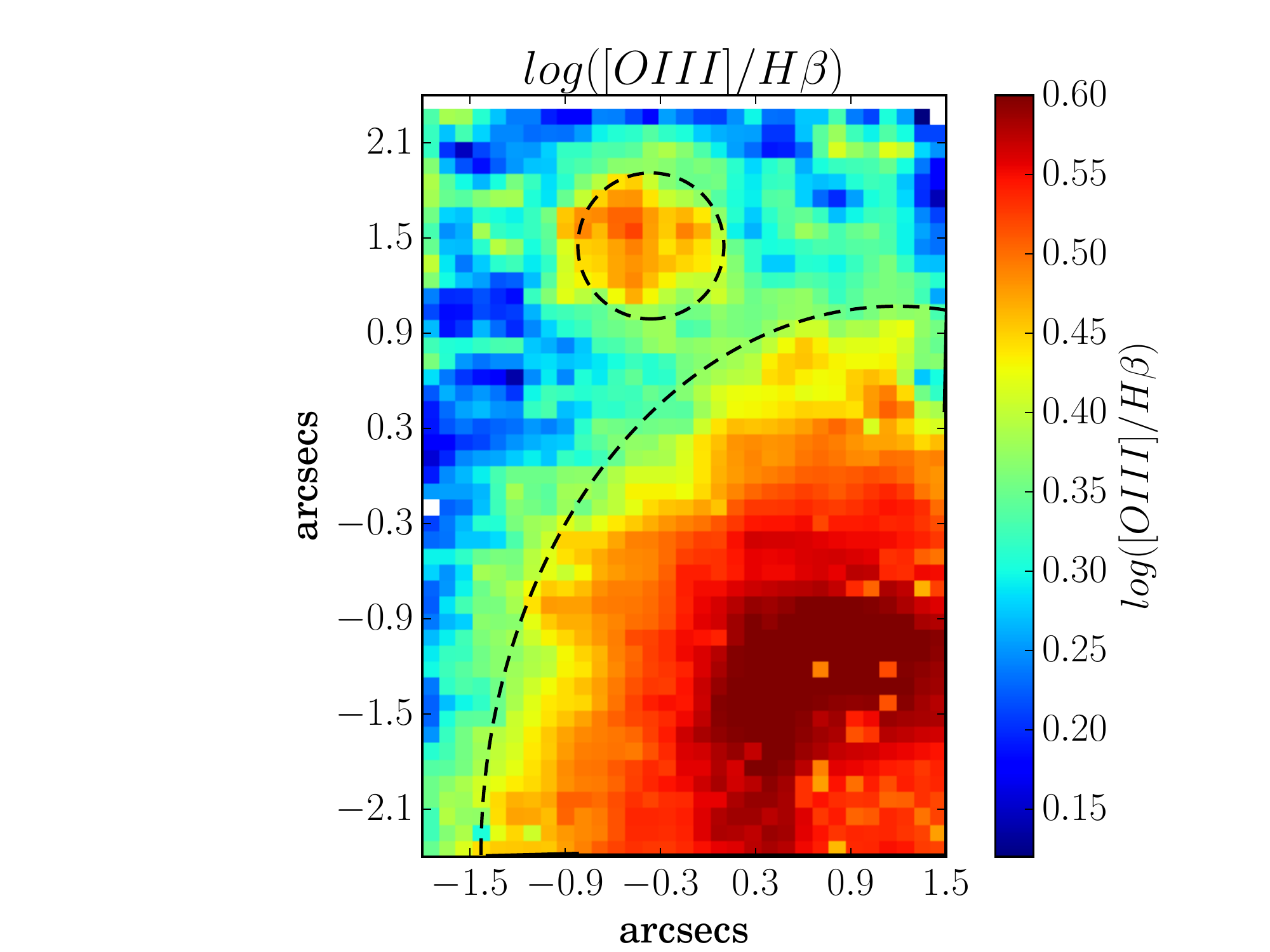}
	\caption{Emission line ratio maps of [N \textsc{ii}] $\lambda$6583/H$\alpha$ (upper-left panel), [S \textsc{ii}] $\lambda \lambda$6717,6731/H$\alpha$ (upper-right panel) and [O \textsc{iii}] $\lambda$5007/H$\beta$ (bottom panel). The dashed quarter ellipse and circle indicate Region 1 and Region 2, respectively. The spaxels in which emission line fluxes had S/N $<$ 3, are shown in white in above line ratio maps. }
	\label{ratio maps}
\end{figure*}


\begin{figure}
	\centering
	\includegraphics[width = 0.45\textwidth, trim={0 2cm 8cm 0},clip]{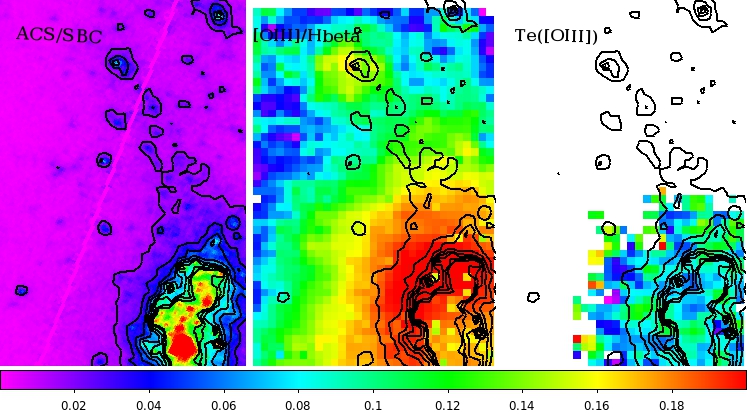}
	\caption{Left panel: ACS/SBC image (0.025\arcsec per pixel) of SBS 1415+437 taken in F125LP filter. Right panel: [O \textsc{iii}]/H$\beta$ map created from the GMOS-IFU data. Black contours are generated from the FUV image on the left panel and are overlaid on the emission line ratio map, to trace the relative position of hot O and B stars and the ionised gas.}
	\label{fig: sbc o3hb}
\end{figure}

\subsection{Gas Kinematics}
\label{section:kinematics}
\indent Figure \ref{fig:kinematics maps} shows the maps of radial velocity (upper panel) and velocity dispersion (lower panel) of the H$\alpha$ emission line, which we obtained from the centroid and the FWHM of the Gaussian fit to the emission line. The radial velocity map is corrected for the systemic velocity of 609 kms$^{-1}$, and the barycentric velocity of $-$6.9 kms$^{-1}$. The velocity dispersion map is corrected for the instrumental broadening of 1.7~\AA ~which corresponds to $\sim$ 78 km s$^{-1}$ at the rest-wavelength of the H$\alpha$ emission line.

\indent Figure \ref{fig:kinematics maps} (upper panel) shows that the radial velocity of the ionised gas in the region of study varies between $\sim$ $-$14 to 19 kms$^{-1}$. \citet{Thuan1999} reports a solid body rotation for this BCD across $\sim$ 30\arcsec. Though our data clearly shows blueshift and redshift at different locations of the FOV, we do not find a definite axis of rotation. This could be because our FOV is comparatively small (3.5\arcsec $\times$ 5\arcsec). 

\indent Figure \ref{fig:kinematics maps} (bottom panel) shows that the  velocity dispersion of the ionised gas in the region of study varies between $\sim$20--70 kms$^{-1}$. Region 1 shows a range of values,  whereas Region 2 has a relatively low velocity dispersion. The ionised gas to the north-east of Region 1 and north-west of Region 2 shows the highest values of velocity dispersion. 
 

\subsection{Emission line ratio diagnostics}
\label{section:ratios}
\indent In Figure \ref{BPT}, we show the classical emission line ratio diagnostic diagrams ([O $\textsc{iii}$]$\lambda$5007/H$\beta$ versus [S $\textsc{ii}$]$\lambda\lambda$6717,6731/H$\alpha$ (left panel) and [O $\textsc{iii}$]$\lambda$5007/H$\beta$ versus [N $\textsc{ii}$]$\lambda$6583/H$\alpha$ (right panel)), which are commonly known as BPT diagrams \citep{Baldwin1981}. In both panels, we show the maximum starburst line also known as the ``Kewley line'' as the solid black curve, which presents a classification based on ionisation/excitation mechanism. On the [N $\textsc{ii}$] diagnostic diagram (Figure \ref{BPT}, right panel), the dashed black curve represents the empirical line derived by \citet{Kauffmann2003} using SDSS spectra of 55 757 galaxies. Our spatially-resolved line ratios are shown by blue, green and magenta markers which correspond to the spaxels of Region 1, Region 2 and the rest of the FOV, respectively. We also show the line ratios obtained from the integrated spectra of Region 1 and Region 2 using larger markers. In the [N \textsc{ii}]-BPT diagram, we find that both spatially-resolved and integrated data lie well below and to the left of the Kewley line, as well as the empirical line of \citet{Kauffmann2003}, while on the [S \textsc{ii}]-BPT diagram, we find that both spatially-resolved and integrated data lie well below and to the left of the Kewley line. This shows that the ionised gas in the region under study is predominantly ionised by the photons from the massive stars.  

\indent Figure \ref{ratio maps} shows the line ratio maps of the region under study, which we use to study the ionisation structure of the ionised gas. As expected, we find that the peak of emission shows the lowest values of [N $\textsc{ii}$] $\lambda$6583/H$\alpha$ (upper-left panel) and [S $\textsc{ii}$] $\lambda\lambda$6717,6731/H$\alpha$ (upper-right panel), and highest values of [O $\textsc{iii}$] $\lambda$5007/H$\beta$ (lower panel). This indicates that the more luminous regions (on H$\alpha$ map) have relatively high excitation, which is due to the presence of young and massive stars producing harder ionising radiation. Figure \ref{fig: sbc o3hb} shows a comparison of ACS/Solar Blind Channel (SBC) FUV (far ultra violet) image of SBS 1415+437 in the F125LP filter (program id: JB7H08010, PI: A Aloisi) and the emission line ratio map [O \textsc{iii}]/H$\beta$ created from GMOS data. The contours generated from ACS/SBC image (left panel) are over-plotted on the [O \textsc{iii}]/H$\beta$ map (right panel). We find that there is an offset between the peak of [O \textsc{iii}]/H$\beta$ and the peak of FUV image which traces the young stars, and that the enhancement of [O \textsc{iii}]/H$\beta$ is on the periphery of the stars traced by FUV. This indicates that the enhancement of [O \textsc{iii}]/H$\beta$ ratio has resulted due to the ionisation of gas by hot O and B stars in the surrounding regions. Note here that the peak of H$\alpha$ map is offset  by $\sim$ 58 pc with respect to the peak of the R-band continuum indicative of the region containing an older stellar population. 


\begin{table}
	\centering
	\caption[Summary of nebular diagnostics, ionic abundances, elemental abundances and abundance ratios obtained from the integrated spectrum, and the spatially-resolved maps of Region 1]{Summary of nebular diagnostics, ionic abundances, elemental abundances and abundance ratios obtained from the integrated spectrum (Value $\pm$ Uncertainty), and the spatially-resolved maps (Median $\pm$  Uncertainty) of Region 1. The uncertainty on the median takes into account the correlation between pixels due to resampling in forming data cube.}
	\begin{tabular}{cccc}
		\hline
		& Integrated spectrum& Spatially-resolved Map\\
		Parameter & Value $\pm$ Uncertainty & Median $\pm$ Uncertainty &\\
		\hline
		Te([O \textsc{iii}]) ($\times$ 10$^4$) & 1.71 $\pm$ 0.07 & 1.72 $\pm$ 0.02\\
		Te([O \textsc{ii}]) ($\times$ 10$^4$) & 1.59 $\pm$ 0.05 &  1.53 $\pm$ 0.02\\
		Te([N \textsc{ii}]) ($\times$ 10$^4$) & 1.42 $\pm$ 0.03 & 1.42  $\pm$ 0.01 \\
		Te([S \textsc{iii}]) ($\times$ 10$^4$) & 1.71 $\pm$ 0.09 & 1.72 $\pm$ 0.03\\
		Ne([SII]) (cm$^{-3}$) &$<$ 50& 50 $\pm$ 9\\
		12 + log(O$^+$/H$^+$) & 7.15 $\pm$ 0.06 & 7.21 $\pm$ 0.03\\
		12 + log(O$^{2+}$/H$^+$) & 7.46 $\pm$ 0.04 & 7.48 $\pm$ 0.01\\
		12+ log(O/H) & 7.63 $\pm$ 0.03  & 7.67 $\pm$ 0.02\\
		\\
		12 + log(N$^+$/H$^+$) & 5.41 $\pm$ 0.05 & 5.39 $\pm$ 0.01\\
		ICF(N$^+$) & 3.03 $\pm$ 0.47 & 2.79 $\pm$ 0.09\\
		12 + log(N/H) & 5.89 $\pm$ 0.08 & 5.84 $\pm$ 0.02\\
		N/O & -1.74 $\pm$ 0.09 & -1.82 $\pm$ 0.02\\
		\\
		S$^+$/H$^+$ ($\times$ 10$^7$) & 1.31 $\pm$ 0.06  & 1.27 $\pm$ 0.06\\
		S$^{2+}$/H$^+$ ($\times$ 10$^7$) & 4.95 $\pm$ 0.69  & 5.02 $\pm$ 0.33\\
		ICF (S$^+$ + S$^{2+}$) & 1.10 $\pm$ 0.02 & 1.08 $\pm$ 0.01\\
		12 + log(S/H) & 5.84 $\pm$ 0.05  & 5.83 $\pm$ 0.02 \\
		log(S/O) & -1.79 $\pm$ 0.06 & -1.83 $\pm$ 0.02 \\
		\\
		Ne$^{2+}$/H$^+$ ($\times$ 10$^5$) & 0.49 $\pm$ 0.08 & 0.55 $\pm$ 0.03 \\
		ICF (Ne$^{2+}$) & 1.10 $\pm$ 0.01 & 1.11 $\pm$ 0.03\\
		12 + log(Ne/H) & 6.73 $\pm$ 0.07  & 6.80 $\pm$ 0.02\\
		log(Ne/O) & -0.90 $\pm$ 0.08 & -0.90 $\pm$ 0.02 \\
		\\
		Ar$^{2+}$/H$^+$ ($\times$ 10$^7$) & 1.23 $\pm$ 0.09 & 1.23 $\pm$ 0.03\\
		ICF (Ar$^{2+}$) & 1.15 $\pm$ 0.01& 1.16 $\pm$ 0.01 \\
		12 + log(Ar/H) & 5.15 $\pm$ 0.03 & 5.16 $\pm$ 0.01 \\
		log(Ar/O) & -2.48 $\pm$ 0.04 & -2.50 $\pm$ 0.01 \\
		\hline
	\end{tabular}
	\label{table:chemical properties}
\end{table}

\begin{figure}
	\centering
	\includegraphics[width=0.48\textwidth]{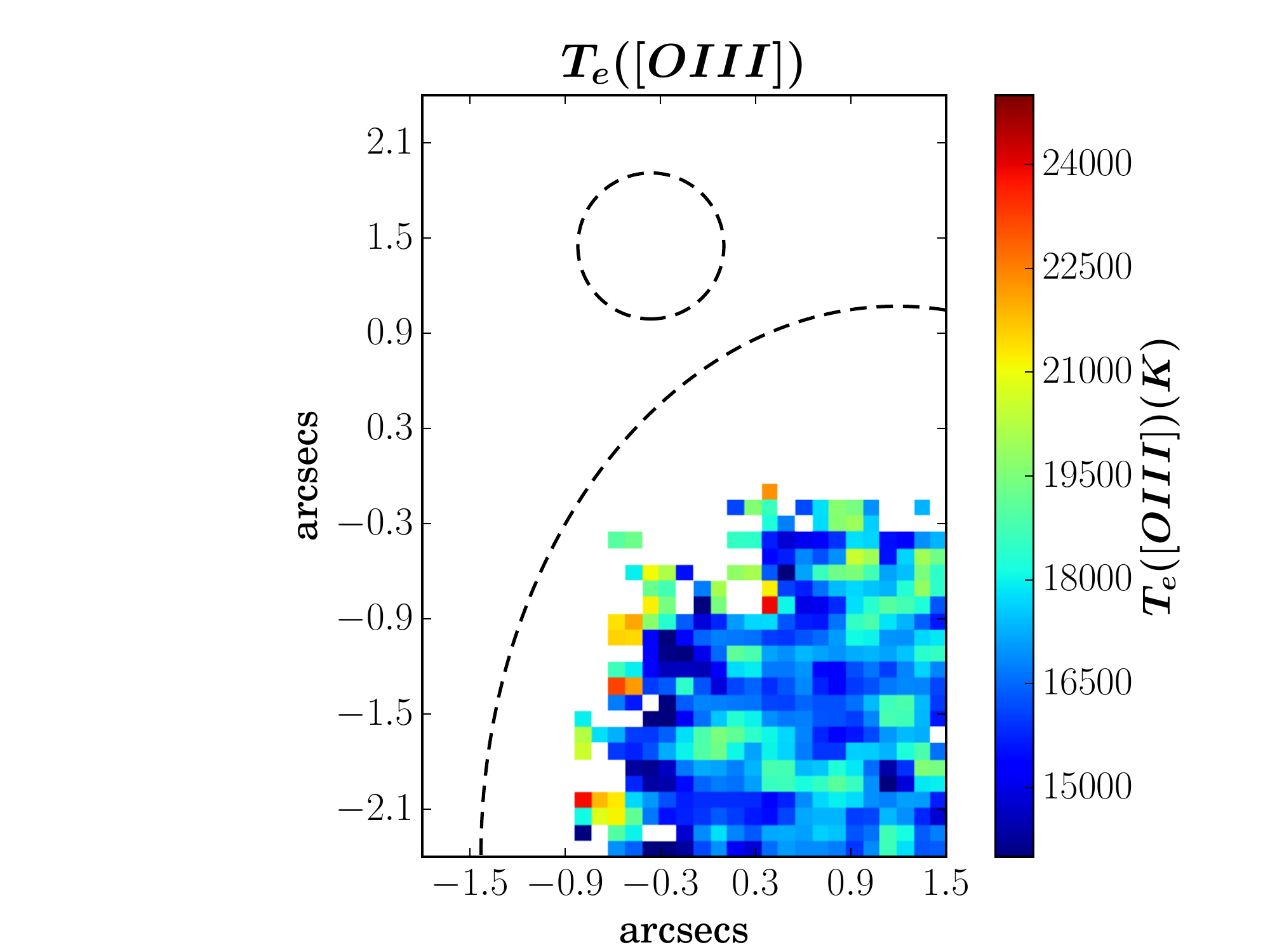}
	\includegraphics[width=0.48\textwidth]{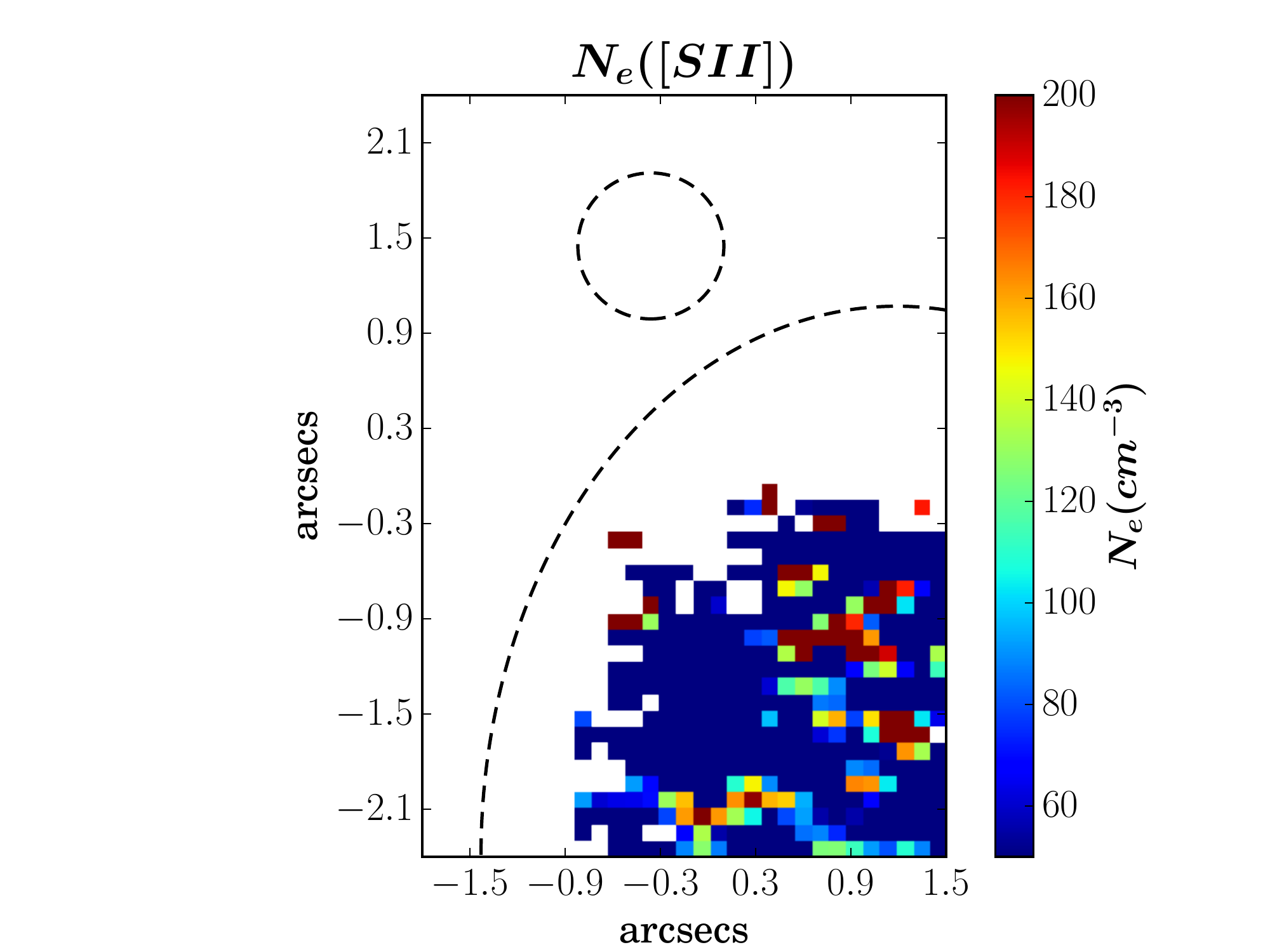}
	\caption{Upper panel: Map of electron temperature of high-ionisation zone (T$_e$([O $\textsc{iii}$])). Lower panel: Electron density (N$_e$([S \textsc{ii}])) map obtained from the [S \textsc{ii}] doublet ratio and T$_e$([O $\textsc{iii}$]) map. The dashed quarter ellipse indicates Region 1. The spaxels in which emission line fluxes had S/N $<$ 3, are shown in white. Single pixel features or few-pixel features should be interpreted with caution since the seeing FWHM (0.6 arcsec) extends over 6 pixels.}
	\label{figure:TeNe}
\end{figure}

\begin{figure*}
	\centering
	\includegraphics[width=0.48\textwidth]{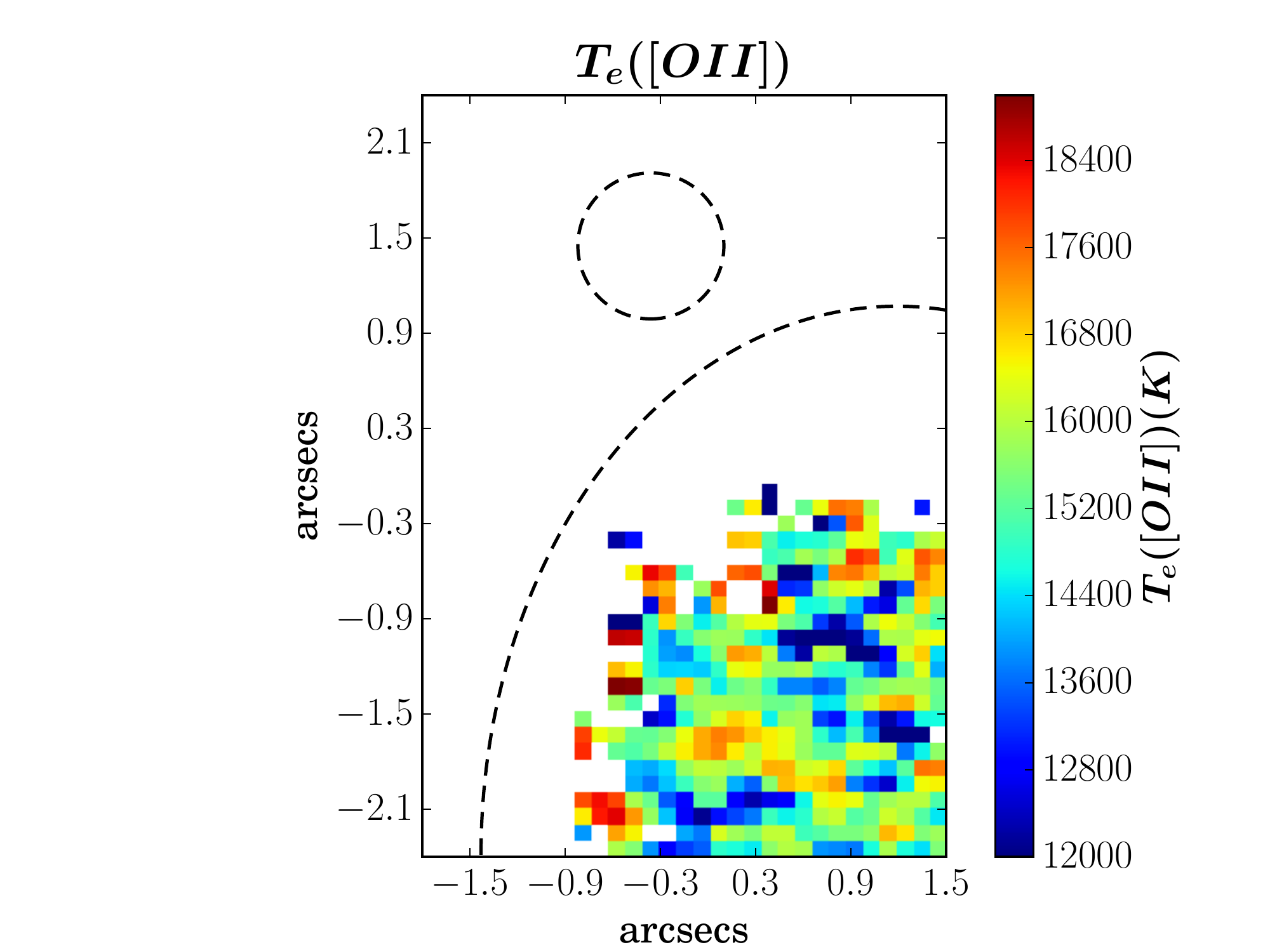}
	\includegraphics[width=0.48\textwidth]{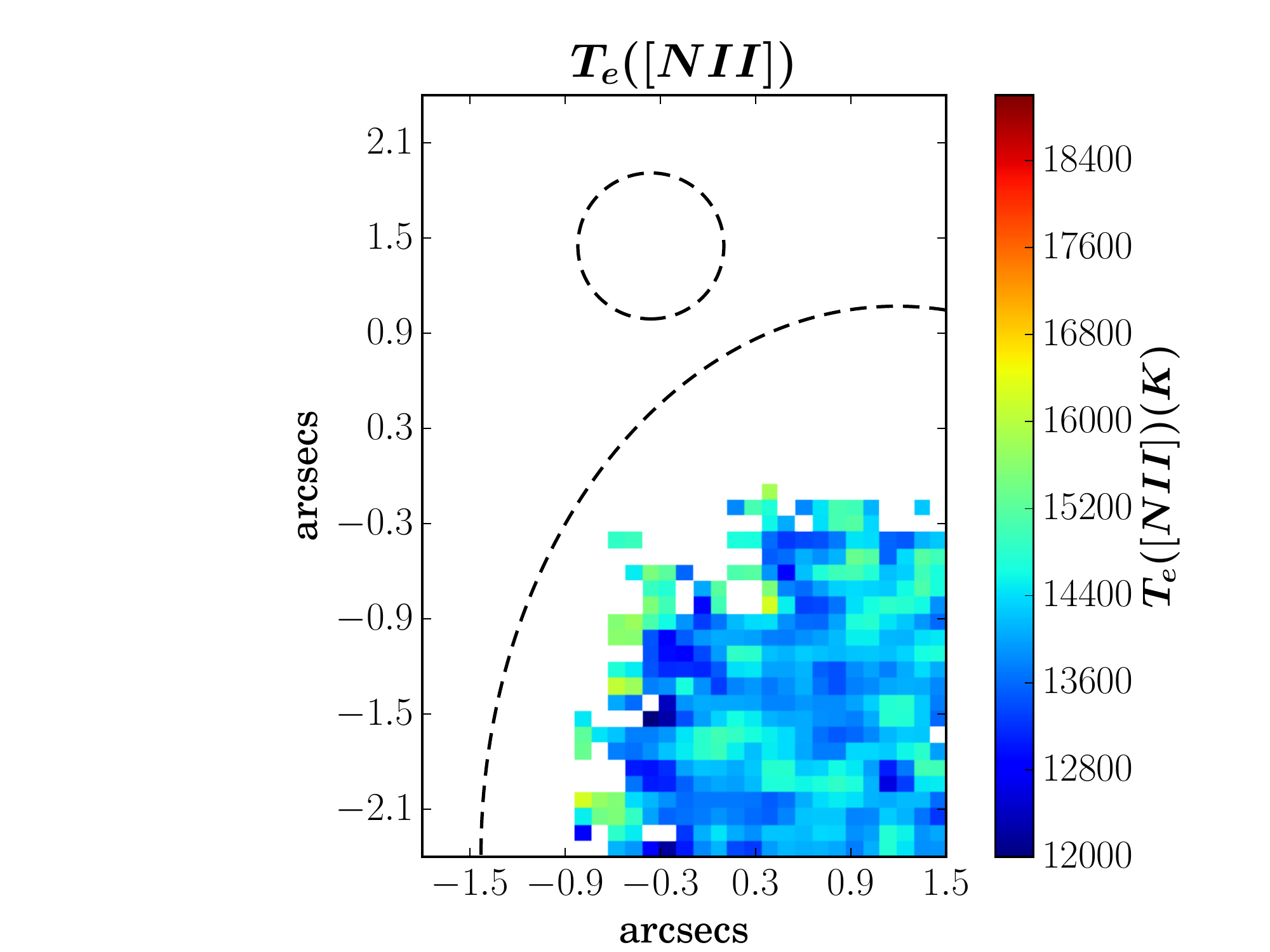}
	\includegraphics[width=0.48\textwidth]{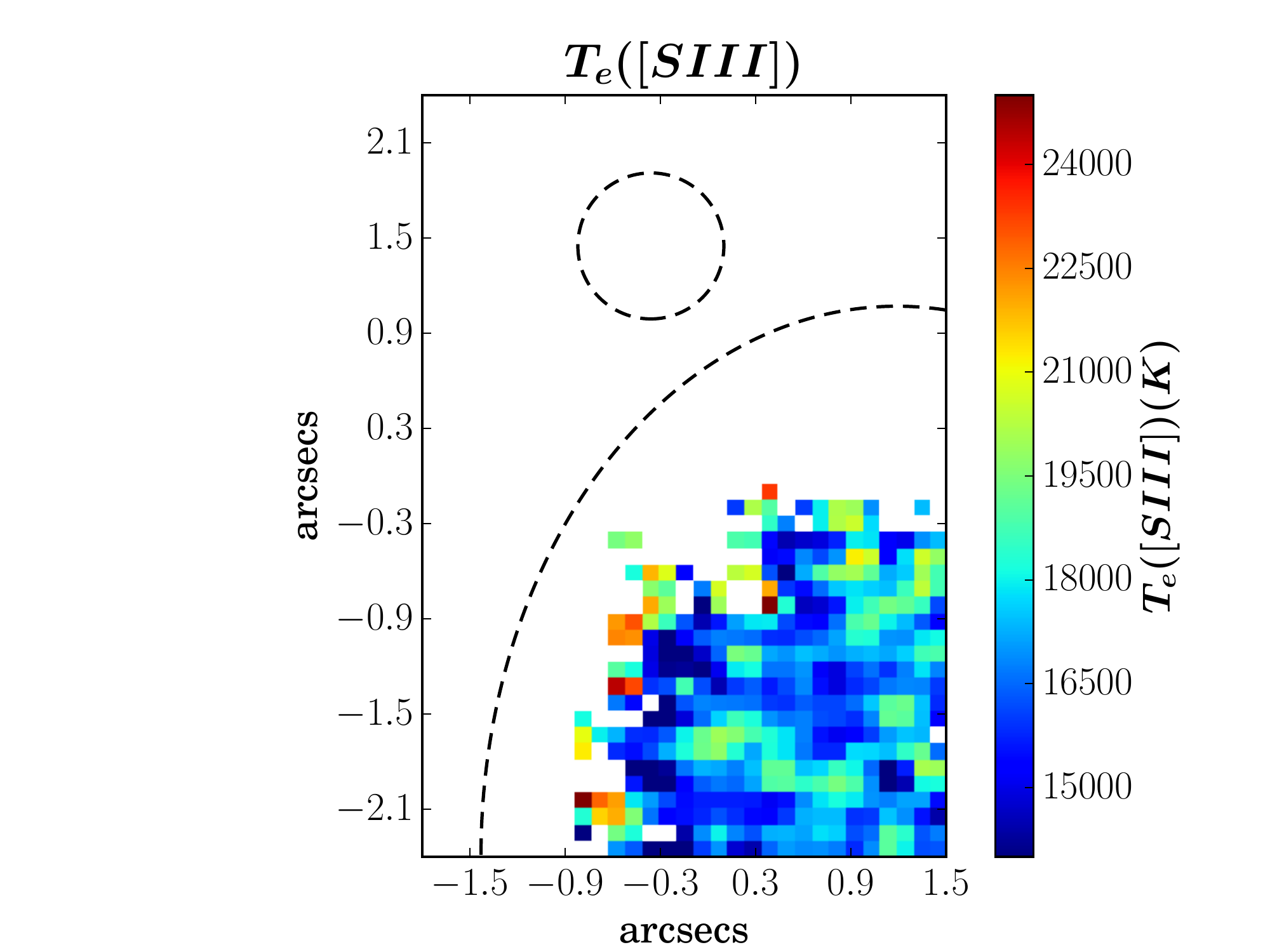}
	\caption{Temperature maps of low- and intermediate ionisation zones: T$_e$([O \textsc{ii}]) (upper-left panel), T$_e$([N \textsc{ii}]) (upper-right panel) and T$_e$([S \textsc{iii}]) (lower panel). The dashed quarter ellipse and circle indicate Region 1 and Region 2. The spaxels in which emission line fluxes had S/N $<$ 3, are shown in white.}
	\label{fig:Te others}
\end{figure*}

\subsection{Electron temperature and density}
\label{section:Te}

\indent An H \textsc{ii} region is a complex structure. In the following analysis, the thermal structure of a H \textsc{ii} region is assumed to be approximated by a three-zone model \citep{Garnett1992}. This model consists of a high-ionisation zone, a low-ionisation zone and an intermediate-ionisation zone. The innermost high-ionisation zone corresponds to species such as O$^{2+}$, Ne$^{2+}$, Fe$^{2+}$, He$^+$ and Ar$^{3+}$. The outermost low-ionisation zone corresponds to species such as O$^+$, N$^+$ and S$^+$. The region between the two is the intermediate-ionisation zone and corresponds to species such as Ar$^{2+}$ and S$^{2+}$. The electron temperature of each zone can be approximated from the temperature of the species related to that zone. 

\indent To estimate electron temperature T$_e$([O \textsc{iii}]), we use the dereddened [O $\textsc{iii}$] line ratio, [O $\textsc{iii}$] ($\lambda$5007 $+\lambda$4959)/[O $\textsc{iii}$]$\lambda$4363 and the calibration from \citet[][hereafter \tPerezMontero]{Perez-Montero2017}, which is derived assuming a five-level atom, using collision strengths from \citet{Aggarwal1999}. Figure \ref{figure:TeNe} (upper panel) shows the derived T$_e$([O \textsc{iii}]) map. From the integrated spectrum of Region 1, we find T$_e$([O \textsc{iii}]) =  17100 $\pm$ 700 K, in agreement with the value found by \tGuseva~ (16490 $\pm$ 140 K).  

\indent To derive the electron density N$_e$, we use [S \textsc{ii}] doublet ratio and T$_e$([O \textsc{iii}]) derived above, along with the theoretical curves of the [S \textsc{ii}] doublet ratio versus N$_e$ at given temperatures. Figure \ref{figure:TeNe} (lower panel) shows the N$_e$ map, where we find that majority of the spaxels have N$_e$ $<$ 50 cm$^{-3}$, indicating that the region under study lies in the low-density regime. From the integrated spectrum of Region 1, we find N$_e$ $<$ 50 cm$^{-3}$ which is in agreement with the value (60 $\pm$ 30 cm$^{-3}$) found by \tGuseva. 

\indent The temperature of low-ionisation zone T$_e$([O \textsc{ii}]) is estimated by using N$_e$ and T$_e$ calculated above and the density-dependent calibration given in \tPerezMontero. The T$_e$([O \textsc{ii}]) map (Figure \ref{fig:Te others}, upper-left panel) shows a mean and standard deviation of $\sim$15000 K and $\sim$1600 K, respectively. 
 We calculated T$_e$([O \textsc{ii}]) $\sim$15900 $\pm$ 500 for the integrated spectrum of Region 1, which is higher than the value found by \tGuseva~ (14430 $\pm$ 110 K). 
 The main difference is that \tGuseva~ have  used the expression from \citet{Izotov1994}, which is independent of density. Using their expression, we find the value to be $\sim$14700 $\pm$ 300 K for the integrated spectrum Region 1, which is in good agreement with their value (\tGuseva). 
The electron density should be taken into account in determining T$_e$([O \textsc{ii}]) because of only a weak correlation between T$_e$([O \textsc{ii}]) and T$_e$([O \textsc{iii}]) \citep{Hagele2006, Hagele2008, Lopez-Hernandez2013}.



\indent We map T$_e$([N \textsc{ii}]) (Figure \ref{fig:Te others}, upper-right panel) by using the T$_e$([O \textsc{iii}]) map in the calibration of \citet{Perez-Montero2009}, which is based on photoionisation models. The map shows a variation of 11900--16300 K. For the integrated of spectrum Region 1, we followed same procedure involving the T$_e$([O \textsc{iii}]) estimated above  and calculated T$_e$([N \textsc{ii}]) = 14200 $\pm$ 300 K.

\indent Though both T$_e$([O \textsc{ii}]) and T$_e$([N \textsc{ii}]) correspond to temperatures of low-ionisation zones, their maps show different structure. This is because the recipe used to obtain T$_e$([O \textsc{ii}]) involves the use of N$_e$([S \textsc{ii}]), which results in similar structures in T$_e$([O \textsc{ii}]) and  N$_e$([S \textsc{ii}]) maps, whereas T$_e$([N \textsc{ii}]) is directly obtained from T$_e$([O \textsc{iii}]). For the integrated spectrum of Region 1 as well, we find that T$_e$([N \textsc{ii}]) is slightly lower than T$_e$([O \textsc{ii}]), even though both ions lie in the same excitation zone of a nebula. Note here that photoionisation models predicted a higher T$_e$([O \textsc{ii}]) than T$_e$([N \textsc{ii}]) for the brightest knot BCD Mrk 209 \citep{Perez-Montero2007b}. It is possible that the temperature of the two ions T$_e$([O \textsc{ii}]) and T$_e$([N \textsc{ii}]) might not be the same even if originating from the same ionisation zone.

\indent We map the temperature of the intermediate-ionization zone T$_e$([S \textsc{iii}]) by using the empirical fitting between T$_e$([O \textsc{iii}]) and T$_e$([S \textsc{iii}]) given by \citet{Hagele2006}, which has a calibration uncertainty of the order of 12\%. The T$_e$([S \textsc{iii}]) map (Figure \ref{fig:Te others}, lower panel) shows a mean and standard deviation of 17000 K and 2000 K, respectively. Following the same procedure for the integrated spectrum of Region 1, we estimated T$_e$([S \textsc{iii}]) = 17100$\pm$900 K, which is higher than reported by \tGuseva ~(15380 $\pm$ 110 K), but is in agreement with their value if we consider the large calibration uncertainty from \citet{Hagele2006}. \tGuseva~  have followed the procedure of \citet{Garnett1992}, who assumed that T$_e$([S \textsc{iii}]) = T$_e$([Ar \textsc{iii}]) and estimated T$_e$([Ar \textsc{iii}]) from T$_e$([O \textsc{iii}]). Using their recipe, we get a value of T$_e$([S \textsc{iii}]) = 15900 $\pm$ 600  which is in good agreement with \tGuseva. 

\indent Table \ref{table:chemical properties}  summarizes the above results for the integrated spectrum of Region 1 and the spatially-resolved data. In this table, we report median and the uncertainty on the median of the values on the entire map and take into account the correlation between pixels due to resampling in forming data cube. 

\begin{figure*}
	\centering
	\includegraphics[width = 0.48\textwidth]{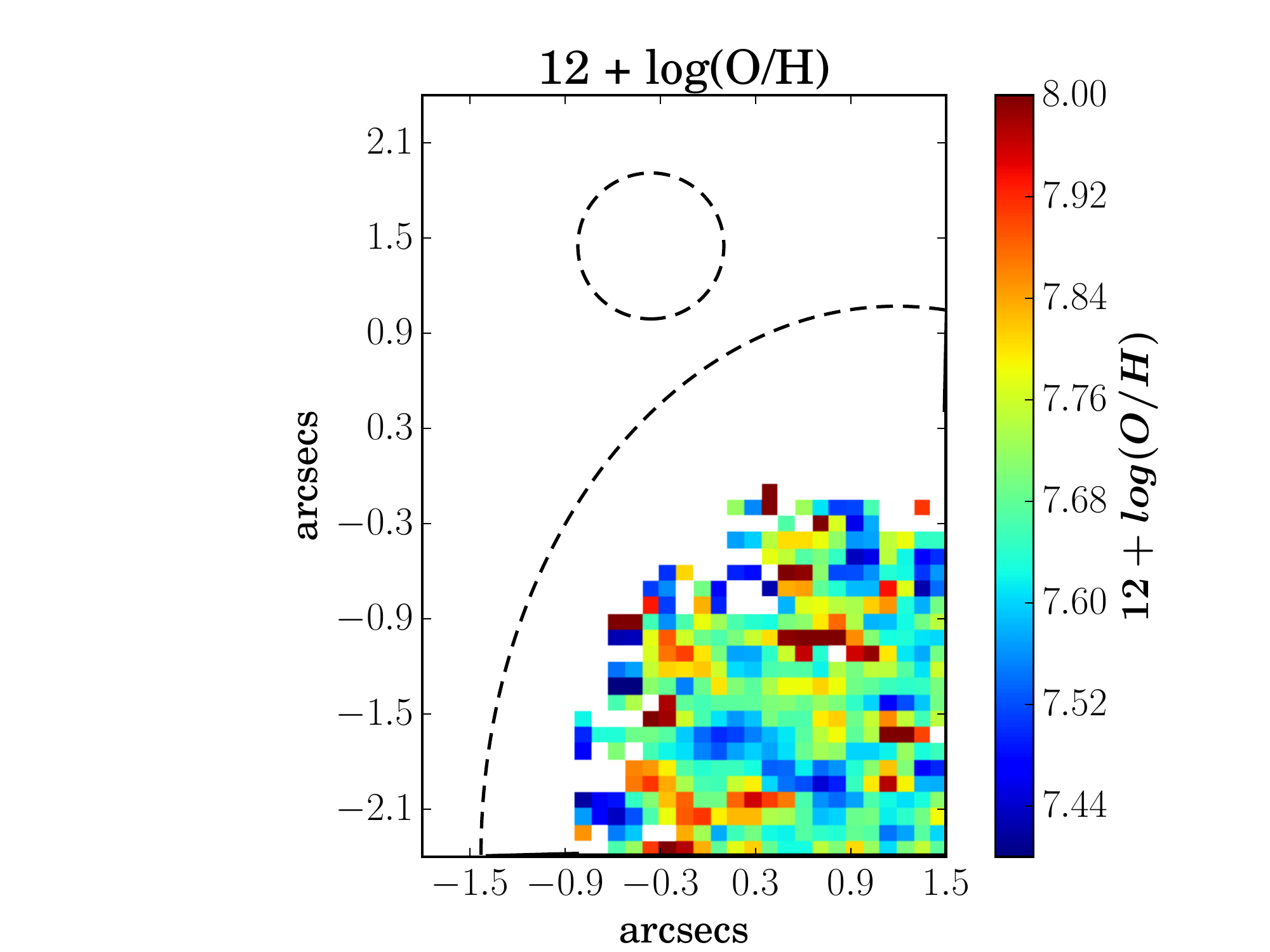}
	\includegraphics[width = 0.48\textwidth]{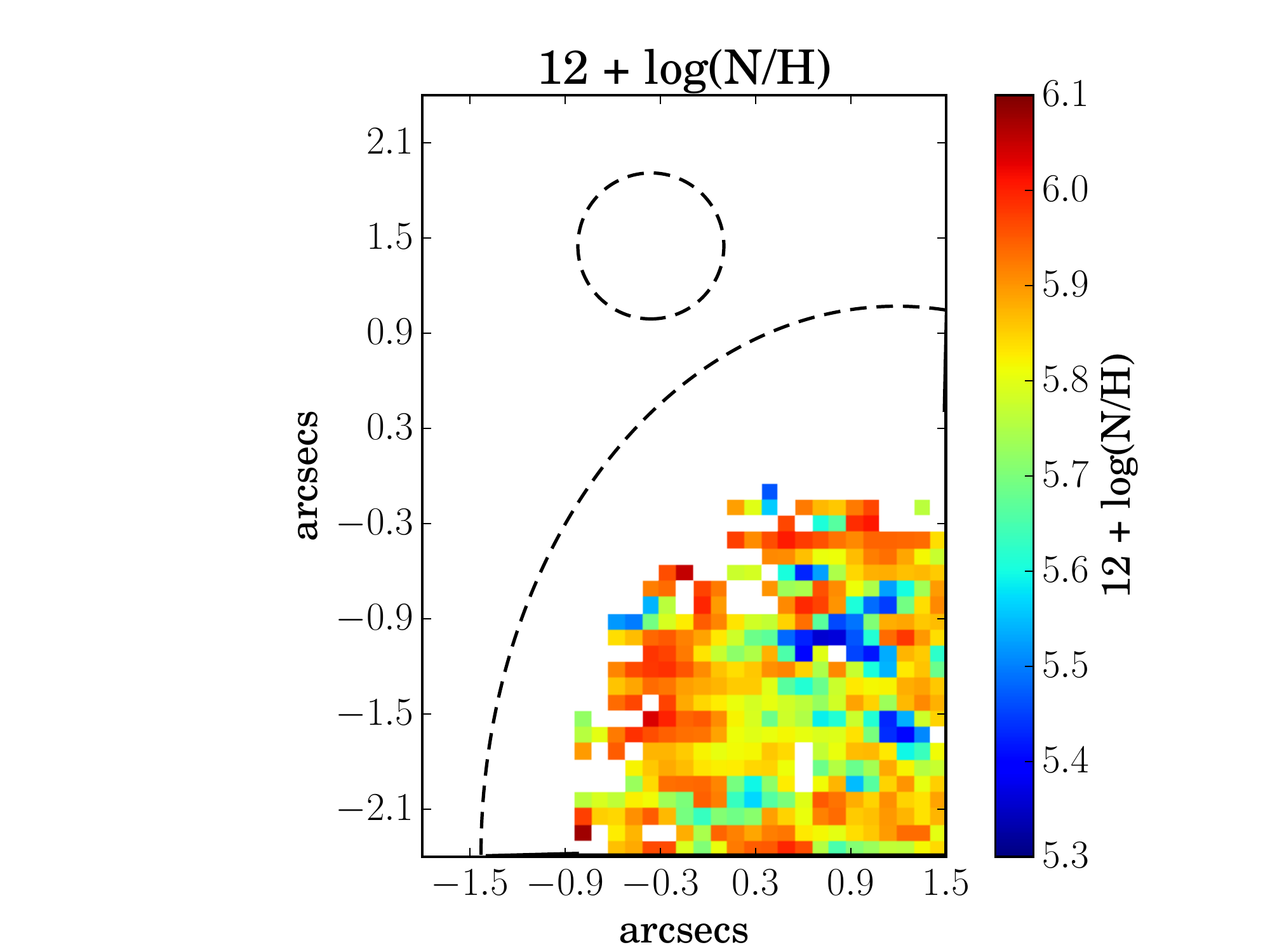}
	\includegraphics[width = 0.48\textwidth]{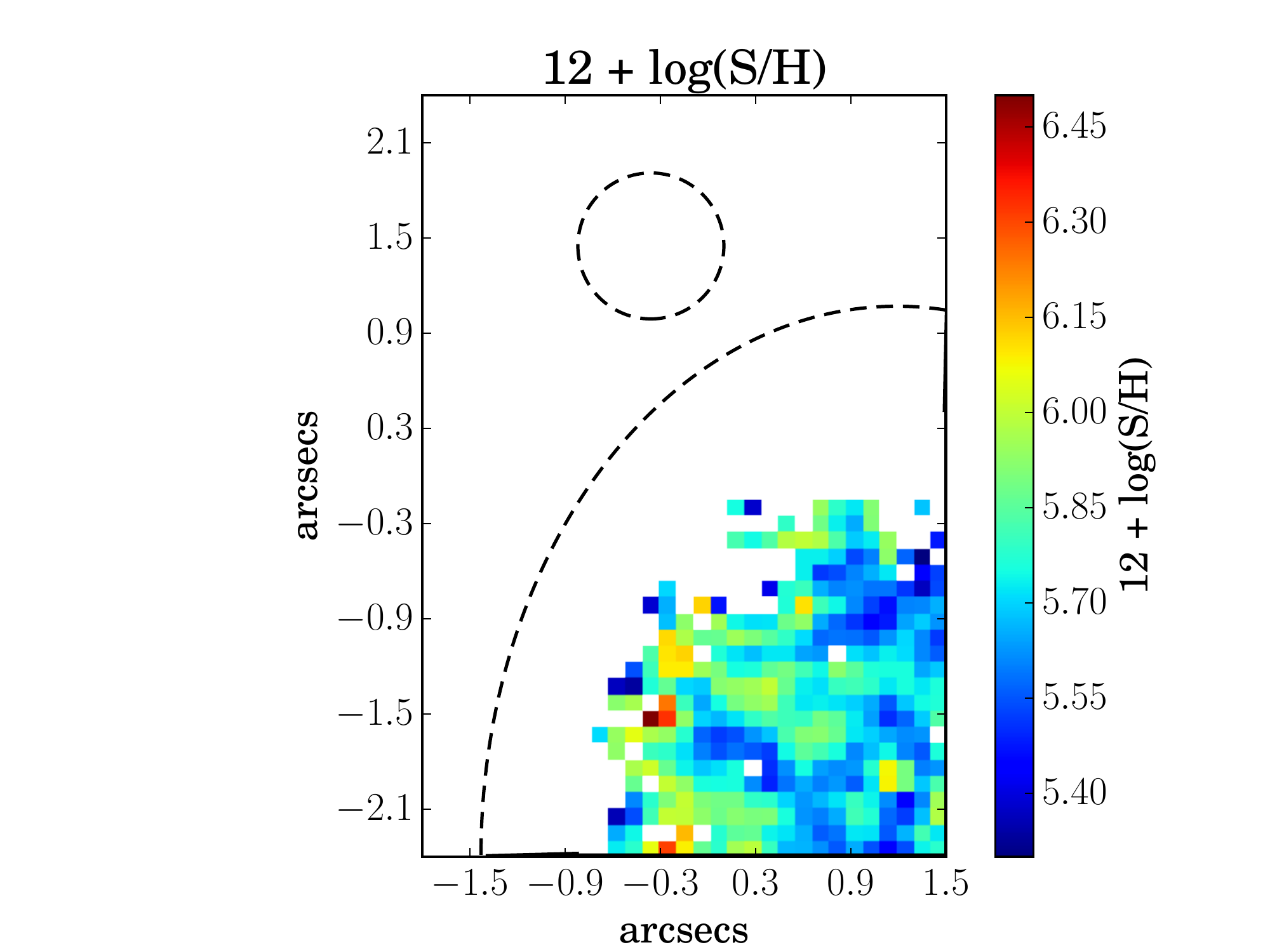}
	\includegraphics[width = 0.48\textwidth]{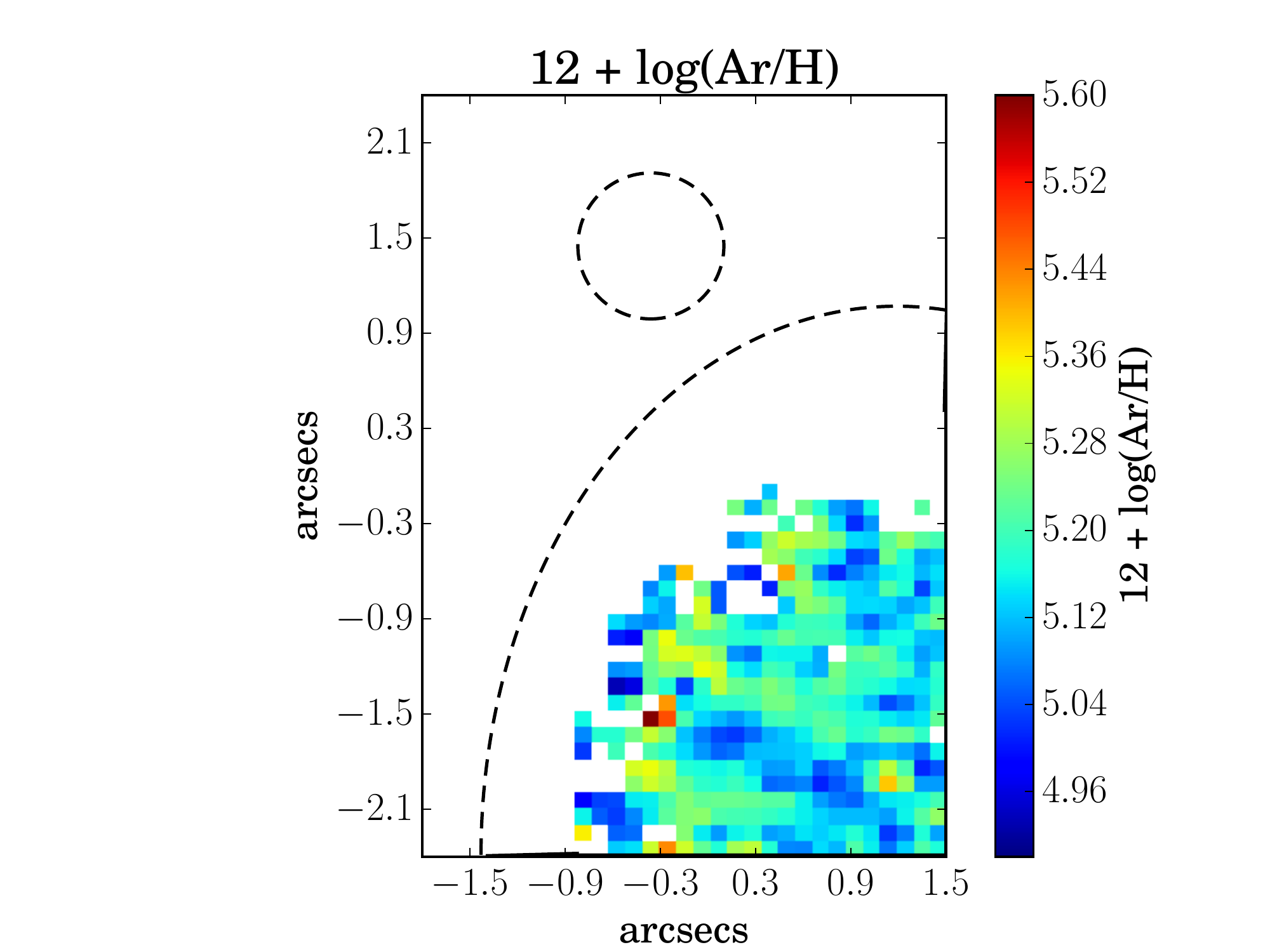}
	\includegraphics[width = 0.48\textwidth]{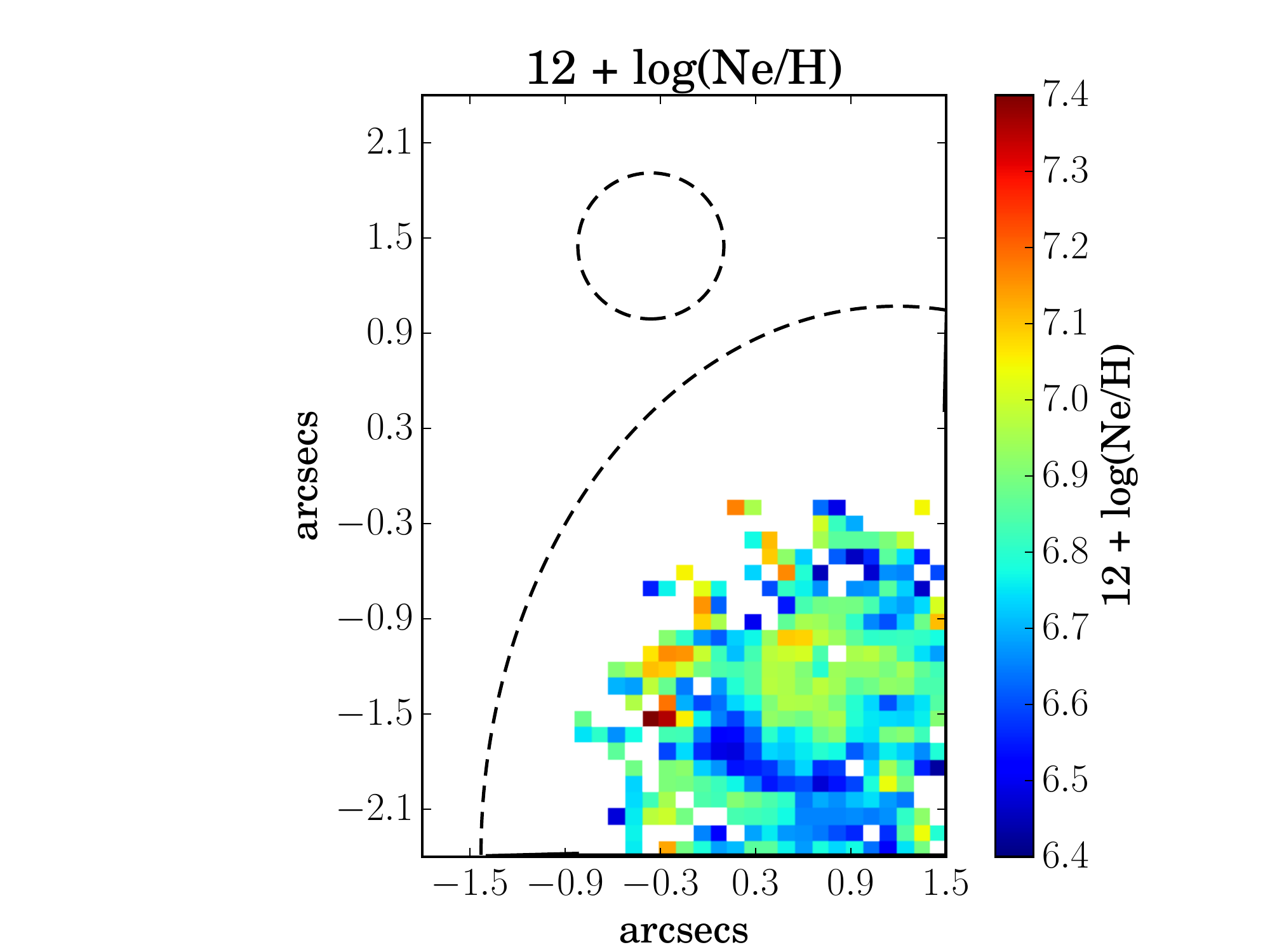}
	\caption{Maps of elemental abundances of oxygen, nitrogen, sulphur, argon and neon. They are obtained by first estimating the ionic abundances and then applying the ICF on a pixel-by-pixel basis. The dashed quarter ellipse and circle indicate Region 1 and Region 2, respectively. Single pixel features or few-pixel features should be interpreted with
			caution since the seeing FWHM (0.6 arcsec) extends over 6 pixels. The median uncertainties on elemental abundance maps of
			oxygen, nitrogen, sulphur, argon and neon are 0.08 dex, 0.19 dex, 0.3
			dex, 0.3 dex and 0.3 dex, respectively, as estimated from the corresponding uncertainty maps given in Figure \ref{fig:uncertainty Z}.}
	\label{figure:elemental}
\end{figure*}

\begin{figure*}
	\centering
	\includegraphics[width = 0.48\textwidth]{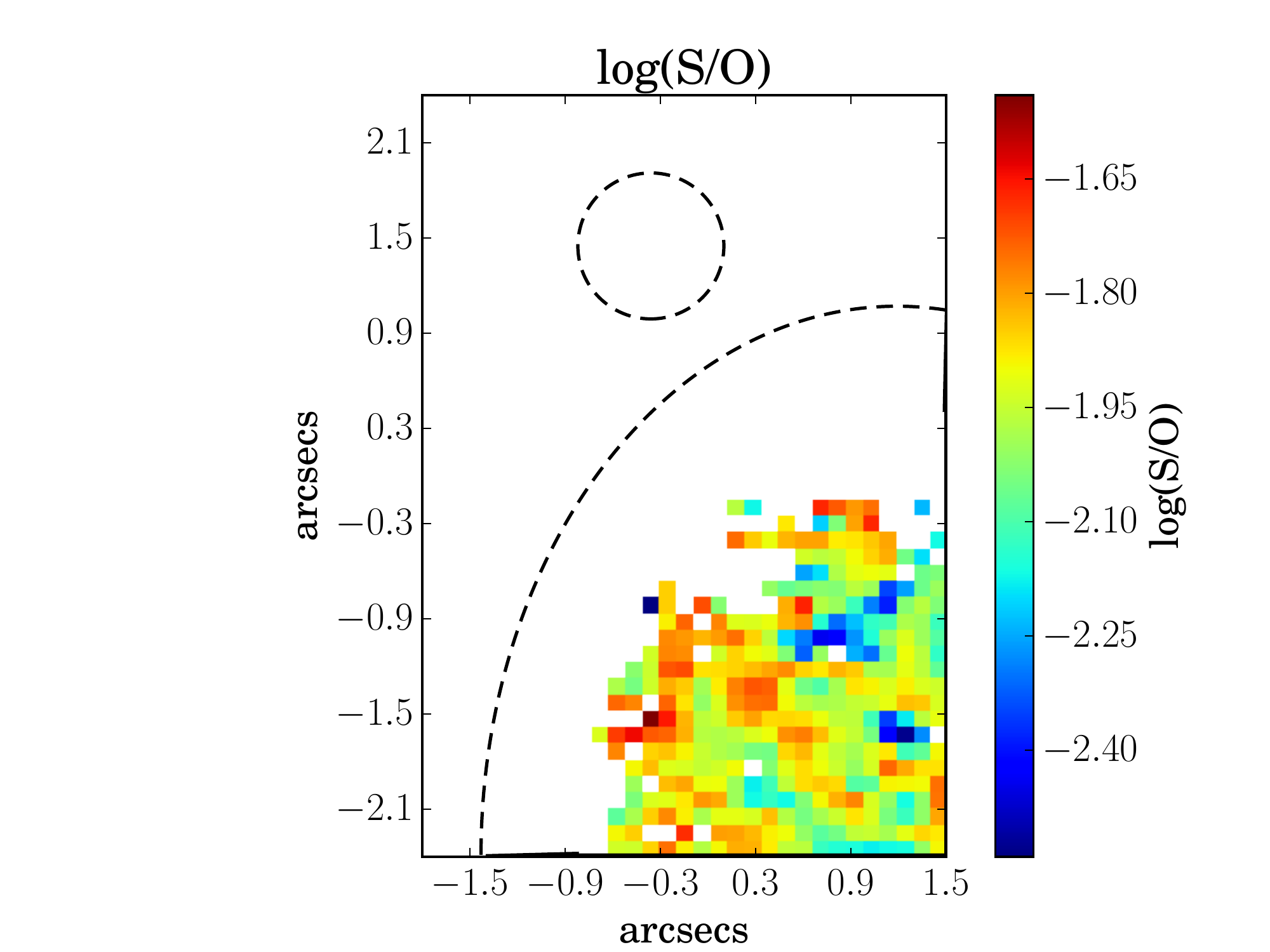}
	\includegraphics[width = 0.48\textwidth]{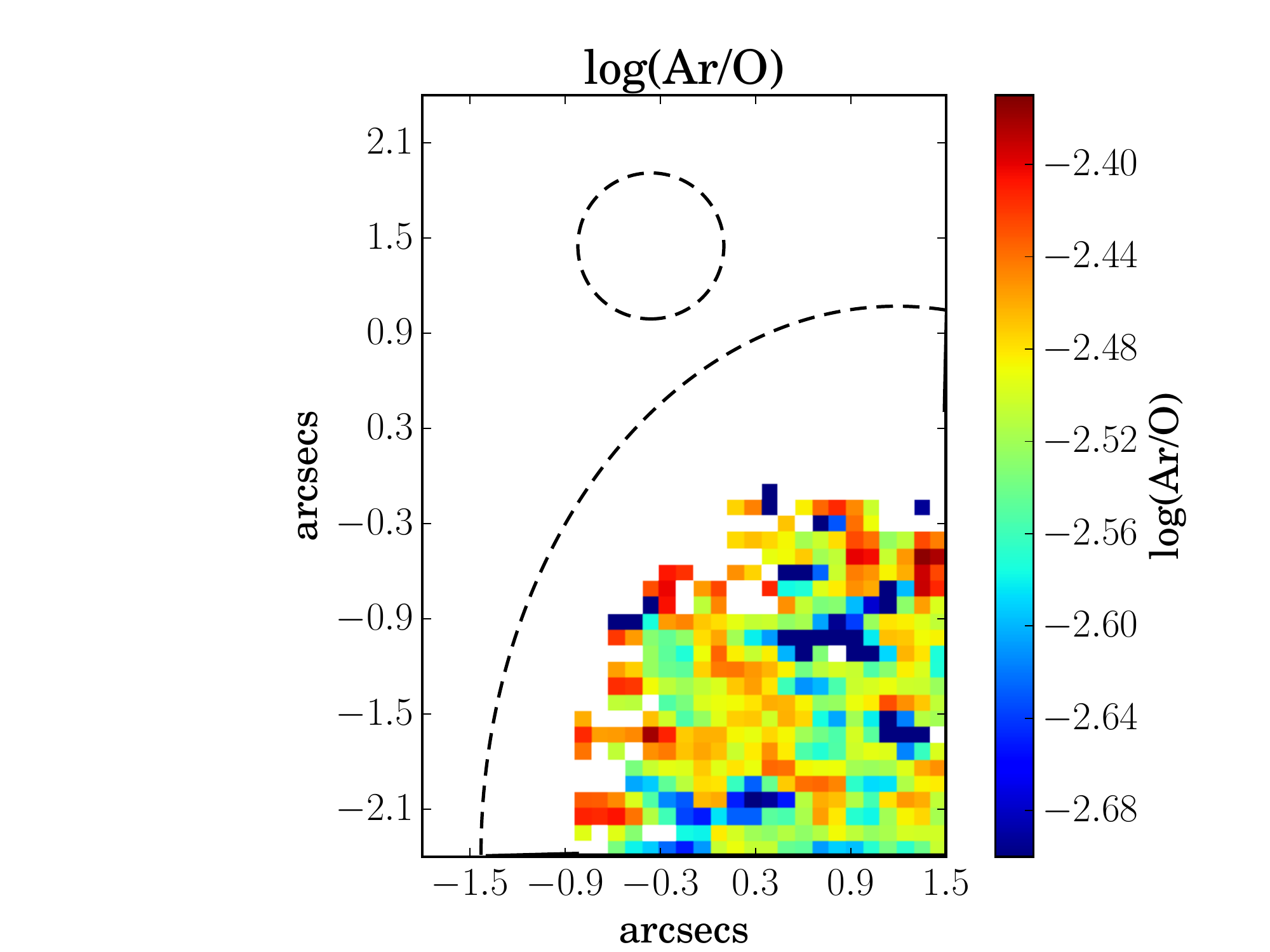}
	\includegraphics[width = 0.48\textwidth]{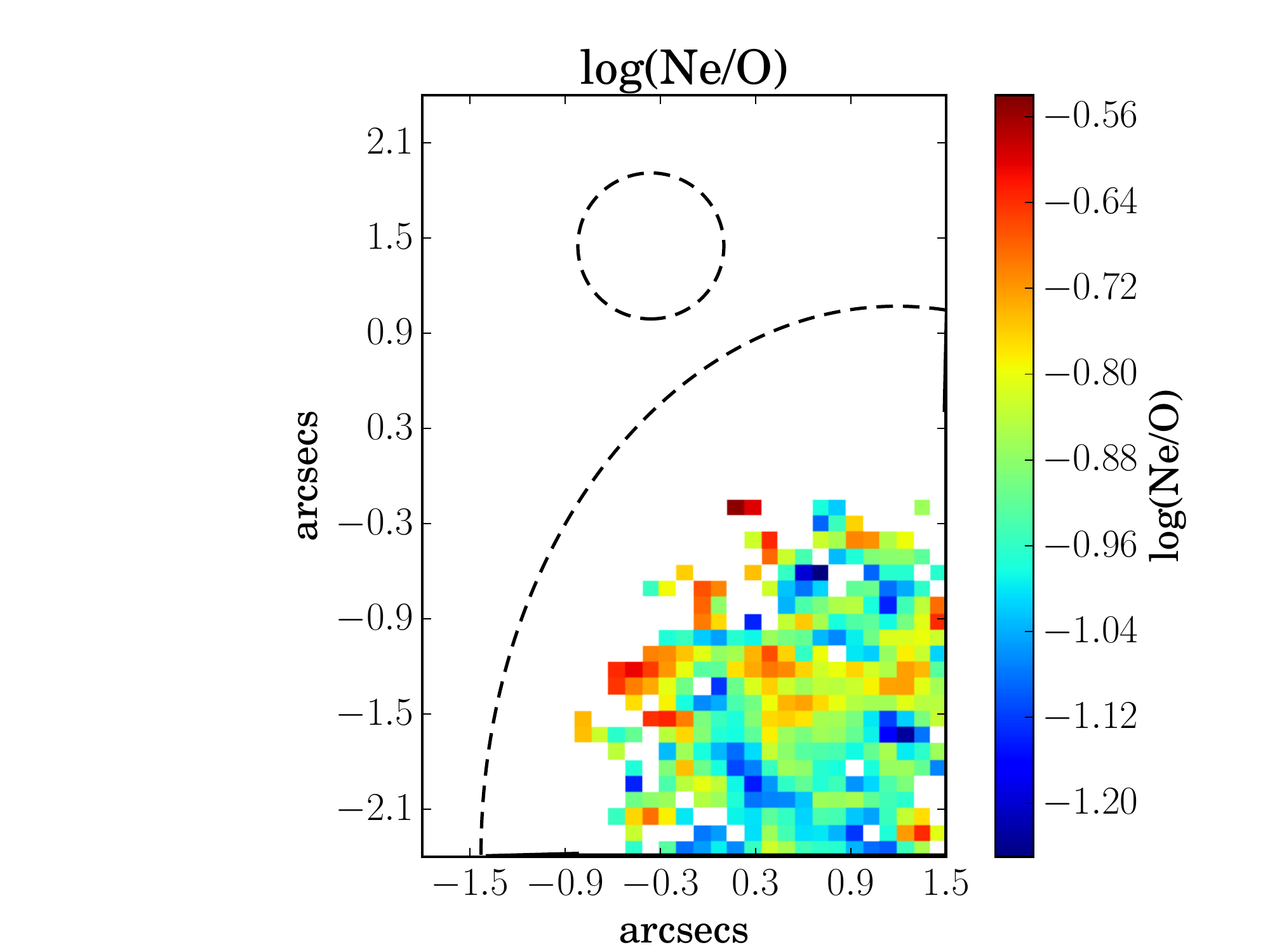}
	\includegraphics[width = 0.48\textwidth]{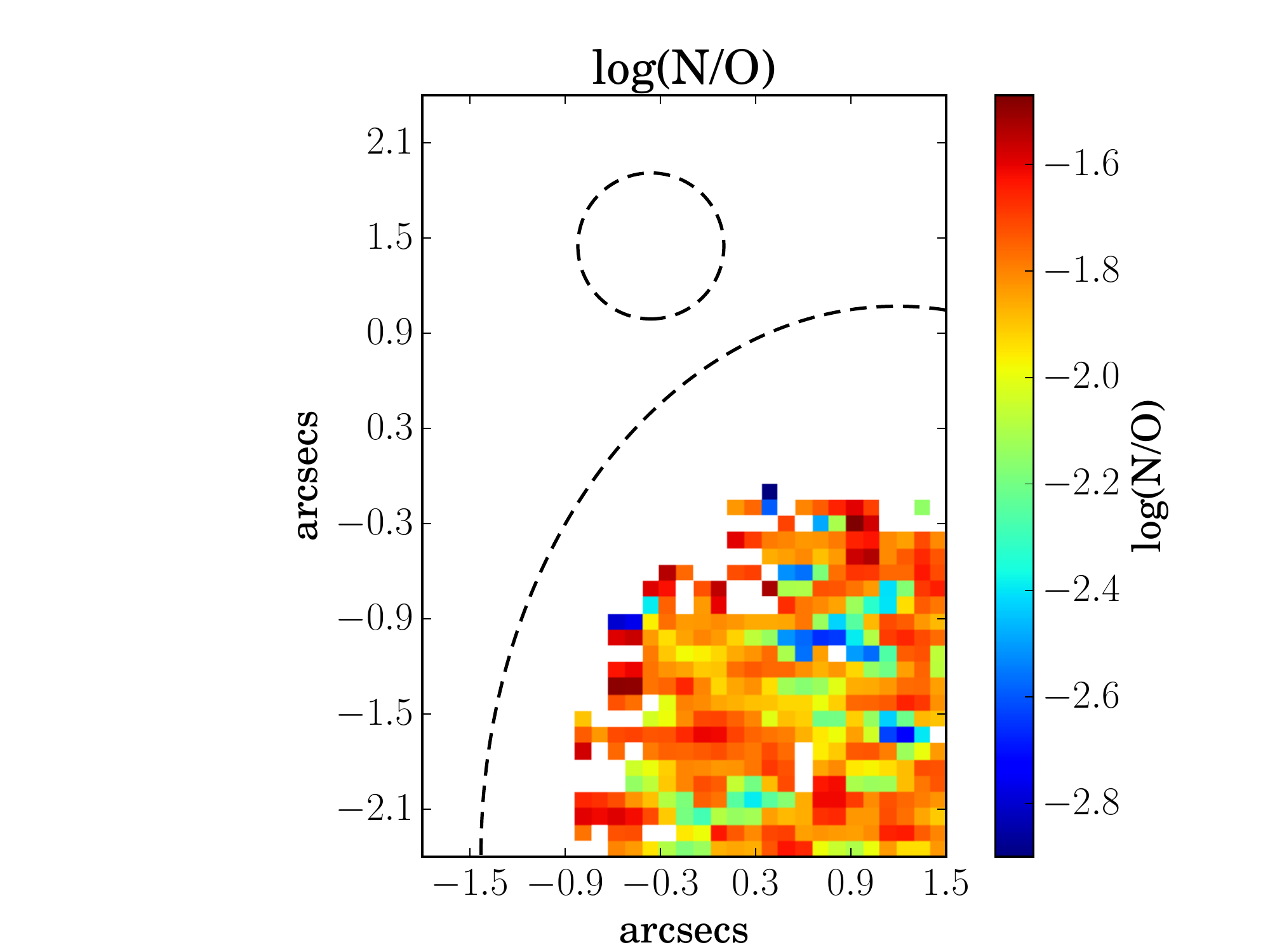}
	\caption{Abundance ratios of S/O, Ar/O, Ne/O and N/O. The dashed quarter ellipse and circle indicate Region 1 and Region 2, respectively. Single pixel feature or few-pixel features should be interpreted with caution since the seeing FWHM (0.6 arcsec) extends over 6 pixels. The median uncertainties on maps of S/O, Ar/O, Ne/O and N/O are 0.3 dex, 0.3 dex, 0.3 dex and 0.2, respectively, as estimated from the corresponding uncertainty maps given in Figure \ref{fig:uncertainty Z/O}.}
	\label{figure:Abundance ratios}
	
\end{figure*}

\subsection{Chemical Abundances}
\label{section:abundance}

\subsubsection{Ionic and Elemental Abundance}
\label{section: ionic elemental}

For both spatially-resolved data and the integrated spectrum of Region 1, we estimate the ionic abundances (O$^+$/H$^+$, O$^{2+}$/H$^+$, N$^+$/H$^+$, S$^+$/H$^+$, S$^{2+}$/H$^+$, Ne$^{2+}$/H$^+$ and Ar$^{2+}$/H$^+$). We calculate the corresponding ionisation correction factor (ICF) for each elemental species. Different ICF recipes exist which may or may not depend on metallicity, depending on the assumptions of the photoionisation models used for deriving ICF expressions \citep[see e.g.][]{Izotov2006, Perez-Montero2007a}. Such details of the ICF prescription for each ionic species are described below. The ICFs are combined with ionic abundances to estimate the elemental abundances of oxygen, nitrogen, sulphur, neon and argon. 
Table \ref{table:chemical properties} presents the ionic and elemental abundances obtained from the integrated spectrum of Region 1 and also the spatially-resolved maps for which we give the median and the uncertainty of all valid data. We find that the inferred values of the quantities for the integrated spectrum and spatially-resolved data are in reasonable agreement with each other. In the following points, we describe the methodology used to derive the chemical abundance maps shown in Figure \ref{figure:elemental}. Corresponding uncertainty maps are presented in Figure \ref{fig:uncertainty Z}.

\begin{itemize}
 \item  \textit{Oxygen abundance:}  \indent The metallicity of the ionised gas is expressed as the oxygen abundance as oxygen is the most prominent heavy element observed in the optical spectrum. The total oxygen abundance (12 + log(O/H)) is calculated from the sum of O$^+$/H$^+$ and O$^{++}$/H$^+$. Generally, O$^+$/H$^+$ is estimated from the oxygen doublets of [O \textsc{ii}] $\lambda\lambda$3727,3729. However, we could not detect these doublets in our data due to the low sensitivity of GMOS-N IFU in the blue end of the optical spectrum. Instead, we use [O \textsc{ii}] $\lambda\lambda$7320,7330 to estimate O$^+/H^+$ at the low-ionisation zone temperature T$_e$([O \textsc{ii}]) by employing the formulation of \citet{Kniazev2003}. 
  We estimated O$^{++}$/H$^+$  using the [O \textsc{iii}] $\lambda\lambda$4959, 5007 at the high ionisation zone temperature T$_e$([O \textsc{iii}]) using the formulation of \tPerezMontero. 
   Finally we  combined O$^+$/H$^+$ and O$^{++}$/H$^+$, and obtained the total metallicity. Map is shown in Figure \ref{figure:elemental} (upper-left panel).
	
	
	\indent 
	
	 \item  \textit{Nitrogen abundance:}  [N \textsc{ii}] $\lambda$6548 is not detected with S/N $>$ 3, but [N \textsc{ii}] $\lambda$6583 is detected at sufficient S/N.  Assuming an intrinsic emission line ratio between the two nitrogen lines [N \textsc{ii}] $\lambda$6548 = (1/2.9)[N \textsc{ii}] $\lambda$6583, we estimated the dereddened [N \textsc{ii}] $\lambda$6548 from the dereddened flux of [N \textsc{ii}]$ \lambda$6583. Using the dereddened flux of [N \textsc{ii}] lines and T$_e$([N \textsc{ii}]) estimated in Section \ref{section:Te}, we employed the formula given in \tPerezMontero~ to map 12 + log(N$^+$/H$^+$). We estimated ICF(N$^+$) from the abundance maps of total oxygen and singly-ionised oxygen, and finally mapped 12 + log(N/H) (Figure \ref{figure:elemental}, upper-right panel).
	
	 \indent

	\item  \textit{Sulphur abundance:} \indent We estimated singly-ionised sulphur abundance (S$^+$/H$^+$) by using the dereddened [S \textsc{ii}] $\lambda\lambda$6717, 6731 and T$_e$([O \textsc{ii}]). The auroral line [S \textsc{iii}] $\lambda$6312 was used along with T$_e$([S~\textsc{iii}]) to estimate S$^{2+}$/H$^+$. All of these estimates were derived by using the corresponding expressions given in \tPerezMontero.  We estimated ICF(S$^+$ + S$^{2+}$) using the classical formula from \citet{Stasinska1978}, but using $\alpha$ = 3.27 derived by \citet{Dors2016}. We finally combined ionic abundance with the ICF correction to derive the elemental sulphur abundance, 12 + log(S/H). Corresponding map is shown in Figure \ref{figure:elemental} (middle-left panel).
	
	\indent
	
	\item \textit{Neon abundance:} \indent We estimated doubly-ionised neon abundance (Ne$^{2+}$/H$^+$) by using the dereddened [Ne \textsc{iii}]  $\lambda$3869 and T$_e$([O \textsc{iii}]), and derived  ICF(Ne$^{2+}$) using the expressions from \citet{Perez-Montero2007a}, which is independent of metallicity. Combining the ionic-abundance with the ICF, we derived the elemental neon abundance, 12 + log(Ne/H). Map is shown in Figure \ref{figure:elemental} (lower panel).
	
	\indent
	
	\item \textit{Argon abundance:} \indent We estimated doubly-ionised argon abundance (Ar$^{2+}$/H$^+$) by using the dereddened [Ar \textsc{iii}]  $\lambda$7135 and T$_e$([S \textsc{iii}]) the temperature of intermediate-ionisation zone, in the expression given in  \tPerezMontero. We estimated the metallicity-independent ICF(Ar$^{2+}$) using expression given in \citet{Perez-Montero2007a} to derive the elemental argon abundance, 12 + log(Ar/H). Map is shown in Figure \ref{figure:elemental} (middle-right panel).	
	
\end{itemize}	

\indent The abundance maps of 12 + log(O/H) and 12 + log(N/H) shows relatively high and low values, respectively in the north-east of the peak H$\alpha$ flux exactly at the same location where we find high electron density (Figure \ref{figure:TeNe}, lower panel). To check this, we estimated electron densities from the integrated spectra of two different regions within this over-density region. Though their electron densities are high, the associated uncertainties are also large, which indicates that the region may well have the same electron density as the majority of the pixels, i.e. N$_e$ $<$ 50 cm$^{-3}$. So we suspect that the abundance patterns on the maps of 12 + log(O/H) and 12 + log(N/H) may not be real variations. 

\indent Our derived values of 12 + log(O/H), 12 + log(O$^{2+}$/H$^+$), 12 + log(N$^+$/H$^{+}$) and Ne$^{2+}$/H$^{+}$ are in excellent agreement with those obtained by \tGuseva. However, the values of 12 + log(O$^+$/H$^+$), S$^+$/H$^+$, S$^{2+}$/H$^+$ and Ar$^{2+}$/H$^+$ do not agree with those from \tGuseva, since these abundance estimates depend on the low-ionisation zone temperature (T$_e$ ([O \textsc{ii}])) and the intermediate zone temperature  (T$_e$ ([S \textsc{iii}])), and these temperature estimates in our work are different from those obtained by \tGuseva~ for the reasons described in section \ref{section:Te}. 


\indent In the above analysis, our derived ICFs do not match those of \tGuseva. As such, we estimated the ICFs for data from \tGuseva~using our adopted recipes for ICF calculation, which now exactly match our derived ICFs.  The total elemental abundances of nitrogen, sulphur, neon and argon do not match those derived from \tGuseva~because of the following two differences: firstly the ionic abundances in the two works are different as a result of the different electron temperatures, secondly different ICF prescriptions have been used in the two works. This results in obvious differences in the abundance ratios between our work and \tGuseva. In addition to the ICF and ionic abundances, difference in ionic and elemental abundances might also be due to spatial effects since the GMOS-IFU FOV does not coincide exactly with the long-slit of \tGuseva.

\subsubsection{Abundance Ratios}
\label{section: abundance ratios}

\indent  Figure \ref{figure:Abundance ratios} shows the maps of abundance ratios, while corresponding uncertainty maps are presented in Figure \ref{fig:uncertainty Z/O}. We find an abundance pattern in the north-east region in the maps of S/O, Ar/O and N/O similar to that of the electron density map. Hence, these may not be real as discussed in Section \ref{section:  ionic elemental}. The abundance ratios for the integrated spectrum and spatially-resolved maps of Region 1 agree with each other within error bars as tabulated in Table \ref{table:chemical properties}. Our abundance ratio estimates are systematically lower than those of a sample of 40  BCDs analysed by \citet{Izotov1999}. However, such low values have been reported in the IFU studies of BCDs. One such BCD is the interaction-induced starburst UM 462 \citep{James2010} where abundance ratios of N/O, S/O and Ar/O are found to be in agreement with SBS 1415+437 from our study. Likewise, the values of N/O, Ne/O and Ar/O in our BCD are in agreement with those of the main body of another metal-poor BCD Tol 65 \citep{Lagos2016} where tidal interaction has also been proposed. Though there have been no study reporting the existence of a companion or an interaction-induced starburst in the current BCD under study SBS 1415+437, the comet-shaped structure of this BCD might be indicative of tidal interaction of this galaxy with the local intergalactic environment.

\begin{figure}
	\centering
	\includegraphics[width=0.2\textwidth]{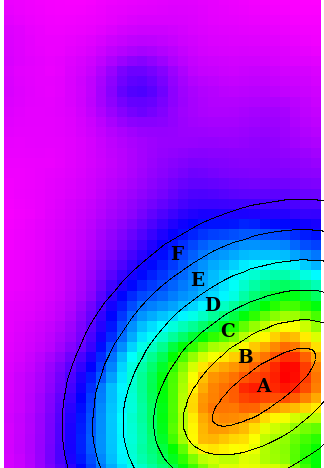}
	\caption{H$\alpha$  flux map is used to separate the H $\textsc{ii}$ region into six equally-spaced segments (A, B, C, D, E, F) such that each of them have similar overall flux within each annulus. These annuli are used for the chemical inhomogeneity investigation described in Section \ref{section:variation abundances}. Please see location of Region 1 in Figure 2.}
	\label{figure:segments}
\end{figure}

 \begin{figure*}
 	\centering
 	\includegraphics[width = 0.45\textwidth]{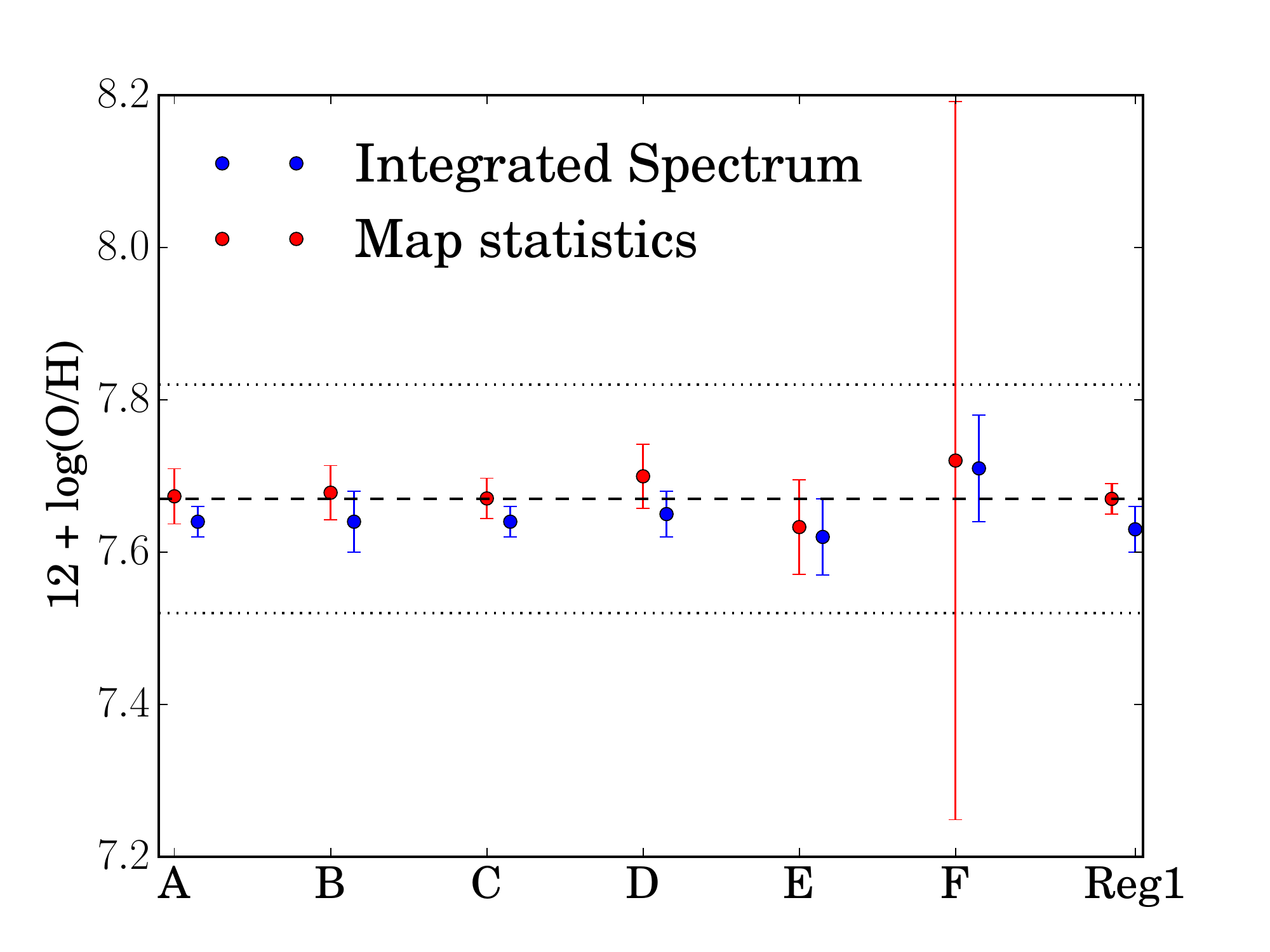}
 	\includegraphics[width = 0.45\textwidth]{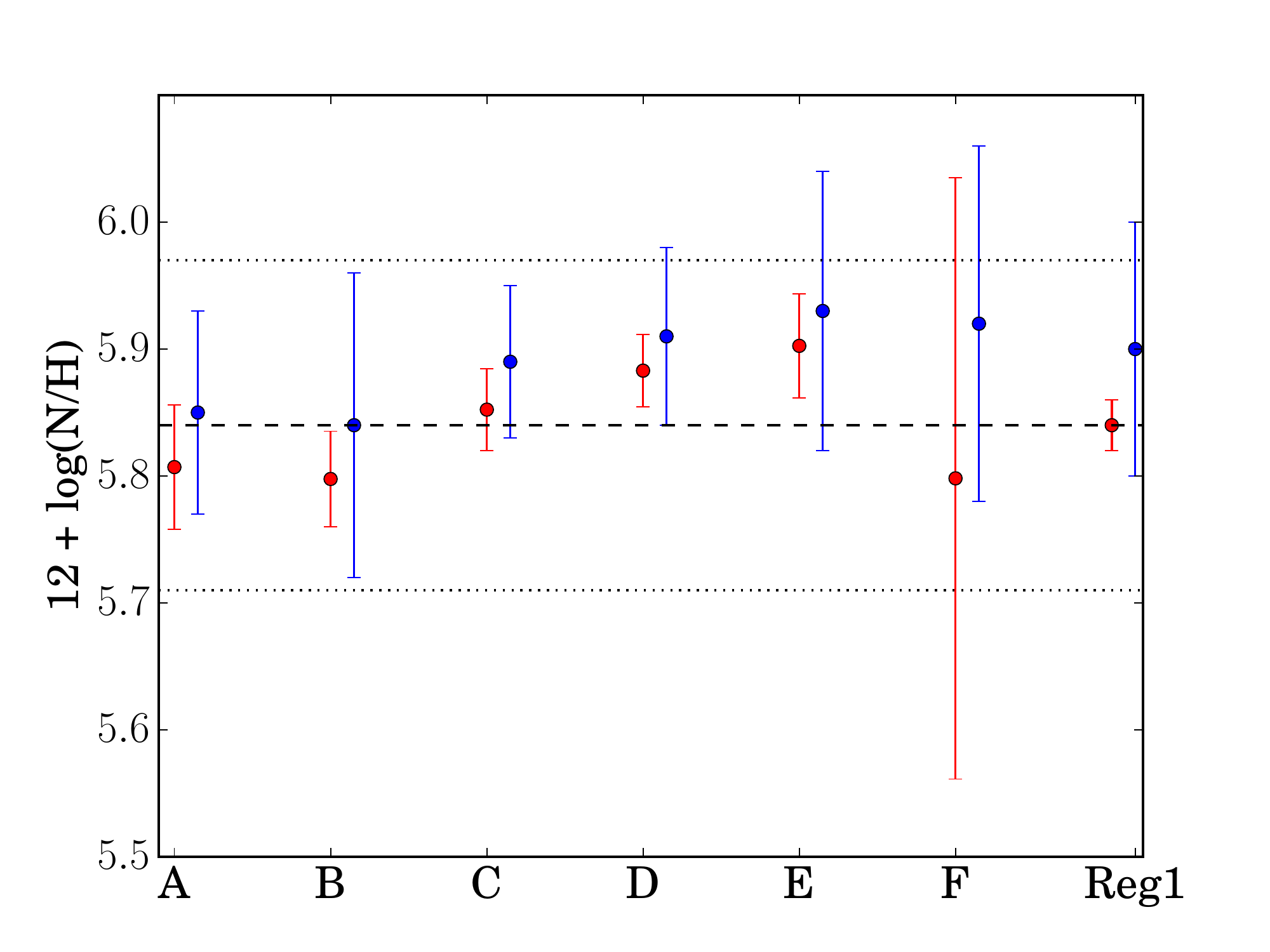}
 	\includegraphics[width = 0.45\textwidth]{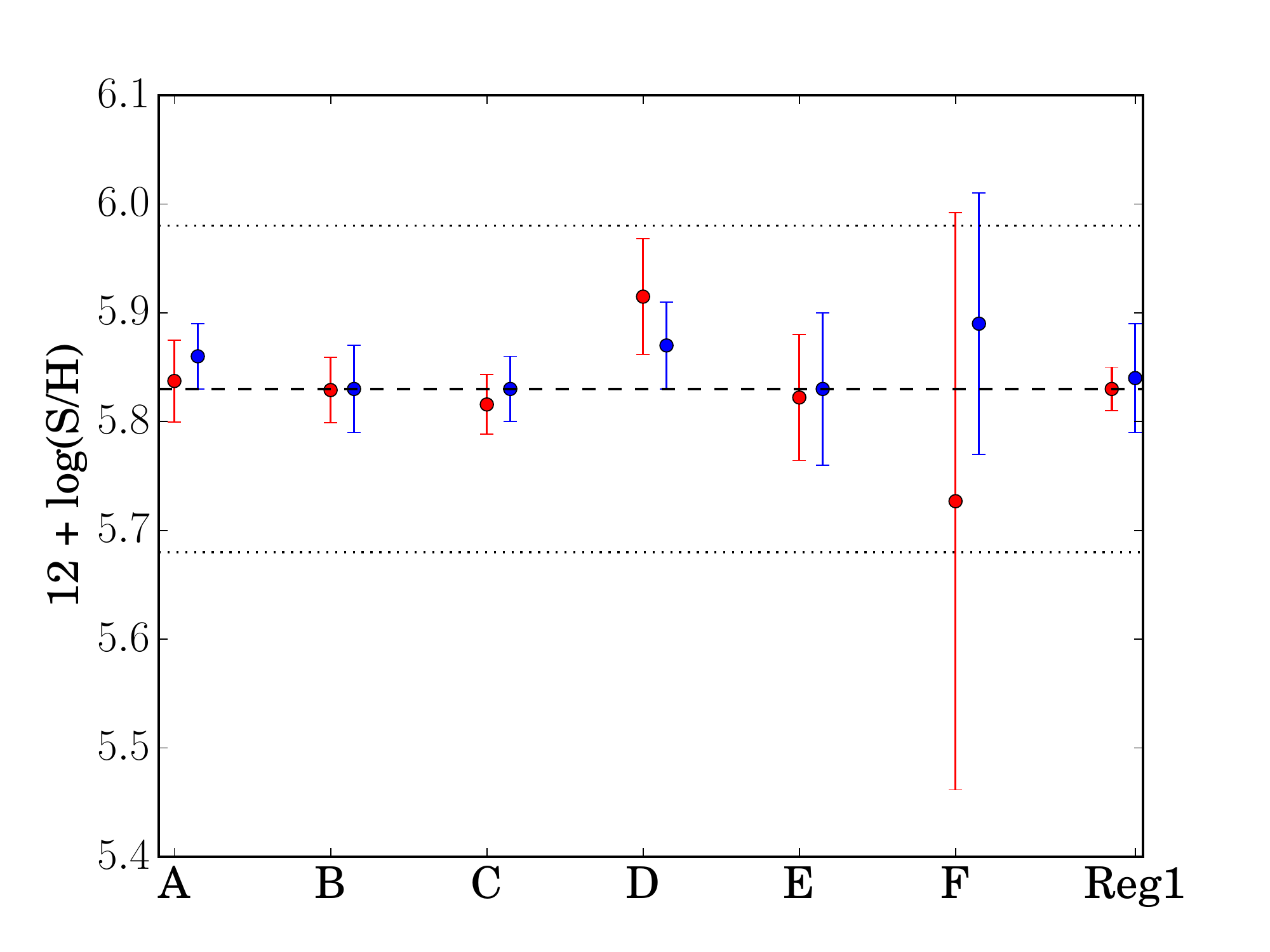}
 	\includegraphics[width = 0.45\textwidth]{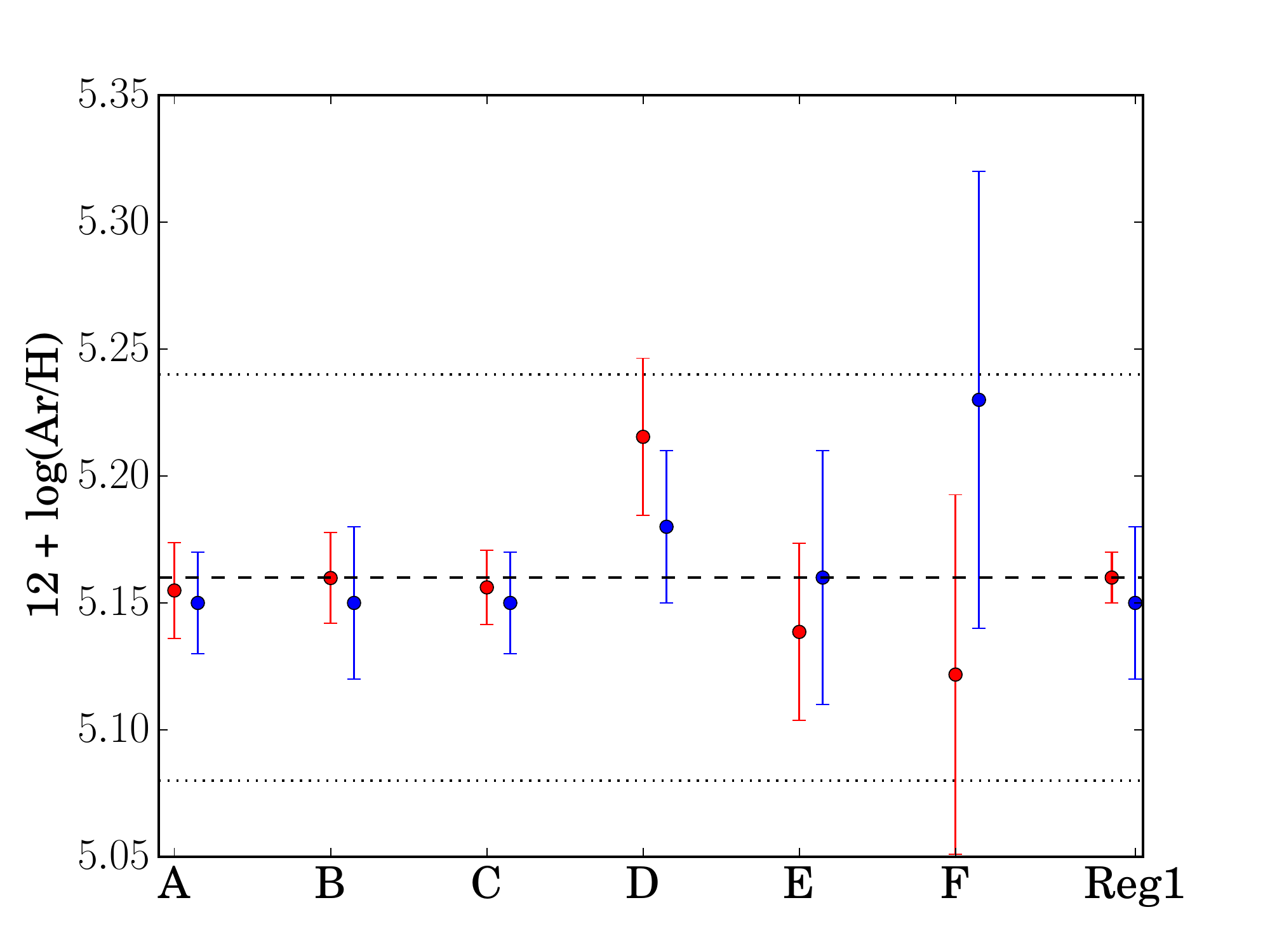}
 	\includegraphics[width = 0.45\textwidth]{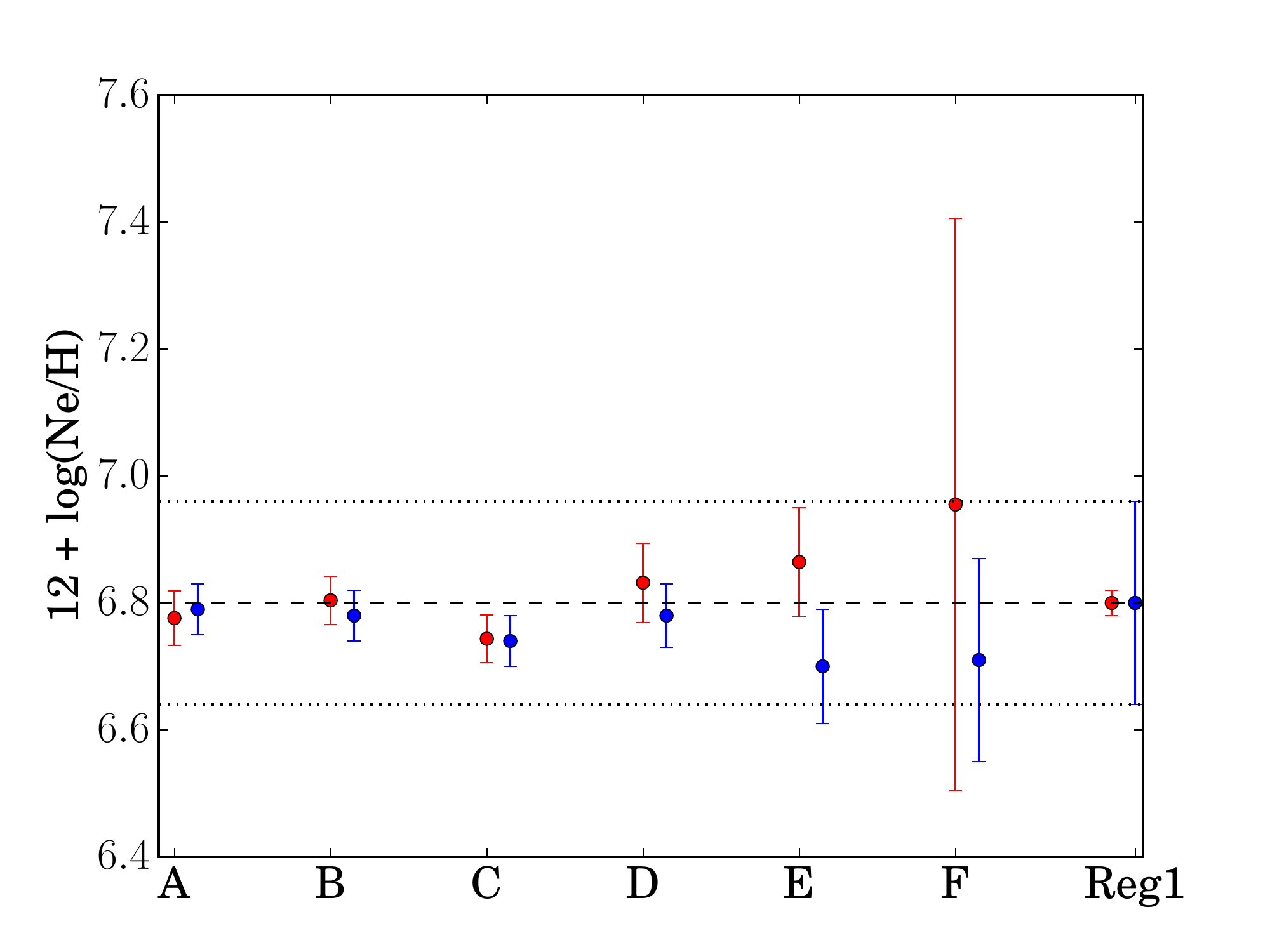}
 	
 	\caption[Variation of elemental abundances across Region 1 in SBS 1415+437]{Variation of elemental abundances across Region 1. In all panels, red points with vertical bars correspond to the median of data and the associated uncertainty within annuli A, B, C, D, E, F (shown in Figure \ref{figure:segments}) of a given chemical abundance map. The error bar corresponding to annulus F is large because of the paucity of data points in this annulus. Blue points are the observables derived from the integrated spectra of each annulus. We also present the corresponding values of entire Region 1 (marked as ``Reg1"). The black dashed line corresponds to the median of all data on an abundance map. The dotted black lines indicate the $\pm 1\sigma$ as determined from the standard deviation of all valid data in a map.}
 	\label{figure:element variation}	
 \end{figure*}
 
 \begin{figure*}
 	\centering	
 	\includegraphics[width = 0.45\textwidth]{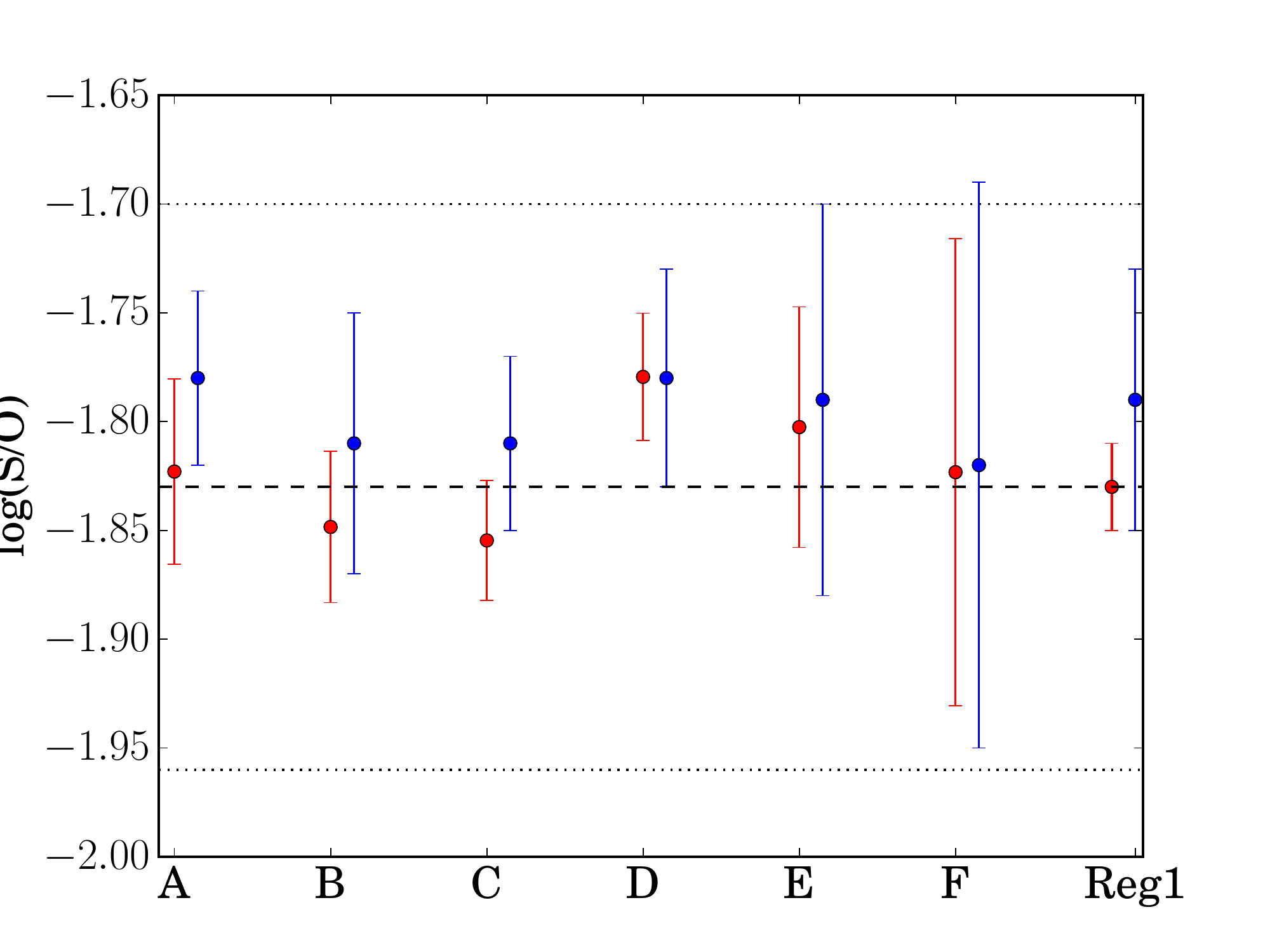}
 	\includegraphics[width = 0.45\textwidth]{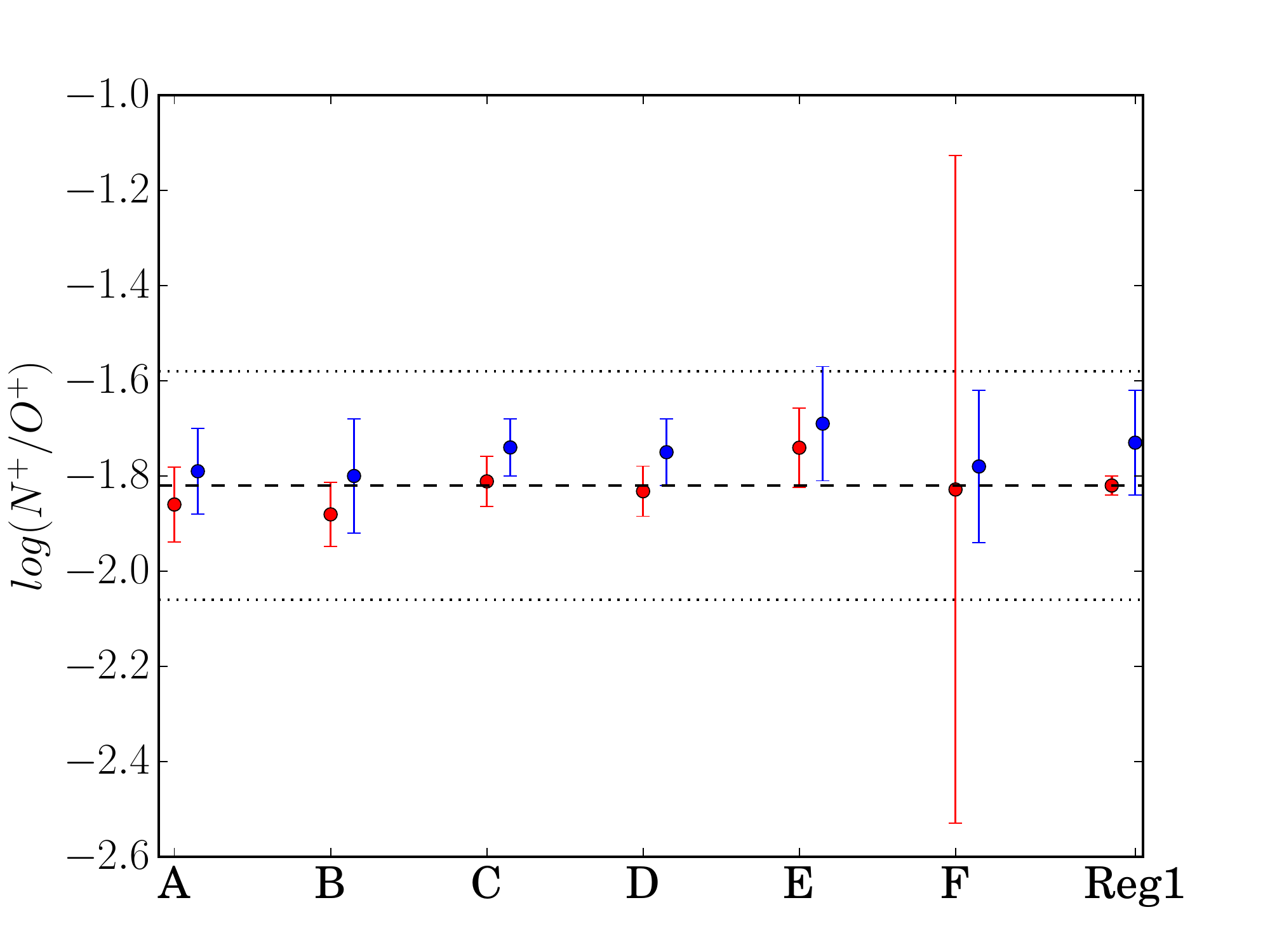}
 	\includegraphics[width = 0.45\textwidth]{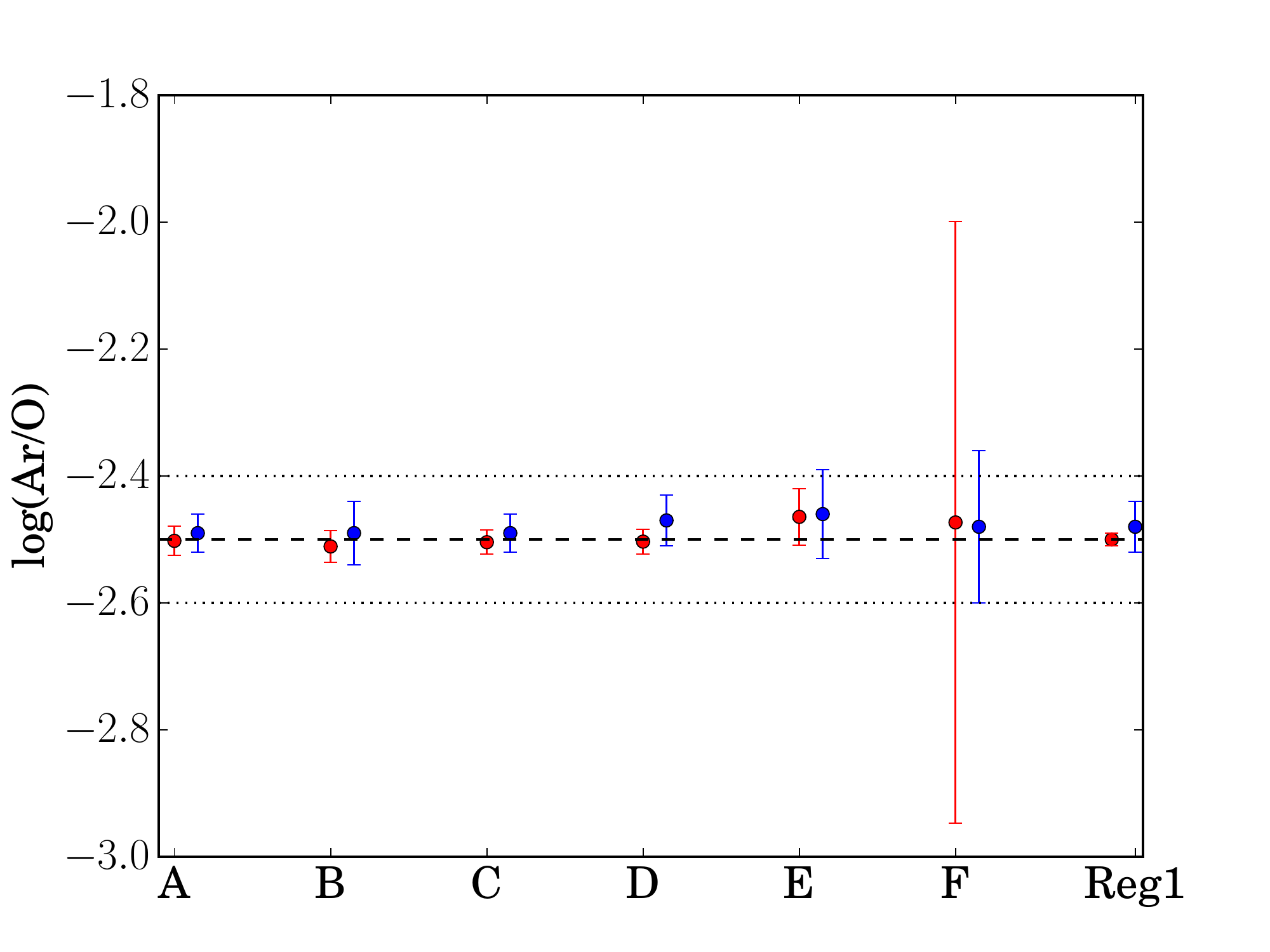}
 	\includegraphics[width = 0.45\textwidth]{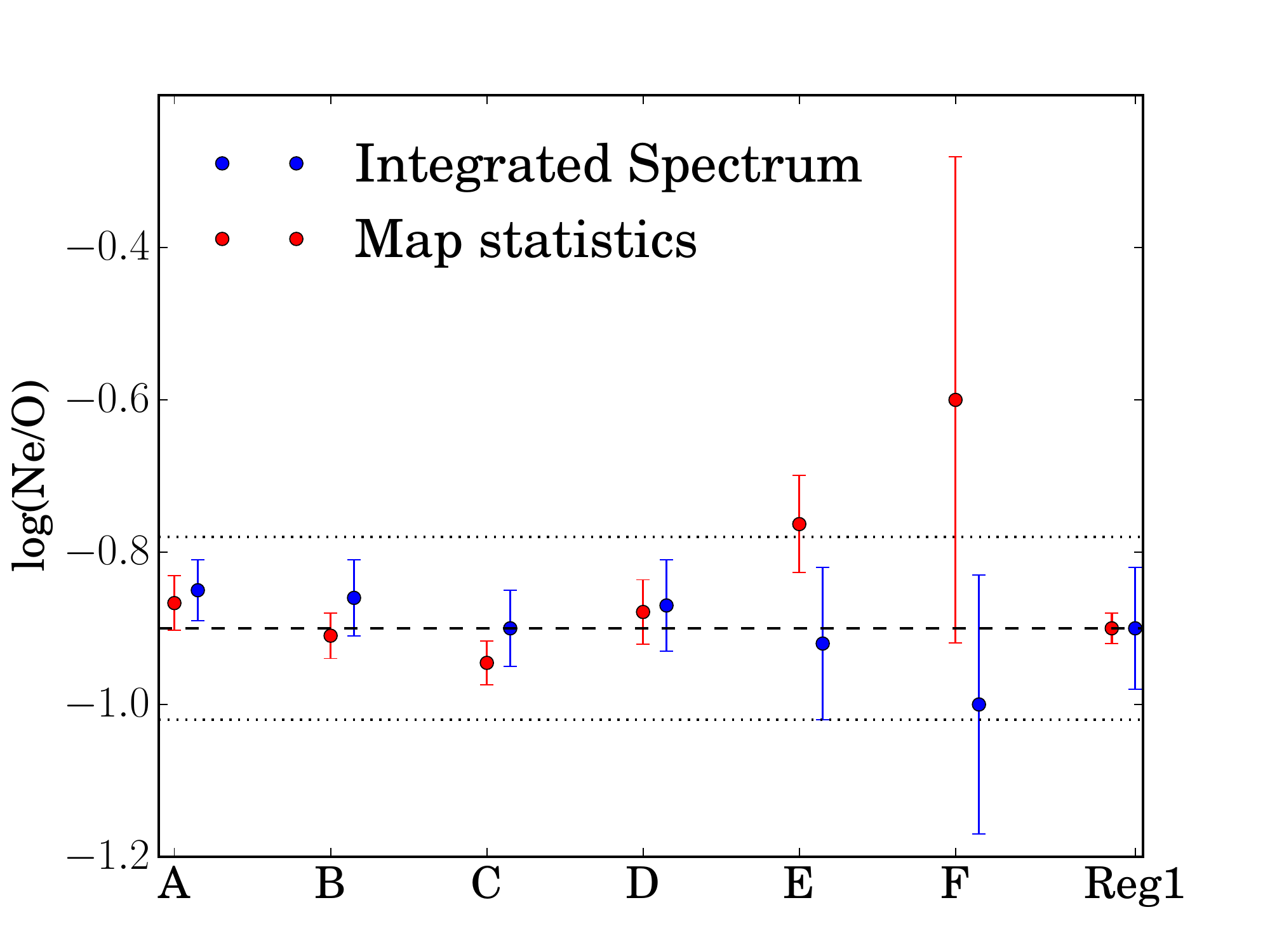}
 	\caption[Variation of abundance ratios across Region 1 in SBS 1415+437]{Variation of abundance ratios across Region 1 - see caption of figure \ref{figure:element variation} for details on legends.}
 	\label{figure:ratio variation}
 \end{figure*}

 \begin{figure*}
 	\includegraphics[width=0.8\textwidth]{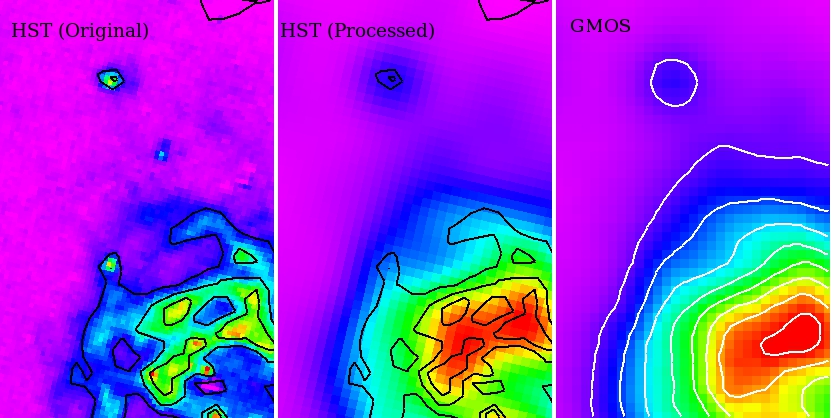}
 	\caption{Left: HST continuum subtracted image in F658N filter with pixel sampling of 0.05 arcsec. Middle: HST continuum subtracted image, convolved to the FWHM seeing of 0.6 arcsec and binned to 0.1 arcsec per pixel. The black contours on this panel are generated on the original HST image shown on the left panel. Right: GMOS H$\alpha$ image, where the white contours are generated on this image itself. Comparison of the three images shows that the region under study hosts multiple H \textsc{ii} regions, whose structures are smoothed out in the GMOS image due to seeing lowering the resolution of GMOS compared to HST.}
 	\label{figure:contours}
 \end{figure*}

\subsubsection{Variation of abundances}
\label{section:variation abundances}

\indent The high-resolution HST image shows distinct structure throughout the region of our study (Figure \ref{fig:hst}). However, the observed H$\alpha$ emission line flux map of Region 1 shows approximately an elliptical distribution (Figure \ref{fig:observed flux}).  No elliptical distribution is obvious in any abundance maps (Figure \ref{figure:elemental}) or abundance ratio maps (Figure \ref{figure:Abundance ratios}), though we note that an over-density is observed in Ne abundance map (Figure \ref{figure:elemental}, bottom panel) in the region north-east of the peak of H$\alpha$ emission. In this Section, we explore the small scale abundance variation to investigate how well-mixed the gas is within the region of study. 

\indent We perform this analysis via three different methods, though we show results from only two methods. The first two methods are based on the segmentation of Region 1 in six elliptical annuli of equal widths selected on the basis of the H$\alpha$ flux distribution such that each of the annuli have roughly similar levels of flux within a given annulus. The position of the chosen elliptical annuli are shown in Figure \ref{figure:segments}, and are marked as A, B, C, D, E and F. In the first method, we estimated abundances and abundance ratios from the emission line fluxes from the integrated spectra within each annulus, which are reported in Table \ref{table:annuli abundances}.  In the second method, we estimate abundances and abundance ratios within each annulus from the spaxel maps instead of integrated spectra. We study the variation of estimates from the two methods from one annulus to another as described later. In the third method, we simply study the variation of abundances in a pixel with respect to its physical distance from the pixel of peak H$\alpha$ flux (i.e. radial distance in elliptical coordinates). This last method was performed to check if our method of segmentation could affect any observed trend.  

\indent Figures \ref{figure:element variation} and \ref{figure:ratio variation} show the variation of elemental abundances and abundance ratios from one elliptical annulus to another. The observables from the integrated spectra of each annulus are shown as blue dots. For all maps of abundances and abundance ratios, we estimate  median and uncertainty on the median  within each annulus, and are shown as red dots and vertical bars, respectively. The uncertainties on the median values for each sample point take
	into account the spatial correlation between adjacent spaxels arising
	from the resampling used in generating the data cubes. The overall average
	level of the correlation was estimated using an autocorrelation analysis of selected regions from the spaxel map after first removing larger
	scale gradients. In practice this provides an estimate of the factor (in this case $\approx 4$) by which the number of apparently independent
	spaxels in each elliptical annulus region should be reduced to allow for the induced spatial correlation. The abundance and abundance ratios of region 1 are also shown on the panels in Figures \ref{figure:element variation} and \ref{figure:ratio variation}, as estimated from the integrated spectrum (blue dots) and the maps (red dots) for a better visual comparison, and should not be considered for abundance variation analysis. In each panel, the dashed black horizontal line indicates the median ($\mu$) of the abundances or abundance ratio distribution of entire maps, while the two dotted black horizontal lines indicate $\mu$ $\pm$ $\sigma$, where $\sigma$ denoted the standard deviation of entire maps.

\indent Inspecting all panels in Figure \ref{figure:element variation} and \ref{figure:ratio variation}, we find that the integrated spectra values and mapped values in each annulus are in reasonable agreement with each other. Results of annulus F should be interpreted with caution where uncertainties on each observable are comparatively large because of a few valid data points ($\sim$ 2--3 depending on the map) in this annulus (see maps in Figures \ref{figure:elemental} and \ref{figure:Abundance ratios}). Though results of annulus F are not particularly informative, we have included them in our analysis as we do not wish to  discard any detection.

\indent No variation is found in 12 + log(O/H). There appears to be a variation in abundances of elements (other than O) and abundance ratios around annuli C and D, though these variations are $<$ 0.1 dex for each observable. Given that there exists a region of over-density in the Ne abundance map (Figure \ref{figure:elemental}), it is possible that the physical conditions are different within the region of study and is supported by the HST image which shows structures throughout this region. Figure \ref{figure:contours} shows a comparison of H$\alpha$ images obtained with the GMOS-IFU and HST (F658N). The corresponding continuum image (taken in the F606W filter of HST) was scaled and subtracted from the narrow-band F658N image, to obtain the continuum-subtracted H$\alpha$ image (Figure \ref{figure:contours}, left panel). The overlaid black contours are generated from this original HST image, showing resolvable structure. To compare the HST image directly with the GMOS H$\alpha$ image, we convolved it to FWHM seeing of 0.6 arcsec and rebinned to 0.1 arcsec per pixel. The resultant smoothed image (Figure \ref{figure:contours}, middle panel) has the black contours generated from the original HST image overlaid for comparison. The processed HST image is remarkably similar to the GMOS H$\alpha$ image (Figure \ref{figure:contours}, right panel), whose contours (white curves) also show a smooth variation of flux.

\indent Since the region under study appears to host multiple H \textsc{ii} regions, we infer that the absence of radial variation is probably because of the effects of seeing and that a considerable number of pixels with lower abundances in annuli A, B and C have suppressed signatures of over-abundance should they exist. This experiment shows that such segmentation analysis may not be a good tool to find localised abundance variation, which results in averaging out such local effects. 

We also performed a Lillefors test of normality for testing homogeneity on the maps of abundances and abundance ratios, which was inconclusive. Peforming the Lillefors test on the abundance and abundance ratio maps normalised by their error maps was also inconclusive. For example, the p-value of Lillefors test was greater than 0.05 for Ne abundance map indicating chemical homogeneity in spite of the map apparently showing some chemical inhomogeneity. Hence, we conclude that tests of chemical inhomogeneity must account for both the spatial information as well as the spaxel error. The current analysis also shows the power of IFS with good spatial-sampling, without which over-abundance in the Ne abundance map would only have appeared as noise. Furthermore, we need to consider the effects of seeing which tends to wash out small spatial scale variation in the abundance maps.

\subsection{Stellar Properties}
\label{section:stellar}

\subsubsection{Age of stellar population}
\label{section: age}

\begin{figure}
	\centering
	\includegraphics[width=0.48\textwidth]{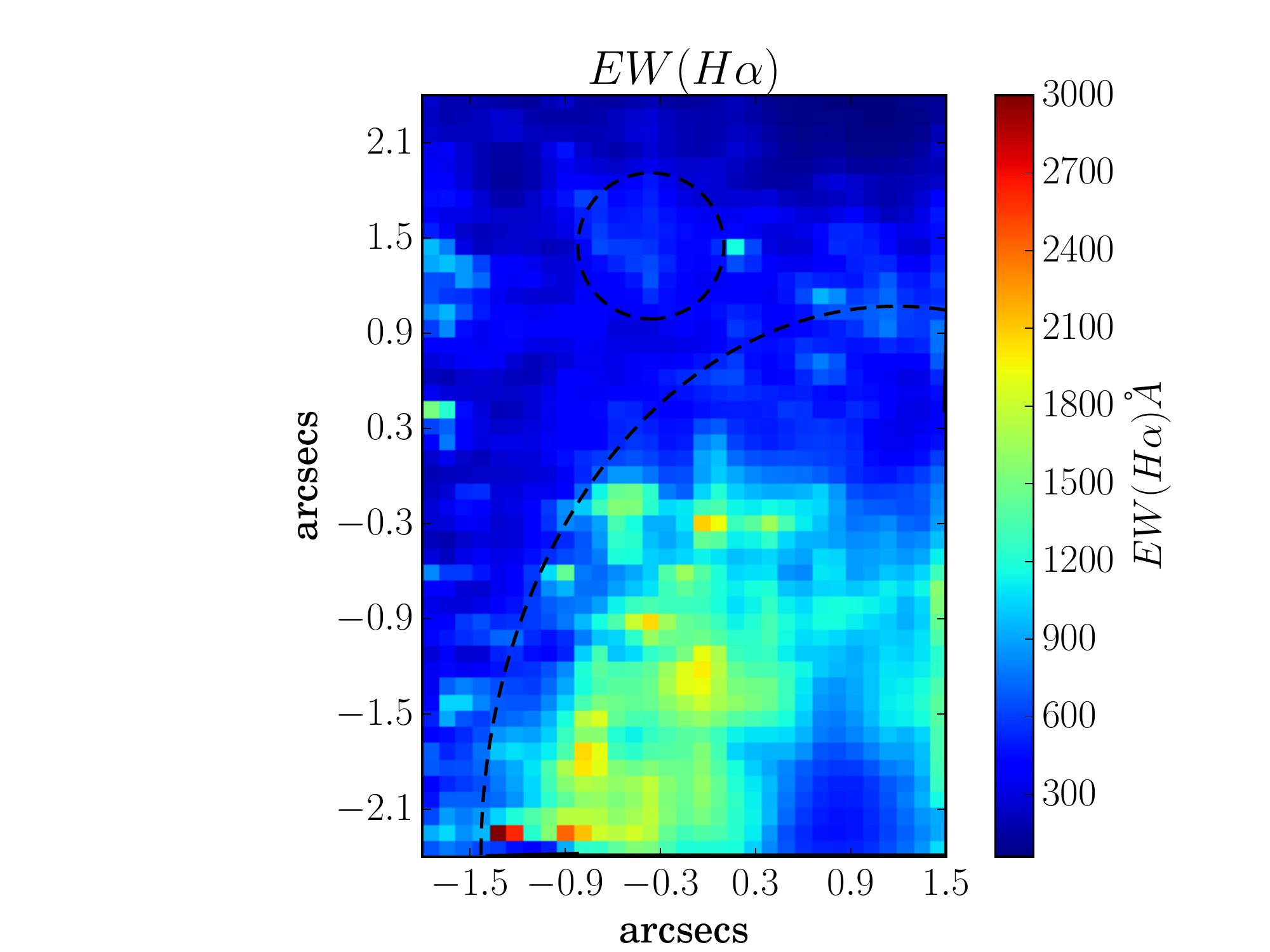}
	\caption{Map of the equivalent width of H$\alpha$ (in \AA).  The dashed quarter ellipse and circle indicate Region 1 and Region 2, respectively.}
	\label{figure:EW}
\end{figure}

\begin{figure}
	\centering
	\includegraphics[width=0.48\textwidth]{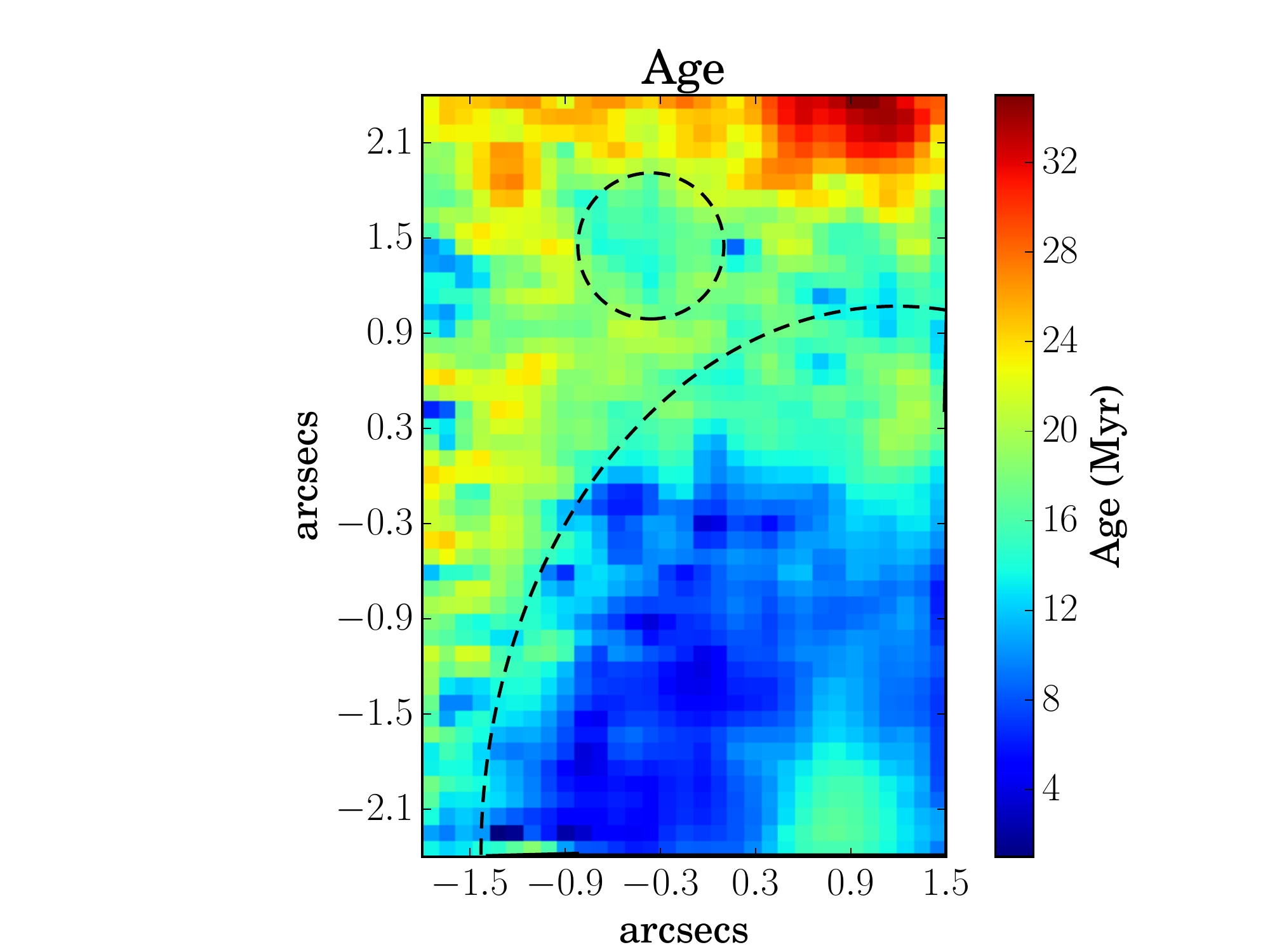}
	\caption{Age map in Myr estimated from Starburst99 models at a constant metallicity of Z = 0.1 Z$_{\odot}$ (mean of the metallicity map in Figure \ref{figure:elemental}, upper-left panel).  The dashed quarter ellipse and circle indicate Region 1 and Region 2, respectively.}
	\label{figure:Age}
\end{figure}

\begin{figure}
	\centering
	\includegraphics[width=0.48\textwidth]{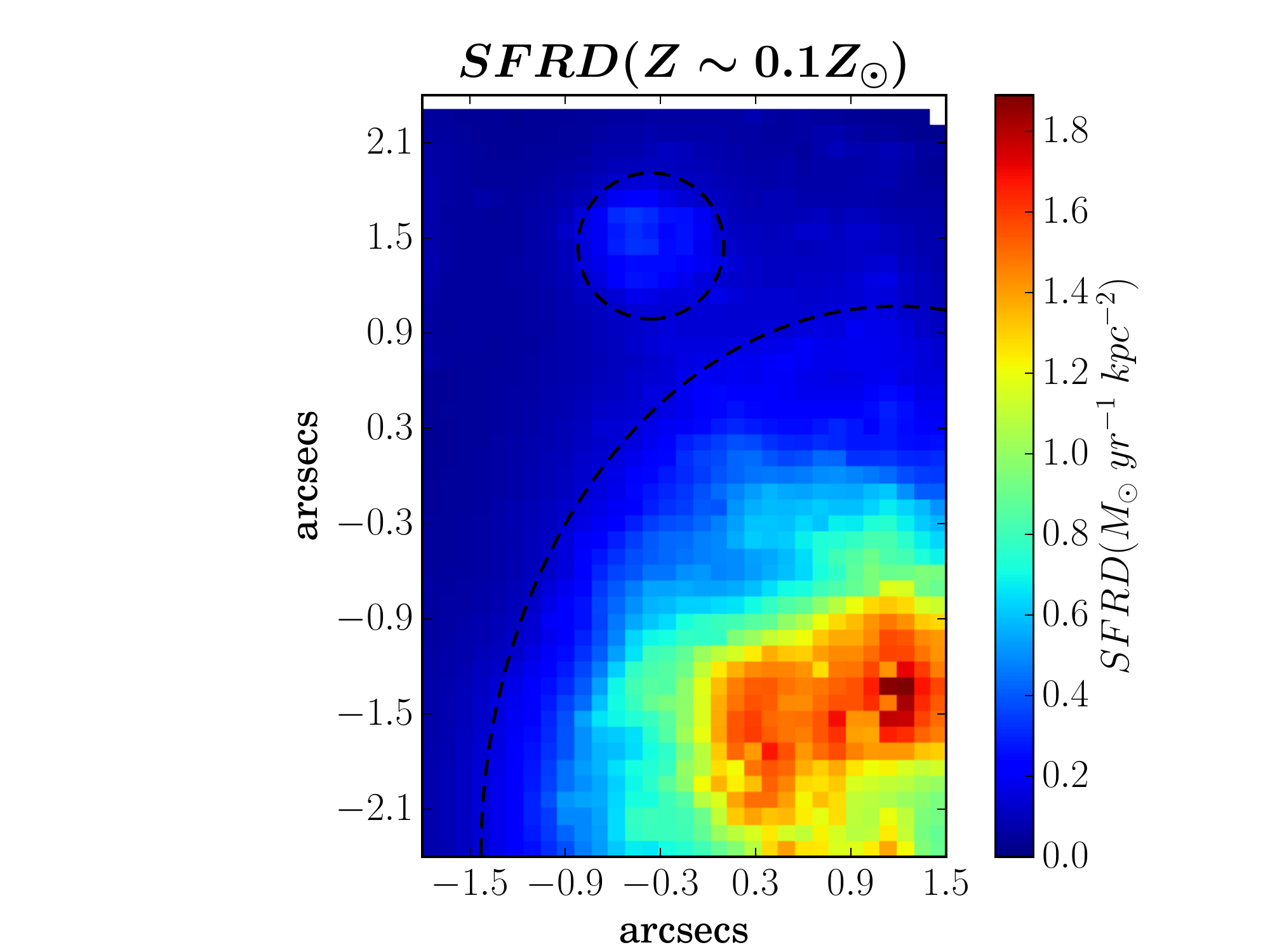}
	\caption{SFRD map assuming a constant metallicity of Z = 0.1 Z$_{\odot}$ (mean of the metallicity map in Figure \ref{figure:elemental}, upper-left panel).  The dashed quarter ellipse and circle indicate Region 1 and Region 2, respectively.}
	\label{figure:sfrd}
\end{figure}

\indent  The integrated spectrum of the FOV (Figure \ref{spectra}) show Balmer emission lines indicating the presence of young, hot and massive O and B stars embedded in gas, while we do not find any Balmer absorption lines showing that the region is mostly composed of younger stellar population. No Wolf-Rayet features are found in the integrated or spatially-resolved spectra of SBS 1415+437, in agreement with the results of \citet{Lopez2010}.

\indent For age-dating the current ionising population, we first map the equivalent width (EW) of the H$\alpha$ recombination line (Figure \ref{figure:EW}).This map shows a mean of $\sim$696 \AA ~and standard deviation of $\sim$440 ~\AA. From the integrated spectrum of Region 1, we find an equivalent width of $\sim$940 $\pm$ 10 \AA, which is in reasonable agreement with the value found by \tGuseva~(997.9 $\pm$ 1.7 \AA).

\indent Next, we calculate the EW(H$\alpha$) from the evolutionary synthesis models of \textsc{starburst99}. Note here that the \textsc{starburst99} models are used here considering that an individual spaxel might constitute an independent H \textsc{ii} region on the basis of the following argument. H$\alpha$ flux map (Figure \ref{fig:observed flux},
	middle-right panel) shows a variation of 4.0$\times$10$^{-17}$ to 1.2$\times$10$^{-15}$ erg s$^{-1}$ cm$^{-2}$,
	which corresponds to luminosities varying between 8.9$\times$10$^{35}$ and 2.7$\times$10$^{37}$
	erg s$^{-1}$ and ionising photons varying between 6.5$\times$10$^{47}$ and 2.0$\times$10$^{49}$
	photons s$^{-1}$. The ionising photons on a spaxel-by-spaxel basis lie
	within the range of ionising photons emitted by individual O and B
	stars, i.e. 10$^{46}$ -- 5$\times$10$^{49}$ photons s$^{-1}$ for dwarfs and 4$\times$
	10$^{47}$ -- 8$\times$10$^{49}$ photons s$^{-1}$ for supergiants \citep{Sternberg2003}. This shows that each spaxel might host O and B stars
	leading to H \textsc{ii} regions, though we can not discard the effects of
	a stochastic sampling of the initial mass function at such small spatial
	scales. The parameters and assumptions for running these models are described in detail in \citet{Kumari2017}, which include the assumptions of instantaneous starburst, Salpeter intial mass function, Geneva tracks with stellar rotation and expanding atmosphere models. The only difference here is on the assumption of metallicity, which is the mean of the metallicity map (Figure \ref{figure:elemental}, upper-left panel), i.e.  Z = 0.1Z$_{\odot}$. Note here that instead of assuming a constant metallicity for the entire FOV, we could use the metallicity map where metallicity varies from spaxel to spaxel. Howeover, the minimum metallicity of the spectra output by \textsc{Starburst99} is 0.05Z$_{\odot}$, whereas our metallicity map shows values $<$ 0.05Z$_{\odot}$. Moreover, the metallicity grid used by \textsc{Starburst99} is far too coarse to correctly reflect the pixel-to-pixel variation seen in our maps. Hence, with the current set of models, it is not possible to create an age map varying as a function of metallicity. By comparing the modelled EW from the evolutionary synthesis models with the observed EW(H$\alpha$), we obtain the age map (Figure \ref{figure:Age}) which shows a mean and standard deviation of 15 Myr and 6 Myr, respectively. We find the age of $\sim$ 10 Myr for Region 1 obtained from its integrated spectrum. Our results are in reasonable agreement with that of  \citet{Thuan1999}, who reports the presence of young stellar population ($\sim$ 5 Myr) along with stars as old as 100 Myr in Region 1. \tGuseva~ estimated an age of 4 Myr for Region 1 from the galactic evolution code PEGASE, which is lower than our value of $\sim$ 10 Myr for two reasons. Firstly the metallicity assumed by \tGuseva~ is 0.05 Z$_{\odot}$ compared to our metallicity assumption of 0.1 Z$_{\odot}$. Secondly, our modelling does not take into account nebular continuum or dust-extinction which will lead to systematic uncertainties in the estimated age \citep{Cantin2010, Perez-Montero2007b}. Note also that \textsc{Starburst99} does not take into account the binary
	population, even though $\sim$ 50\% of stars are found in
	binaries. Ignoring the binary population is another source of
	systematic uncertainty in the determined age.
 These age estimates should be further interpreted with caution as suggested by \citet{Aloisi2005}, who report the presence of stars older than $\sim$ 1.3 Gyr (e.g. red giant branch stars) via a photometric analysis involving colour-magnitude diagrams. 
 
\subsubsection{Star Formation Rate}
\label{section:sfr}
\indent We estimated the star-formation rate (SFR) from the dereddened H$\alpha$ luminosity assuming solar and sub-solar metallicity by using the recipes of \citet{Ly2016} and \citet{Kennicutt1998}, respectively assuming a Chabrier IMF. We normalise the SFR map by the area of each pixel to obtain the SFR density (SFRD) map. 
The SFRD map created assuming sub-solar metallicity (Figure \ref{figure:sfrd}) shows a variation of 0.03--1.89 M$_{\odot}$ yr$^{-1}$ kpc$^{-2}$. The SFRD map created assuming solar metallicity shows a variation of 0.06--3.79 M$_{\odot}$ yr$^{-1}$ kpc$^{-2}$ and has the same appearance as the map at sub-solar metallicity. SFR values on pixel-by-pixel basis should be interpreted with caution because of the failure of SFR recipes due to the stochastic sampling of the IMF at such small spatial scales. For Region 1, we find a SFR of $\sim$ 22.0 $ \pm$ 0.3 $\times$ 10$^{-3}$ M$_{\odot}$ yr$^{-1}$ assuming metallicity of 12 + log(O/H) = 7.63 , and $\sim$ 44.9 $\pm$ 0.5 $\times$ 10$^{-3}$ M$_{\odot}$ yr$^{-1}$ assuming solar metallicity. For Region 1, we estimate a SFRD $\sim$ 0.7 M$_{\odot}$ yr$^{-1}$ kpc$^{-2}$  from our data assuming a metallicity of 12+log(O/H) = 7.63. For comparison, if we were to use the MMT spectroscopic observation of \citet{Thuan1999} extracted over a region of 1.5 arcsec $\times$ 5 arcsec, we estimate a SFRD $\sim$ 0.5  M$_{\odot}$ yr$^{-1}$ kpc$^{-2}$ assuming 12 + log(O/H) = 7.60 \citep{Thuan1999}.  Our value of SFRD agrees with that of \citet{Thuan1999} within a factor of 2. From the H$\alpha$ flux value of \tGuseva~, we estimate a SFRD $\sim$ 0.17 M$_{\odot}$ yr$^{-1}$ kpc$^{-2}$ assuming 12 + log(O/H) = 7.61 \pGuseva over a region of 2 arcsec $\times$ 4.6 arcsec. The low value of SFRD obtained from the data of \tGuseva~ is due to their estimate of c(H$\beta$) = 0.00, which results in a lower value of dereddened H$\alpha$ flux. It is also possible that the GMOS-FOV covers brighter star-forming regions than the long-slit data of \tGuseva.

\indent 

\section{Summary \& Conclusion}
\label{section:summary}
\indent  Using GMOS-N IFS data, we carried out a spatially-resolved analysis of the ionised gas in a star-forming region within the BCD SBS 1415+437 at scales of $\sim$6.5 pc. What follows is a summary of our main results.

\begin{enumerate}
	\item The radial velocity of the ionised gas varies between $-$12  to 18 km s$^{-1}$, with no particular axis of rotation in the central region. The velocity dispersion varies between 20--70 km s$^{-1}$. The gas is predominantly photoionised as inferred from the emission line ratio diagnostic diagrams.

	\item The IFS data allows us to map the weak auroral line [O \textsc{iii}] $\lambda$4363 across a region of 143 $\times$ 143 pc$^2$ thereby enabling us to map the electron temperature T$_e$([O \textsc{iii}]) and density N$_e$([S \textsc{ii}]) in this region. This allows us to use the direct T$_e$-method to map the ionic and elemental abundances of various elements.
	
	
	\item  We map the ionic and elemental abundances of O, N, Ne, Ar and S, and also the abundance ratios, N/O, Ne/O, S/O and Ar/O. We also estimate these observables from the integrated spectrum of the main emission region (Region 1), which are in reasonable agreement with the median of corresponding values in the maps. The oxygen abundance from our IFS data is in agreement with that obtained from the long-slit spectroscopy, i.e. 12 + log(O/H) = 7.63 $\pm$ 0.03.
	

	\item  The age map of the region under study shows a mean and standard deviation of 15 Myr and 6Myr, respectively, as inferred from a comparison of the EW(H$\alpha$) map to the values determined from the \textsc{starburst99} models at a uniform metallicity of 0.1 Z$_{\odot}$. The SFRD is found to vary between 0.03--1.89 M$_{\odot}$ yr$^{-1}$ kpc$^{-2}$ across the region of study. The SFR of Region 1 is estimated to be $\sim$ 22.0 $\pm$ 0.3 $\times$ 10$^{-3}$ M$_{\odot}$yr$^{-1}$ assuming a metallicity of 12 + log(O/H) = 7.63.  
	
	\item We performed a radial profile analysis on the maps of chemical abundances and their ratios, where we chose elliptical annuli of equal widths following the H$\alpha$ flux distribution in the FOV.  No significant radial variation was found in either elemental abundance maps or the abundance ratio maps, despite the map of Ne/H exhibiting signatures of chemical inhomogeneity
	   
\end{enumerate}
\indent

\indent In conclusion, this study not only answers questions related to kinematics, ionisation conditions, chemical variations and stellar properties of SBS 1415+437 stated in the beginning of the paper, but also poses further questions on the most effective way to study chemical variations at small scales from IFS studies. We found that the radial profile analysis does not show significant chemical variation despite local enhancements in the neon abundance map. In the current study, seeing prevented us from studying such variations at spatial scales exploitable at the high spatial-sampling of GMOS-IFU. There will not be such problem with the implementation of adaptive optics in the current IFS instruments, e.g. MUSE on the Very Large Telescope, and space-based IFUs, e.g. the Near-Infrared Spectrograph on the James Webb Space Telescope (JWST) and Wide Field Infrared Survey Telescope. Above analysis shows that a test of chemical inhomogeneity must take into account various factors such as spaxel uncertainty, spatial information and the effects of seeing, which we will explore on a larger sample of galaxies in an upcoming work. From this study, we learnt that the abundances estimated from the integrated spectrum of a star-forming region is in reasonable agreement with the average value derived from the abundance maps across the star-forming region. This consistency adds confidence to our chemical abundance measurements of star-forming systems in the high-redshift Universe, where no spatial resolution is available, allowing us to accurately study the chemical evolution across different star-forming epochs. With the long-awaited JWST, we will be able to detect the distant and hence faint star-forming systems and study the chemical evolution of Universe with relatively less biases.

\section*{Acknowledgements}

\indent We thank Elisabeth Stanway and Paul Hewett for useful comments on an earlier draft of this paper. We also thank the referee {\'A} D\'{\i}az for useful comments on determining age of stellar populations. NK thanks the Institute of Astronomy, Cambridge and the Nehru Trust for Cambridge University for the financial support during her PhD, and the Schlumberger Foundation for funding her post-doctoral research. BLJ thanks support from the European Space Agency (ESA).  This research made use of the NASA/IPAC Extragalactic Database (NED) which is operated by the Jet Propulsion Laboratory, California Institute of Technology, under contract with the National Aeronautics and Space Administration; SAOImage DS9, developed by Smithsonian Astrophysical Observatory"; Astropy, a community-developed core Python package for Astronomy \citep{Astropy2013}. Based on observations obtained at the Gemini Observatory (processed using the Gemini IRAF package), which is operated by the Association of Universities for Research in Astronomy, Inc., under a cooperative agreement with the NSF on behalf of the Gemini partnership: the National Science Foundation (United States), the National Research Council (Canada), CONICYT (Chile), Ministerio de Ciencia, Tecnolog\'{i}a e Innovaci\'{o}n Productiva (Argentina), and Minist\'{e}rio da Ci\^{e}ncia, Tecnologia e Inova\c{c}\~{a}o (Brazil).



\bibliographystyle{mnras}
\bibliography{biblio} 




\section*{Supporting Information}

Figures and Tables presented in appendices are available as supplementary material at $MNRAS$ online.

\appendix

\section{Uncertainty maps}
\begin{figure*}
	\centering
	\includegraphics[width=0.45\textwidth]{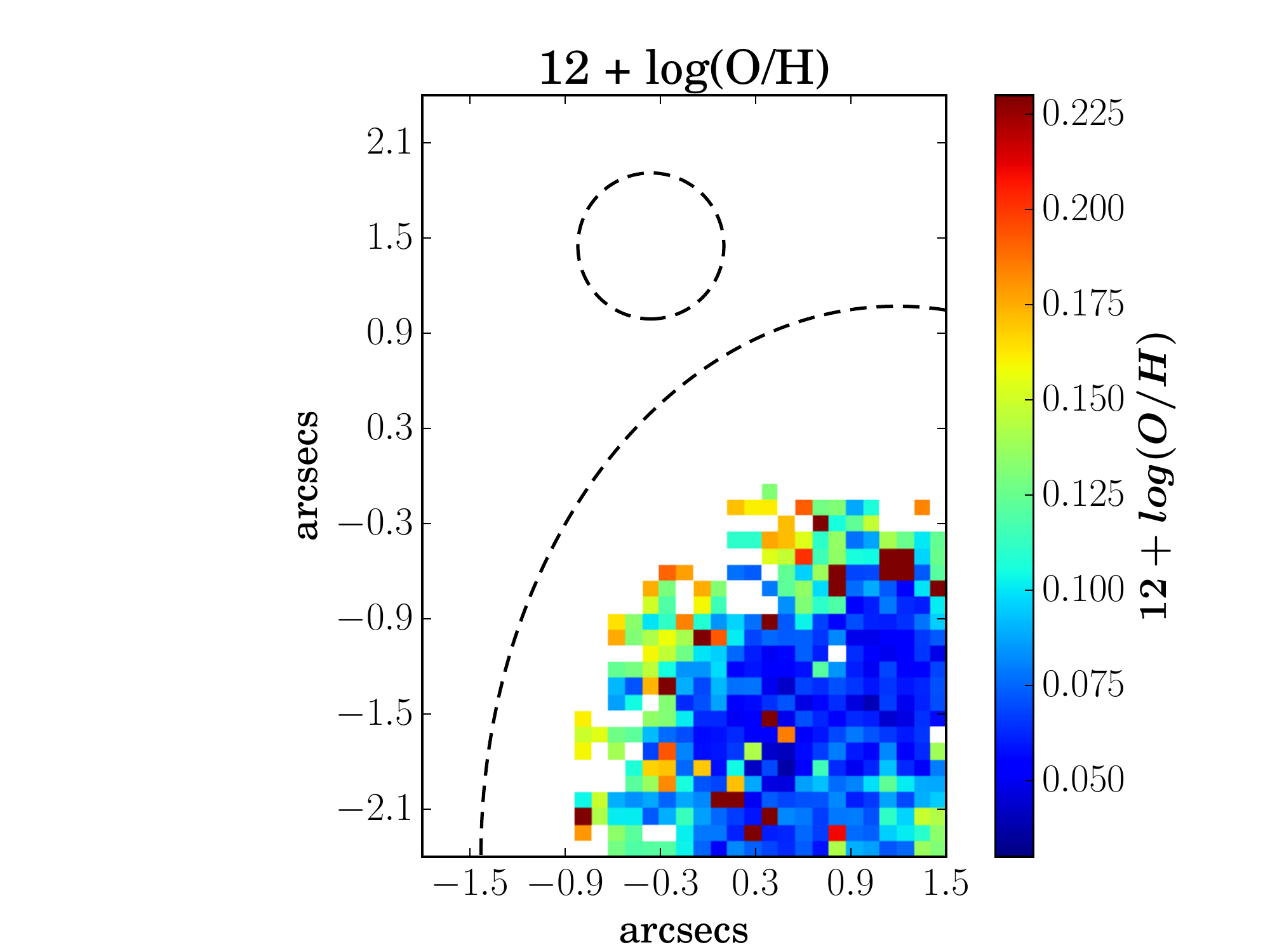}
	\includegraphics[width=0.45\textwidth]{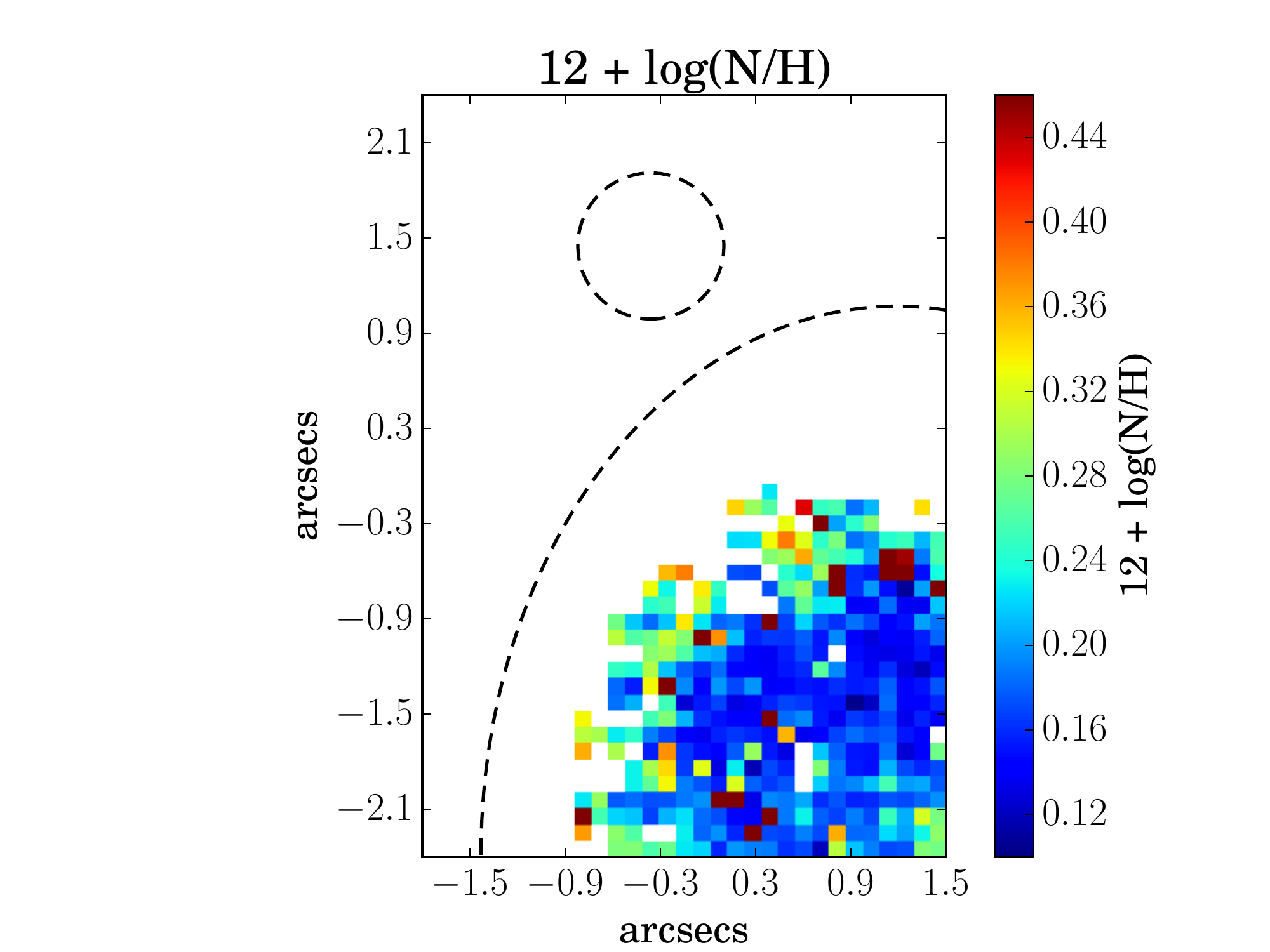}
	\includegraphics[width=0.45\textwidth]{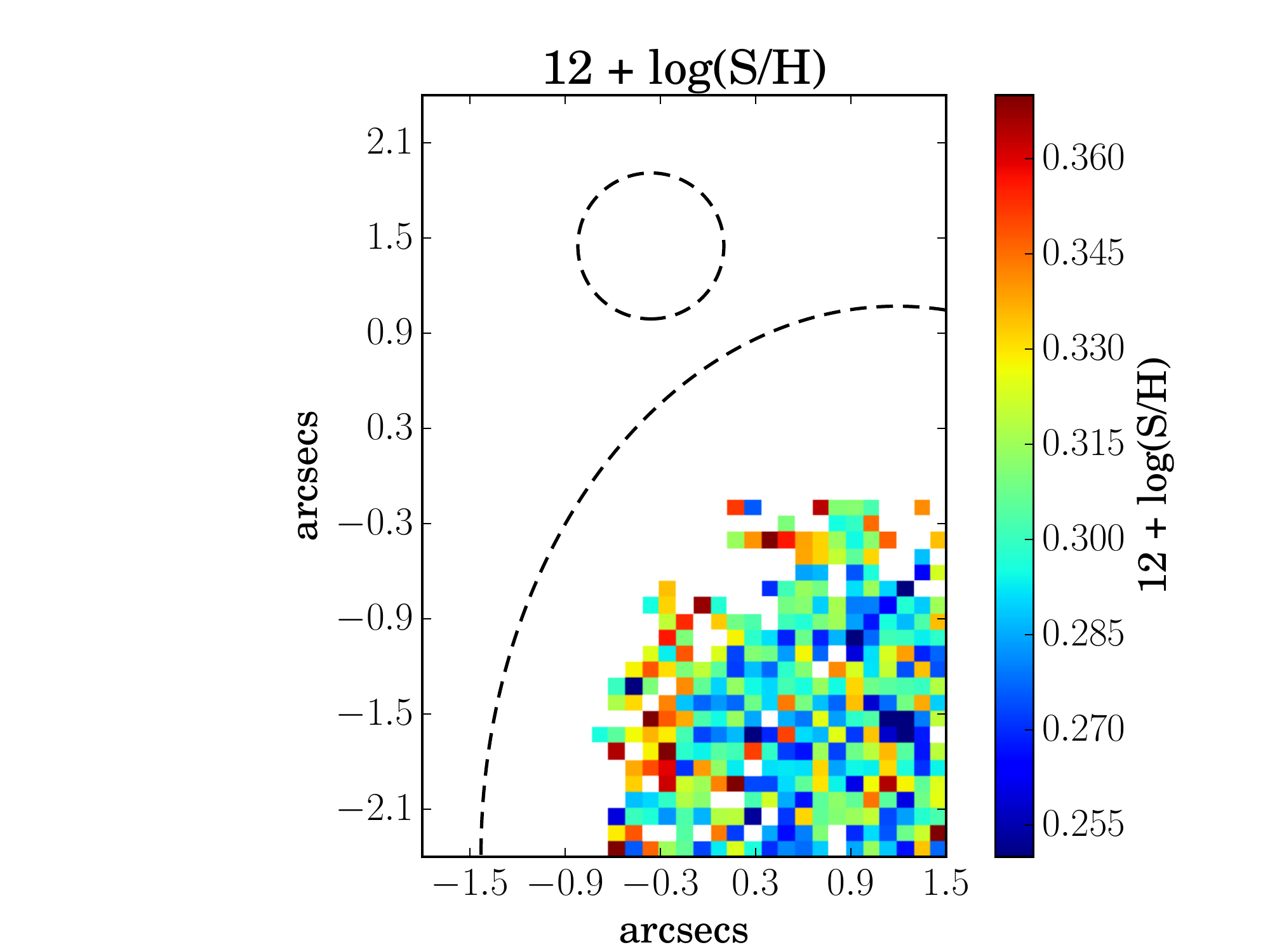}
	\includegraphics[width=0.45\textwidth]{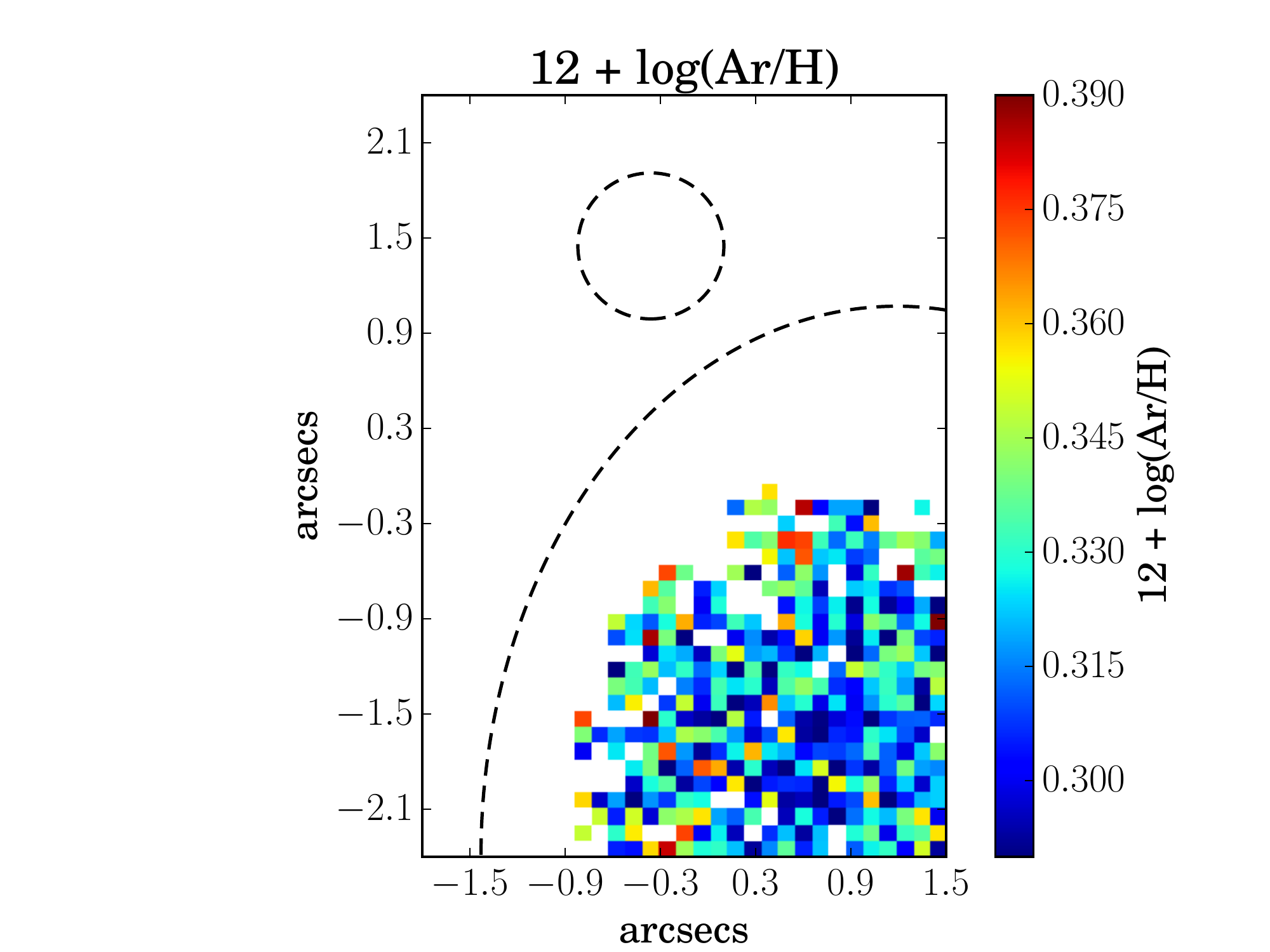}
	\includegraphics[width=0.45\textwidth]{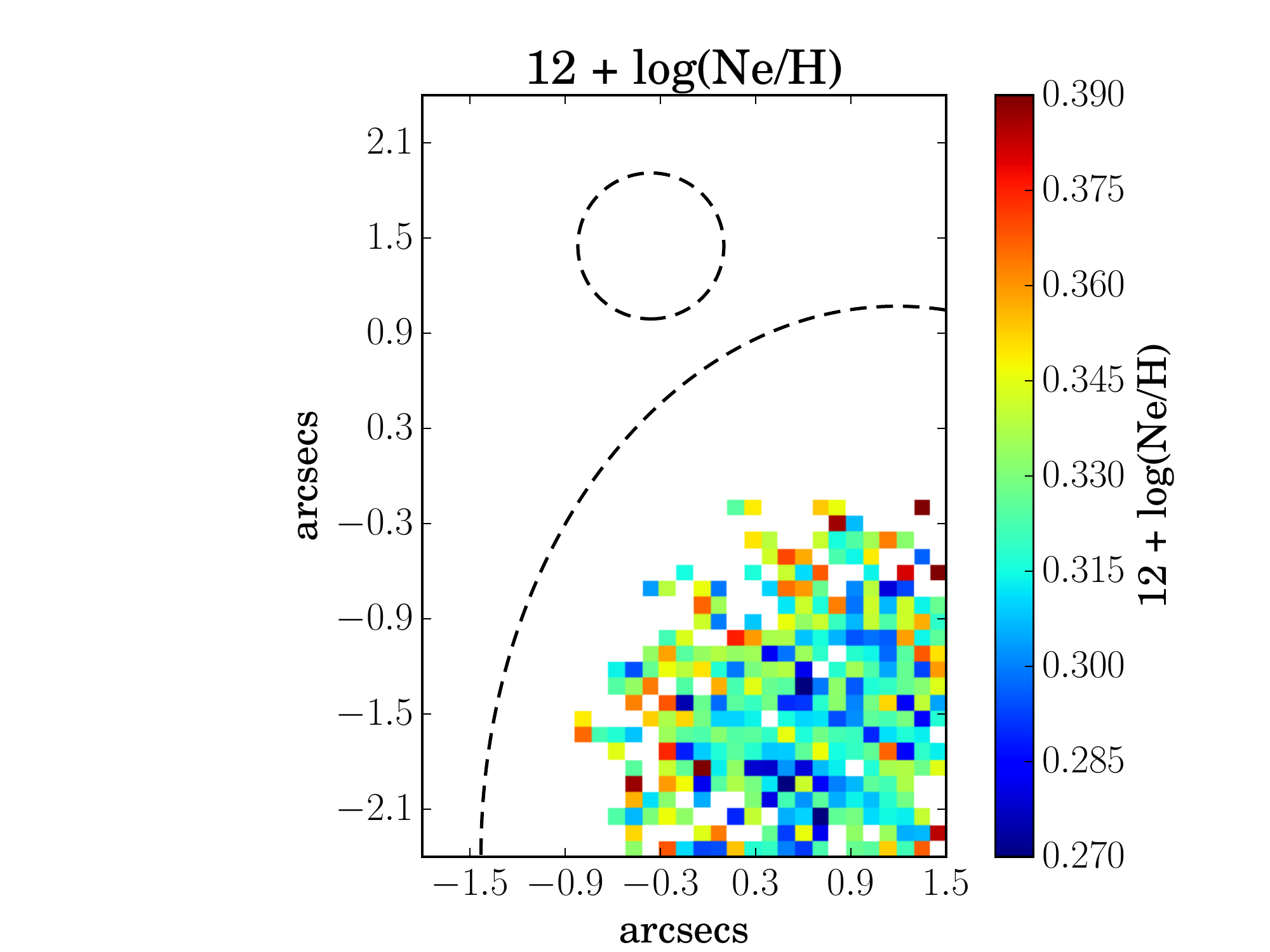}
	\caption{Uncertainty maps for the abundance maps presented in Figure \ref{figure:element variation}.}
	\label{fig:uncertainty Z}
\end{figure*}

\begin{figure*}
	\centering
	\includegraphics[width=0.45\textwidth]{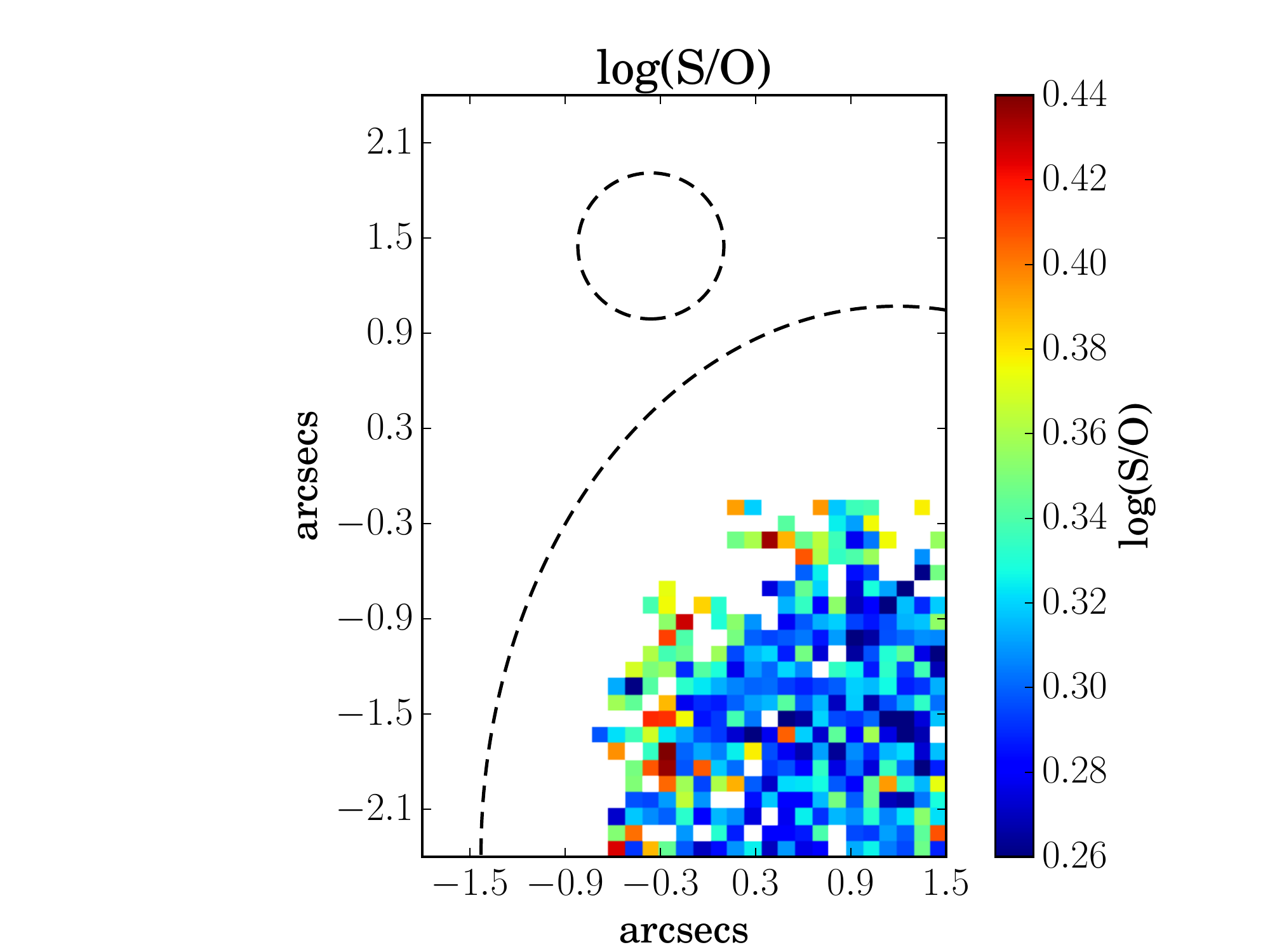}
	\includegraphics[width=0.45\textwidth]{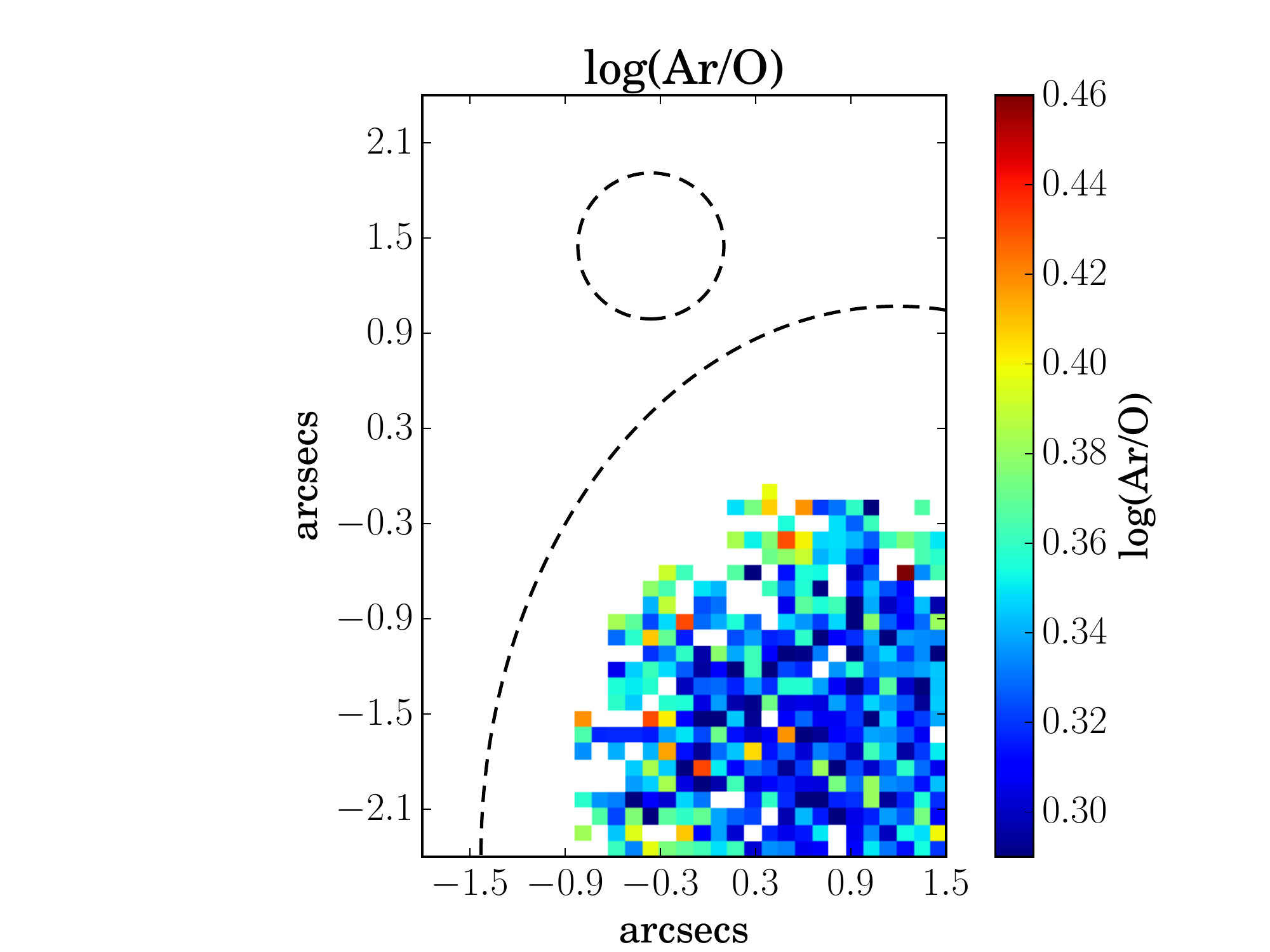}
	\includegraphics[width=0.45\textwidth]{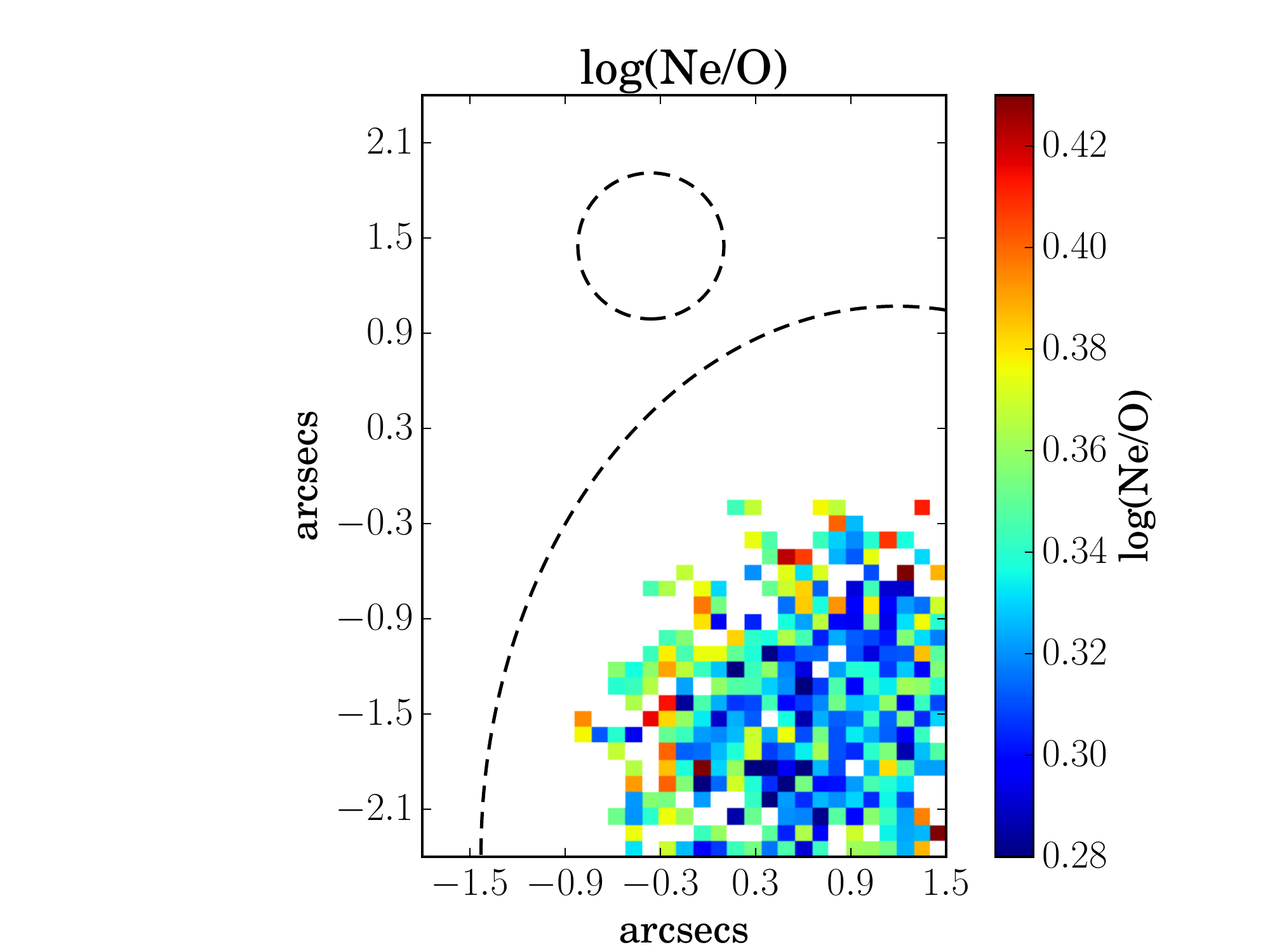}
	\includegraphics[width=0.45\textwidth]{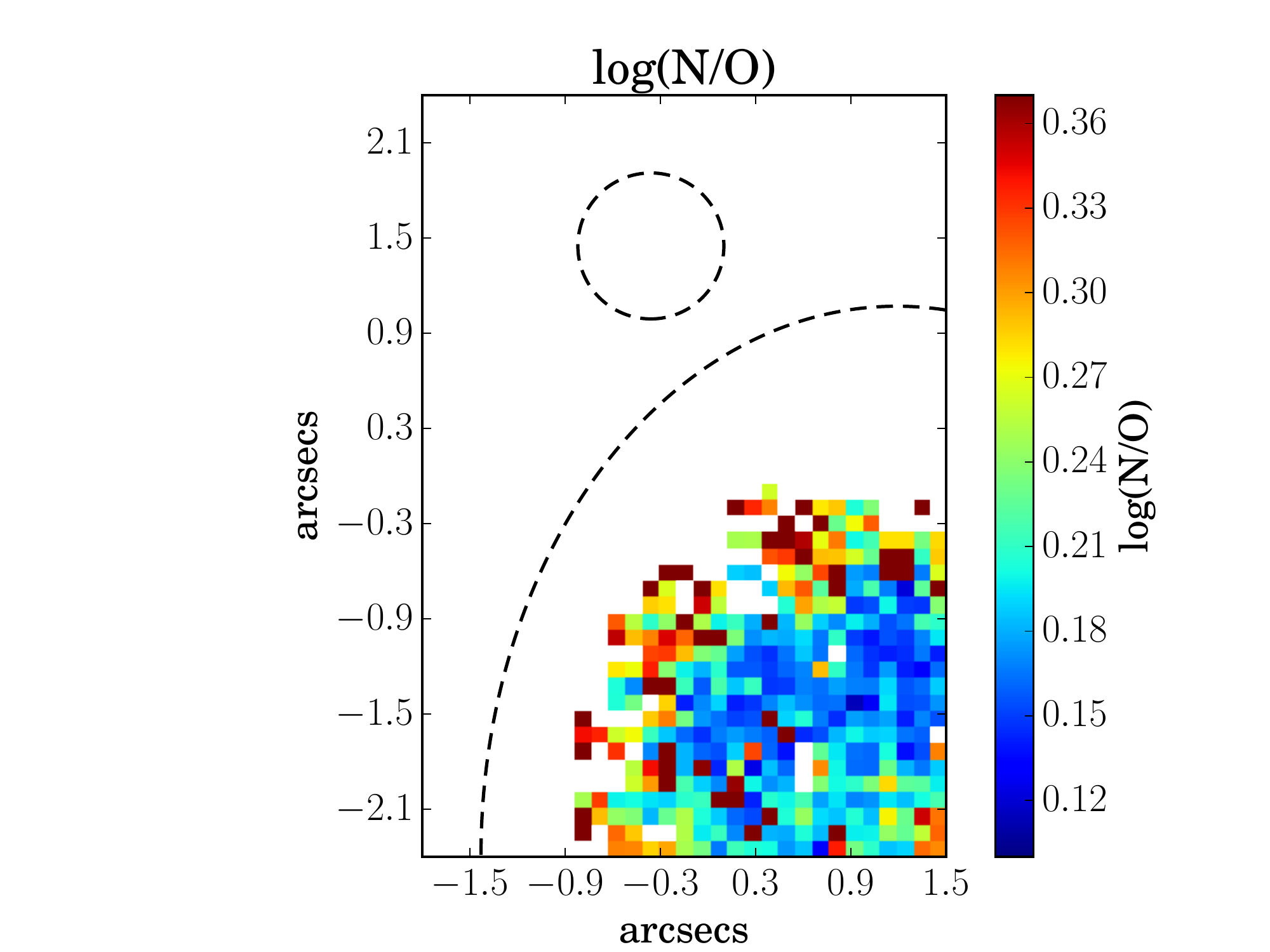}
		\caption{Uncertainty maps for the abundance ratio maps presented in Figure \ref{figure:ratio variation}.}
	\label{fig:uncertainty Z/O}
\end{figure*}

\indent Figures \ref{fig:uncertainty Z} \& \ref{fig:uncertainty Z/O} present the uncertainty maps for abundance maps (Figure \ref{figure:element variation}) and abundance ratio maps (Figure \ref{figure:ratio variation}), respectively. These maps have been obtained via Monte-Carlo simulations.

\section{Flux and abundance tables corresponding to each annulus}
\indent  Table \ref{table: annuli flux} presents the observed and intrinsic flux values from integrated spectra of each annulus described in Section \ref{section:variation abundances}. The abundances and abundance ratios within each annulus are presented in Table \ref{table:annuli abundances}.

\begin{landscape}

	\begin{table}
	\centering
	\caption[Emission line measurements from the integrated spectrum of each annulus]{Emission line measurements (relative to H$\beta$ = 100) for the integrated spectrum of each annulus shown in Figure \ref{figure:segments}. Line fluxes (F$_{\lambda}$) are extinction corrected using E(B-V) to calculate F$_{\lambda}$.}
	\label{table: annuli flux}
	\resizebox{\columnwidth}{!}{%
		\begin{tabular}{cccccccccccccc}
			\toprule
			Line & $\lambda_{air}$ & $F(A)_{\lambda}$ & $I(A)_{\lambda}$ & $F(B)_{\lambda}$ & $I(B)_{\lambda}$ & $F(C)_{\lambda}$ & $I(C)_{\lambda}$ & $F(D)_{\lambda}$ & $I(D)_{\lambda}$ & $F(E)_{\lambda}$ & $I(E)_{\lambda}$ & $F(F)_{\lambda}$ & $I(F)_{\lambda}$ \\
			\midrule
			
			[NIII] & 3868.76 & $27.11 \pm 1.75 $ & $28.82 \pm 2.04 $ & $26.63 \pm 1.23 $ & $28.18 \pm 1.52 $ & $24.13 \pm 1.50 $ & $25.55 \pm 1.72 $ & $26.22 \pm 1.88 $ & $27.34 \pm 2.09 $ & $23.62 \pm 2.74 $ & $24.94 \pm 3.01 $ & $19.29 \pm 4.12 $ & $19.98 \pm 4.32 $ \\
			$H\gamma$ & 4340.47 & $44.05 \pm 0.56 $ & $45.42 \pm 1.36 $ & $44.79 \pm 0.55 $ & $46.07 \pm 1.34 $ & $44.50 \pm 0.58 $ & $45.78 \pm 1.27 $ & $45.53 \pm 0.76 $ & $46.48 \pm 1.40 $ & $46.47 \pm 0.85 $ & $47.75 \pm 1.71 $ & $47.81 \pm 1.04 $ & $48.65 \pm 1.94 $ \\
			$[OIII]$ & 4363.21 & $9.38 \pm 0.30 $ & $9.66 \pm 0.40 $ & $9.46 \pm 0.44 $ & $9.72 \pm 0.52 $ & $9.21 \pm 0.31 $ & $9.46 \pm 0.40 $ & $8.53 \pm 0.41 $ & $8.70 \pm 0.48 $ & $8.58 \pm 0.86 $ & $8.81 \pm 0.93 $ & $6.71 \pm 1.17 $ & $6.82 \pm 1.22 $ \\
			$H\beta$ & 4861.33 & $100.00 \pm 0.49 $ & $100.00 \pm 1.86 $ & $100.00 \pm 0.48 $ & $100.00 \pm 1.81 $ & $100.00 \pm 0.43 $ & $100.00 \pm 1.67 $ & $100.00 \pm 0.47 $ & $100.00 \pm 1.72 $ & $100.00 \pm 0.58 $ & $100.00 \pm 2.11 $ & $100.00 \pm 0.65 $ & $100.00 \pm 2.31 $ \\
			$[OIII]$ & 4958.92 & $130.21 \pm 0.95 $ & $129.57 \pm 3.38 $ & $129.92 \pm 0.90 $ & $129.32 \pm 3.29 $ & $126.28 \pm 0.83 $ & $125.70 \pm 2.96 $ & $118.76 \pm 0.74 $ & $118.36 \pm 2.85 $ & $111.56 \pm 0.87 $ & $111.07 \pm 3.27 $ & $100.63 \pm 0.83 $ & $100.34 \pm 3.22 $ \\
			$[OIII]$ & 5006.84 & $388.09 \pm 2.75 $ & $385.12 \pm 9.97 $ & $385.75 \pm 2.64 $ & $383.02 \pm 9.68 $ & $374.54 \pm 2.35 $ & $371.88 \pm 8.68 $ & $353.61 \pm 2.30 $ & $351.76 \pm 8.43 $ & $330.75 \pm 2.41 $ & $328.50 \pm 9.57 $ & $300.52 \pm 2.35 $ & $299.20 \pm 9.50 $ \\
			$HeI$ & 5875.67 & $10.67 \pm 0.17 $ & $10.18 \pm 0.28 $ & $10.78 \pm 0.15 $ & $10.32 \pm 0.27 $ & $10.64 \pm 0.18 $ & $10.19 \pm 0.27 $ & $10.67 \pm 0.20 $ & $10.33 \pm 0.29 $ & $10.65 \pm 0.23 $ & $10.21 \pm 0.35 $ & $10.40 \pm 0.30 $ & $10.12 \pm 0.41 $ \\
			$[OI]$ & 6300.3 & $1.44 \pm 0.07 $ & $1.36 \pm 0.07 $ & $1.36 \pm 0.05 $ & $1.28 \pm 0.05 $ & $1.74 \pm 0.06 $ & $1.64 \pm 0.06 $ & $2.09 \pm 0.08 $ & $2.00 \pm 0.09 $ & $2.77 \pm 0.10 $ & $2.62 \pm 0.12 $ & $3.35 \pm 0.16 $ & $3.23 \pm 0.18 $ \\
			$[SIII]$ & 6312.1 & $1.54 \pm 0.07 $ & $1.45 \pm 0.08 $ & $1.47 \pm 0.05 $ & $1.39 \pm 0.06 $ & $1.43 \pm 0.05 $ & $1.35 \pm 0.06 $ & $1.51 \pm 0.07 $ & $1.45 \pm 0.07 $ & $1.51 \pm 0.07 $ & $1.42 \pm 0.08 $ & $1.25 \pm 0.10 $ & $1.20 \pm 0.11 $ \\
			$[NII]$ & 6548.03 & $0.81 \pm 0.45 $ & $0.76 \pm 0.42 $ & $0.90 \pm 0.44 $ & $0.84 \pm 0.42 $ & $1.04 \pm 0.43 $ & $0.98 \pm 0.40 $ & $1.07 \pm 0.37 $ & $1.02 \pm 0.36 $ & $1.20 \pm 0.41 $ & $1.13 \pm 0.39 $ & $1.58 \pm 0.37 $ & $1.51 \pm 0.36 $ \\
			$H\alpha$ & 6562.8 & $307.16 \pm 1.73 $ & $286.00 \pm 6.42 $ & $305.53 \pm 1.68 $ & $286.00 \pm 6.27 $ & $305.63 \pm 1.55 $ & $286.00 \pm 5.79 $ & $300.28 \pm 1.56 $ & $286.00 \pm 5.94 $ & $304.73 \pm 1.94 $ & $286.00 \pm 7.27 $ & $297.93 \pm 2.07 $ & $286.00 \pm 7.93 $ \\
			$[NII]$ & 6583.41 & $2.14 \pm 0.45 $ & $1.99 \pm 0.42 $ & $2.23 \pm 0.44 $ & $2.08 \pm 0.42 $ & $2.62 \pm 0.43 $ & $2.45 \pm 0.40 $ & $2.99 \pm 0.37 $ & $2.85 \pm 0.36 $ & $3.82 \pm 0.41 $ & $3.59 \pm 0.40 $ & $3.95 \pm 0.37 $ & $3.79 \pm 0.37 $ \\
			$HeI$ & 6678.15 & $2.92 \pm 0.07 $ & $2.71 \pm 0.08 $ & $2.87 \pm 0.05 $ & $2.68 \pm 0.07 $ & $2.88 \pm 0.06 $ & $2.69 \pm 0.08 $ & $2.87 \pm 0.07 $ & $2.73 \pm 0.09 $ & $2.85 \pm 0.09 $ & $2.66 \pm 0.10 $ & $2.76 \pm 0.15 $ & $2.64 \pm 0.16 $ \\
			$[SII]$ & 6716.47 & $6.05 \pm 0.09 $ & $5.61 \pm 0.15 $ & $6.13 \pm 0.07 $ & $5.71 \pm 0.14 $ & $7.21 \pm 0.07 $ & $6.72 \pm 0.15 $ & $8.82 \pm 0.10 $ & $8.38 \pm 0.19 $ & $11.04 \pm 0.12 $ & $10.32 \pm 0.28 $ & $12.79 \pm 0.17 $ & $12.25 \pm 0.36 $ \\
			$[SII]$ & 6730.85 & $4.42 \pm 0.09 $ & $4.09 \pm 0.12 $ & $4.57 \pm 0.07 $ & $4.25 \pm 0.11 $ & $5.15 \pm 0.07 $ & $4.79 \pm 0.11 $ & $6.11 \pm 0.09 $ & $5.80 \pm 0.15 $ & $7.79 \pm 0.11 $ & $7.27 \pm 0.21 $ & $9.12 \pm 0.16 $ & $8.72 \pm 0.28 $ \\
			$[ArIII]$ & 7135.8 & $4.11 \pm 0.07 $ & $3.76 \pm 0.10 $ & $4.11 \pm 0.05 $ & $3.79 \pm 0.09 $ & $4.09 \pm 0.09 $ & $3.77 \pm 0.11 $ & $4.16 \pm 0.07 $ & $3.91 \pm 0.10 $ & $4.32 \pm 0.08 $ & $3.99 \pm 0.12 $ & $4.13 \pm 0.11 $ & $3.92 \pm 0.15 $ \\
			$[OII]$ & 7318.92 & $1.31 \pm 0.07 $ & $1.19 \pm 0.07 $ & $1.31 \pm 0.08 $ & $1.20 \pm 0.07 $ & $1.49 \pm 0.08 $ & $1.36 \pm 0.08 $ & $1.66 \pm 0.10 $ & $1.56 \pm 0.09 $ & $2.01 \pm 0.12 $ & $1.85 \pm 0.12 $ & $2.18 \pm 0.19 $ & $2.06 \pm 0.18 $ \\
			$[OII]$ & 7329.66 & $1.11 \pm 0.07 $ & $1.01 \pm 0.06 $ & $1.10 \pm 0.07 $ & $1.01 \pm 0.07 $ & $1.17 \pm 0.07 $ & $1.08 \pm 0.07 $ & $1.31 \pm 0.08 $ & $1.22 \pm 0.08 $ & $1.63 \pm 0.11 $ & $1.50 \pm 0.11 $ & $1.74 \pm 0.17 $ & $1.65 \pm 0.16 $ \\
			E(B-V) &&$0.069 \pm 0.005$& &$0.063 \pm 0.005$&&$0.064 \pm 0.005$&&$0.047 \pm 0.005$&&$0.061 \pm 0.006$ &&$0.039 \pm 0.007$\\ 
			F(H$\beta$)&&$12.61 \pm 0.06$& $15.83 \pm 0.29$&$ 28.6 \pm 0.14$&$35.25 \pm 0.64$& $26.83 \pm 0.12$& $33.14 \pm 0.57$& $17.07 \pm 0.08$&$19.94 \pm 0.34$& $13.50 \pm 0.08$& $16.53 \pm 0.35$& $10.26 \pm 0.07$& $11.68 \pm 0.27$\\ \bottomrule
		\end{tabular}%
	}
\end{table}
\clearpage
\begin{table}
	\centering
	\caption[Summary of nebular diagnostics, ionic abundances, elemental abundances and abundance ratios obtained from the integrated spectra of all annuli]{Summary of nebular diagnostics, ionic abundances, elemental abundances and abundance ratios obtained from the integrated spectra of annuli (A, B, C, D, E, F) shown in Figure \ref{figure:segments}.}
	\label{table:annuli abundances}
	\scalebox{0.9}{
		\begin{tabular}{ccccccc}
			\toprule
			Parameter & A & B & C & D & E & F \\
			\midrule
			Te([OIII]) ($X$ 10$^4$ K) & $1.69 \pm 0.04 $ & $1.70 \pm 0.04 $ & $1.70 \pm 0.04 $ & $1.67 \pm 0.05 $ & $1.75 \pm 0.11 $ & $1.60 \pm 0.15 $ \\
			Te([OII]) ($X$ 10$^4$ K) & $1.58 \pm 0.02 $ & $1.57 \pm 0.09 $ & $1.59 \pm 0.03 $ & $1.58 \pm 0.04 $ & $1.62 \pm 0.07 $ & $1.53 \pm 0.09 $ \\
			Te([NII]) ($X$ 10$^4$ K) & $1.41 \pm 0.01 $ & $1.41 \pm 0.02 $ & $1.41 \pm 0.01 $ & $1.40 \pm 0.02 $ & $1.43 \pm 0.04 $ & $1.38 \pm 0.06 $ \\
			Te([SIII]) ($X$ 10$^4$ K) & $1.69 \pm 0.04 $ & $1.70 \pm 0.05 $ & $1.70 \pm 0.05 $ & $1.67 \pm 0.06 $ & $1.76 \pm 0.13 $ & $1.59 \pm 0.19 $ \\
			Ne([SII]) (cm$^{-3}$)& $<$ 50 &$ 60 \pm 14 $& $<$ 50&$<$ 50&$<$ 50&$<$ 50\\
			12 + log(O$^+$/H$^+$) & $7.07 \pm 0.03 $ & $7.10 \pm 0.11 $ & $7.11 \pm 0.04 $ & $7.18 \pm 0.05 $ & $7.21 \pm 0.08 $ & $7.36 \pm 0.11 $ \\
			12 + log(O$^{2+}$/H$^+$) & $7.51 \pm 0.02 $ & $7.50 \pm 0.03 $ & $7.49 \pm 0.03 $ & $7.48 \pm 0.03 $ & $7.40 \pm 0.06 $ & $7.45 \pm 0.10 $ \\
			12 + log(O/H) & $7.64 \pm 0.02 $ & $7.64 \pm 0.04 $ & $7.64 \pm 0.02 $ & $7.65 \pm 0.03 $ & $7.62 \pm 0.05 $ & $7.71 \pm 0.07 $ \\
			
			12 + log(N+/H+) & $5.28 \pm 0.08 $ & $5.29 \pm 0.04 $ & $5.36 \pm 0.04 $ & $5.44 \pm 0.04 $ & $5.52 \pm 0.04 $ & $5.58 \pm 0.04 $ \\
			ICF (N$^{+}$) & $3.73 \pm 0.34 $ & $3.53 \pm 0.94 $ & $3.39 \pm 0.33 $ & $2.96 \pm 0.36 $ & $2.56 \pm 0.57 $ & $2.22 \pm 0.67 $ \\
			12 + log(N/H) & $5.85 \pm 0.08 $ & $5.84 \pm 0.12 $ & $5.89 \pm 0.06 $ & $5.91 \pm 0.07 $ & $5.93 \pm 0.11 $ & $5.92 \pm 0.14 $ \\
			log(N/O) & $-1.79 \pm 0.09 $ & $-1.80 \pm 0.12 $ & $-1.74 \pm 0.06 $ & $-1.75 \pm 0.07 $ & $-1.69 \pm 0.12 $ & $-1.78 \pm 0.16 $ \\
			S$^+$/H$^+$ ($X$ 10$^7$) & $0.93 \pm 0.03 $ & $0.97 \pm 0.10 $ & $1.10 \pm 0.04 $ & $1.37 \pm 0.06 $ & $1.62 \pm 0.13 $ & $2.14 \pm 0.23 $ \\
			S$^{2+}$/H$^+$ ($X$ 10$^7$) & $5.33 \pm 0.49 $ & $5.02 \pm 0.51 $ & $4.84 \pm 0.42 $ & $5.45 \pm 0.62 $ & $4.66 \pm 0.97 $ & $5.24 \pm 1.89 $ \\
			ICF (S$^+$ + S$^{2+}$) & $1.15 \pm 0.01 $ & $1.13 \pm 0.04 $ & $1.13 \pm 0.02 $ & $1.10 \pm 0.02 $ & $1.07 \pm 0.02 $ & $1.05 \pm 0.03 $ \\
			12 + log(S/H) & $5.86 \pm 0.03 $ & $5.83 \pm 0.04 $ & $5.83 \pm 0.03 $ & $5.87 \pm 0.04 $ & $5.83 \pm 0.07 $ & $5.89 \pm 0.12 $ \\
			log(S/O) & $-1.78 \pm 0.04 $ & $-1.81 \pm 0.06 $ & $-1.81 \pm 0.04 $ & $-1.78 \pm 0.05 $ & $-1.79 \pm 0.09 $ & $-1.82 \pm 0.13 $ \\
			Ne$^{2+}$/H$^+$ ($X$ 10$^5$) & $0.57 \pm 0.05 $ & $0.55 \pm 0.05 $ & $0.50 \pm 0.05 $ & $0.55 \pm 0.06 $ & $0.45 \pm 0.09 $ & $0.45 \pm 0.15 $ \\
			ICF (Ne$^{2+}$) & $1.09 \pm 0.00 $ & $1.09 \pm 0.01 $ & $1.10 \pm 0.00 $ & $1.10 \pm 0.01 $ & $1.12 \pm 0.02 $ & $1.14 \pm 0.04 $ \\
			12 + log(Ne/H) & $6.79 \pm 0.04 $ & $6.78 \pm 0.04 $ & $6.74 \pm 0.04 $ & $6.78 \pm 0.05 $ & $6.70 \pm 0.09 $ & $6.71 \pm 0.16 $ \\
			log(Ne/O) & $-0.85 \pm 0.04 $ & $-0.86 \pm 0.05 $ & $-0.90 \pm 0.05 $ & $-0.87 \pm 0.06 $ & $-0.92 \pm 0.10 $ & $-1.00 \pm 0.17 $ \\
			Ar$^{2+}$/H$^+$ ($X$ 10$^7$) & $1.24 \pm 0.06 $ & $1.23 \pm 0.07 $ & $1.23 \pm 0.06 $ & $1.31 \pm 0.08 $ & $1.23 \pm 0.14 $ & $1.43 \pm 0.30 $ \\
			ICF (Ar$^{2+}$) & $1.14 \pm 0.00 $ & $1.14 \pm 0.01 $ & $1.14 \pm 0.00 $ & $1.15 \pm 0.01 $ & $1.17 \pm 0.03 $ & $1.20 \pm 0.05 $ \\
			12 + log(Ar/H) & $5.15 \pm 0.02 $ & $5.15 \pm 0.03 $ & $5.15 \pm 0.02 $ & $5.18 \pm 0.03 $ & $5.16 \pm 0.05 $ & $5.23 \pm 0.09 $ \\
			log(Ar/O) & $-2.49 \pm 0.03 $ & $-2.49 \pm 0.05 $ & $-2.49 \pm 0.03 $ & $-2.47 \pm 0.04 $ & $-2.46 \pm 0.07 $ & $-2.48 \pm 0.12 $ \\
			\bottomrule
		\end{tabular}%
	}
\end{table}	
\end{landscape}


\bsp	
\label{lastpage}
\end{document}